\documentclass[a4paper,11pt]{article}
\usepackage{amsmath}
\usepackage{braket}
\usepackage{amssymb}
\usepackage{slashed}
\pdfoutput=1 

\usepackage{jheppub} 

\usepackage[T1]{fontenc} 

\newcommand{\LQCD}{\Lambda_{\rm QCD}}

\title{\boldmath The next-to-leading order Higgs impact factor in the infinite top-mass limit}

\author[a,b,c]{Francesco Giovanni Celiberto,}
\author[d,e,f,1]{Michael Fucilla,\note{Corresponding author.}}
\author[g]{Dmitry Yu. Ivanov,}
\author[d,e]{Mohammed M.A. Mohammed,}
\author[d,e]{Alessandro Papa}

\affiliation[a]{European Centre for Theoretical Studies in Nuclear Physics and Related Areas (ECT*), I-38123 Villazzano, Trento, Italy}
\affiliation[b]{Fondazione Bruno Kessler (FBK), I-38123 Povo, Trento, Italy}
\affiliation[c]{INFN-TIFPA Trento Institute of Fundamental Physics and Applications, I-38123 Povo, Trento, Italy}
\affiliation[d]{Dipartimento di Fisica, Universit\`a della Calabria, I-87036 Arcavacata di Rende, Cosenza, Italy}
\affiliation[e]{Istituto Nazionale di Fisica Nucleare, Gruppo collegato di Cosenza, I-87036 Arcavacata di Rende, Cosenza, Italy}
\affiliation[f]{Université Paris-Saclay, CNRS, IJCLab, 91405 Orsay, France}
\affiliation[g]{Sobolev Institute of Mathematics, 630090 Novosibirsk, Russia}

\emailAdd{fceliberto@ectstar.eu}
\emailAdd{michael.fucilla@unical.it}
\emailAdd{d-ivanov@math.nsc.ru}
\emailAdd{mohammed.maher@unical.it}
\emailAdd{alessandro.papa@fis.unical.it}

\abstract{We calculate the next-to-leading order correction to the impact factor (vertex) for the 
production of a forward Higgs boson, obtained in the infinite top-mass limit. We present the result both in the momentum representation and as superposition of the eigenfunctions of the leading-order BFKL kernel.
This impact factor is a necessary ingredient for the description of the inclusive hadroproduction of a forward Higgs in the limit of small Bjorken $x$, as well as for the study of inclusive forward emissions of a Higgs boson in association with a backward identified object.}


\arxivnumber{2205.02681}

\begin{document} 
\maketitle
\flushbottom

\clearpage

\section{Introduction}
\label{sec:intro}

The discovery of the Higgs boson at the Large Hadron Collider (LHC)~\cite{ATLAS:2012yve,CMS:2012qbp} marked the turn of a new era for particle physics.
Over the last ten years a remarkable effort has been made by both theoretical and experimental collaborations to shed light on the Higgs sector.
Here, the hunt for accurate benchmarks of theories underlying the Standard Model as well as for long-awaited signals of New Physics has faced many challenges.
From a formal perspective, higher-order perturbative calculations of the Higgs production in proton collisions {\it via} the gluon and the vector-boson fusion subprocesses have traced a path towards precision.

The analytic description of inclusive cross sections in gluon fusion is an arduous task, since it relies on the massive top-quark loop even at leading order (LO)~\cite{Georgi:1977gs}, and radiative corrections coming from Quantum ChromoDynamics (QCD) are known to be sizeable.
Next-to-leading order (NLO) QCD studies on the gluon-fusion channel were performed~\cite{Dawson:1990zj,Djouadi:1991tka} by making use of an effective Higgs field theory, valid in the infinite top-mass limit~\cite{Wilczek:1977zn}.
The perturbative accuracy was later pushed to the next-to-NLO (NNLO)~\cite{Harlander:2002wh,Anastasiou:2002yz,Ravindran:2003um} and the next-to-next-to-NLO (N$^3$LO) level~\cite{Anastasiou:2014lda,Anastasiou:2015vya,Anastasiou:2016cez,Mistlberger:2018etf}.
Finite top-mass effects on NLO, NNLO and N$^3$LO calculations were estimated in Refs.~\cite{Spira:1995rr},~\cite{Harlander:2009my,Harlander:2009mq,Harlander:2009bw,Pak:2009bx,Pak:2009dg} and~\cite{Davies:2019wmk}, respectively.
Inclusive distributions for the emission of Higgs-plus-jet systems were investigated in the NLO~\cite{Bonciani:2016qxi,Jones:2018hbb} and NNLO~\cite{Boughezal:2013uia,Chen:2014gva,Boughezal:2015dra,Boughezal:2015aha} QCD.

All these higher-order calculations have served as a testing ground for the well established \emph{collinear-factorization} approach (see Ref.~\cite{Collins:1989gx} for a review), where hadronic cross sections are obtained as a one-dimensional convolution between perturbative hard-scattering factors and collinear parton density functions (PDFs).
Although the collinear formalism has achieved many successes in the description of high-energy hadronic reactions, there exist phase-space regions where one or more kinds of logarithmic contributions are enhanced due to kinematics. These logarithms are genuinely neglected by a standard fixed-order treatment. They can be large enough to compensate the smallness of the strong coupling, $\alpha_s$, up to spoiling the convergence of the perturbative series.
All-order techniques, known as \emph{resummations}, have been developed to include these corrections. They depend on the kinematic limit considered and generally rely upon peculiar properties of factorization of QCD matrix elements and the phase space.

The inclusion of large logarithms originating by small observed transverse momenta is a required ingredient to consistently describe transverse-momentum differential distributions for the inclusive production of colorless final states in hadronic collisions.
It can be achieved {\it via} the so-called \emph{transverse-momentum} (TM) resummation (see, \emph{e.g.}, Refs.~\cite{Catani:2000vq,Bozzi:2005wk,Catani:2010pd,Catani:2013tia}).
TM fiducial distributions for the Higgs spectrum were studied within a third-order resummation accuracy in Refs.~\cite{Bizon:2017rah,Bizon:2018foh,Chen:2018pzu,Billis:2021ecs,Re:2021con}.
The same formalism was employed to access doubly differential TM distributions for the simultaneous Higgs-plus-jet detection~\cite{Monni:2019yyr,Buonocore:2021akg}.

Another kinematic regime where gluon-fusion-induced Higgs final states exhibit a strong sensitivity to logarithmic contributions is the large-$x$ sector.
Here, physical cross sections are probed close to edges of their phase space, where Sudakov double logarithms coming from soft and collinear gluons emitted with a large longitudinal fraction $x$ become relevant and must be resummed to all orders.
A \emph{threshold} resummation formalism, originally developed in Refs.~\cite{Sterman:1986aj,Catani:1989ne,Catani:1996yz,Bonciani:2003nt,deFlorian:2005fzc,Ahrens:2009cxz,deFlorian:2012yg,Forte:2021wxe} for rapidity-inclusive productions and then extended to include large-$x$ effects also in rapidity-differential observables~\cite{Mukherjee:2006uu,Bolzoni:2006ky,Becher:2006nr,Becher:2007ty,Bonvini:2010tp,Ahmed:2014era,Banerjee:2018vvb}, was used to provide us with accurate predictions for the scalar Higgs in gluon fusion~\cite{Kramer:1996iq,Catani:2003zt,Moch:2005ky,Bonvini:2012an,Bonvini:2014joa,Catani:2014uta,Bonvini:2014tea,Bonvini:2016frm,Beneke:2019mua,Ajjath:2020sjk,Ajjath:2020lwb,Ahmed:2020nci,Ajjath:2021bbm} (the pseudo-scalar counterpart was addressed in Refs.~\cite{deFlorian:2007sr,Schmidt:2015cea,Ahmed:2016otz,Bhattacharya:2019oun,Bhattacharya:2021hae}) and in bottom-quark annihilations~\cite{Bonvini:2016fgf,Ajjath:2019neu,Ahmed:2020nci}. 
Large-$x$ improved collinear PDFs were extracted in Ref.~\cite{Bonvini:2015ira}.

Energy logarithms arise in the so-called Regge--Gribov limit~\cite{Gribov:1983ivg,Celiberto:2017ius}, also known as \emph{semi-hard} regime.
Here, a strong scale hierarchy given by $\LQCD \ll \{Q\} \ll \sqrt{s}$, with $\LQCD$ the QCD hadronization scale, $\{Q \}$ one or a set of process-dependent perturbative scales, and $s$ the squared center-of-mass energy, lead to the growth of logarithmic contributions of the form $\ln s$, which have to be resummed to all orders.
The Balitsky--Fadin--Kuraev--Lipatov (BFKL) approach~\cite{Fadin:1975cb,Kuraev:1976ge,Kuraev:1977fs,Balitsky:1978ic} allows us to perform such a high-energy resummation in the leading logarithmic approximation (LL), which means considering all the terms proportional to $(\alpha_s \ln s)^n$, and in the next-to-leading logarithmic approximation (NLL), which means accounting for all the contributions of the form $\alpha_s (\alpha_s \ln s)^n$.

BFKL-resummed cross sections are elegantly portrayed by a high-energy convolution between a universal, energy-dependent Green's function, and two \emph{impact factors} depicting the transition from each incoming particle to the outgoing object(s) produced in its fragmentation region (see Section~\ref{ssec:BFKL_general} for technical details). They depend on the final state and on the (set of) scale(s) $\{Q\}$, but not on $s$.
To get a full NLL description, impact factors must be calculated within the NLO accuracy in the $\alpha_s$ perturbative expansion.
They are known at NLO only for a limited selection of reactions: (a) quark and gluon impact factors~\cite{Fadin:1999de,Fadin:1999df,Ciafaloni:1998kx,Ciafaloni:1998hu,Ciafaloni:2000sq}, which represent the common basis to calculate (b) forward jet~\cite{Bartels:2001ge,Bartels:2002yj,Caporale:2011cc,Ivanov:2012ms,Colferai:2015zfa} and (c) forward light-hadron~\cite{Ivanov:2012iv} impact factors, then (d) virtual-photon to vector-meson~\cite{Ivanov:2004pp} and (e) virtual-photon to virtual-photon~\cite{Bartels:2000gt,Bartels:2001mv,Bartels:2002uz,Fadin:2001ap,Balitsky:2012bs} impact factors.

NLO impact factors were used, together the NLL BFKL Green's function, to investigate the high-energy behavior of azimuthal-angle and rapidity distributions for inclusive final states featuring the emission of two objects with transverse masses well above the QCD scale and widely separated in rapidity.
They fall in a special class of semi-hard reactions where energy logarithms are heightened by large rapidity distances ($\Delta Y$), independently of the typical values of longitudinal fractions at which parton distributions are probed.
More in particular, large $\Delta Y$-values lead to non-negligible transverse-momentum exchanges in the $t$-channel and thus to the enhancement ($\ln s$)-type contributions.
Therefore, proton-initiated two-particle semi-hard processes can be described by the hands of a \emph{hybrid} high-energy and collinear formalism, where the BFKL resummation determines the general structure of the high-energy cross section, and impact factors encode collinear inputs coming from PDFs.\footnote{Another hybrid approach, close in spirit to ours, was defined in Refs.~\cite{Deak:2009xt,vanHameren:2022mtk}.}

A non-inclusive list of hadronic processes studied {\it via} the hybrid factorization and with full NLL accuracy includes: Mueller--Navelet jets~\cite{Mueller:1986ey}, for which several phenomenological analyses were provided~\cite{Colferai:2010wu,Caporale:2012ih,Ducloue:2013hia,Ducloue:2013bva,Caporale:2013uva,Caporale:2014gpa,Caporale:2015uva,Celiberto:2015yba,Celiberto:2015mpa,Celiberto:2016ygs,Celiberto:2017ius,Caporale:2018qnm,Celiberto:2020wpk} and compared with 7 TeV LHC data~\cite{Khachatryan:2016udy}, then two-hadron~\cite{Celiberto:2016hae,Celiberto:2017ptm,Celiberto:2017ius,Celiberto:2020wpk,Celiberto:2020rxb,Celiberto:2021dzy,Celiberto:2021fdp,Celiberto:2022rfj} and hadron-jet~\cite{Bolognino:2018oth,Bolognino:2019yqj,Celiberto:2020wpk,Celiberto:2020rxb,Celiberto:2021dzy,Celiberto:2021fdp,Celiberto:2022dyf,Celiberto:2022rfj} inclusive tags.
NLL analyses on leptonic-initiated reactions were mainly focused on the exclusive electroproduction of light vector mesons~\cite{Ivanov:2005gn,Ivanov:2006gt} and on the light-by-light scattering~\cite{Brodsky:2002ka,Chirilli:2014dcb,Ivanov:2014hpa}.
A partial NLL treatment was employed in the description of multi-jet detections~\cite{Caporale:2016soq,Caporale:2016zkc,Celiberto:2017ius}, and on final states where a $J/\psi$~\cite{Boussarie:2017oae}, a Drell--Yan pair~\cite{Golec-Biernat:2018kem}, or a heavy-flavored jet~\cite{Bolognino:2021mrc} is inclusively emitted in association with a light-flavored jet, the two objects being well separated in rapidity. Similar analyses on inclusive heavy-quark pair photo- and hadroproductions were respectively proposed in Refs.~\cite{Celiberto:2017nyx,Bolognino:2019ccd} and~\cite{Bolognino:2019yls,Bolognino:2019ccd}.

Concerning Higgs phenomenology at high energies, all studies conducted so far relied upon a pure LL or a partial NLL setup. This is due to difficulties emerging in the analytic computation of corresponding NLO impact factors, which contain the top-quark loop and at least one off-shell gluon leg.
LL studies on Higgs production in mini-jet events were performed in Refs.~\cite{DelDuca:1993ga,DelDuca:2003ba}.
Ref.~\cite{DelDuca:2003ba} contains the calculation of the LO forward-Higgs impact factor in momentum space with the mass of the top quark being kept finite. Its representative Feynman diagram embodies an on-shell gluon stemming from a collinear PDF, which is connected to the top-quark loop. The outgoing object is a scalar Higgs, whereas an off-shell gluon in nonsense polarization is exchanged in the $t$-channel (see Section~\ref{ssec:LO_IF}).

Azimuthal imprints from Higgs angular correlations were studied in Ref.~\cite{Cipriano:2013ooa}.
The exclusive hadroproduction of Higgs bosons {\it via} the two-Pomeron fusion channel was investigated in Ref.~\cite{Bartels:2006ea}.
The relevance of small-$x$ double-logarithmic terms in Higgs production rates in the infinite top-mass limit was assessed in Ref.~\cite{Hautmann:2002tu}.
High-energy effects from BFKL and Sudakov contributions were combined together to describe cross sections for the Higgs-plus-jet hadroproduction in almost back-to-back configurations~\cite{Xiao:2018esv}.
Azimuthal correlations between a single-charmed hadron emitted in ultra-forward directions of rapidity and a Higgs boson were investigated with a partial NLL accuracy in Ref.~\cite{Celiberto:2022zdg}.

A key ingredient to cast high-energy distributions in the BFKL factorized form is the Mellin representation of impact factors. The Mellin projection of LO forward-Higgs impact factor onto LO eigenfunctions of the BFKL kernel was given for the first time in Ref.~\cite{Celiberto:2020tmb}.
In the same work BFKL predictions for azimuthal-angle correlations and transverse-momentum distributions for the inclusive Higgs-plus-jet reaction in the general finite top-mass case were calculated at partial NLL level {\it via} the {\tt JETHAD} multi-modular interface~\cite{Celiberto:2020wpk}, namely including NLL contributions coming from the Green's function and LO impact factors plus NLO universal terms guessed by a renormalization-group (RG) analysis.

Two major results came out from that study.
First, azimuthal correlations offered a fair and solid stability under next-to-leading corrections as well as under renormalization and factorization scale variations, thus bracing the message that the large transverse masses typical of Higgs emissions act as \emph{natural stabilizers} for BFKL. This result is remarkable, since previous studies on light jets and hadrons~\cite{Celiberto:2015yba,Celiberto:2015mpa,Celiberto:2020wpk} highlighted how higher-order contributions, noticeably large and with opposite sign with respect to LL terms, are the source of strong instabilities that have prevented any attempt at making precision studies of those processes so far. Soon later, an analogous stabilization pattern was observed also in NLL distributions sensitive to heavy-flavored emissions~\cite{Celiberto:2021dzy,Celiberto:2021fdp,Celiberto:2022dyf,Celiberto:2022rfj}.
Second, differential cross sections in the Higgs transverse momentum turned out to be suitable observables whereby discriminating BFKL-driven predictions from fixed-order results, their mutual separation becoming more and more sizeable when the momentum increases.

The correct description of the gluon content in the proton at small $x$ relies on the so-called \emph{unintegrated gluon distribution} (UGD), whose evolution is regulated by the BFKL Green's function and its use is driven by the $\kappa_T$-factorization theorems~\cite{Catani:1990xk,Catani:1990eg,Catani:1993ww,Collins:1991ty,Ball:2007ra}.
Conversely, moderate-$x$ gluons need to be described in pure collinear factorization.
A powerful method, developed by Altarelli, Ball and Forte (ABF)~\cite{Ball:1997vf,Altarelli:2001ji,Altarelli:2003hk,Altarelli:2005ni,Altarelli:2008aj} allows us to coherently embody both small-$x$ and DGLAP inputs through a stabilization procedure of the high-energy series, a symmetrization of the Green's function in collinear and anti-collinear regions of the phase space, and an improvement of the $\alpha_s$ treatment to consistently handle small-$x$ singularities.
ABF-inspired predictions for the central Higgs production at LHC energies were presented in Refs.~\cite{Marzani:2008az,Caola:2010kv,Caola:2011wq,Forte:2015gve}. A numeric implementation of small-$x$ resummed Higgs observables can be obtained from the {\tt HELL} code~\cite{Bonvini:2018iwt}, which was then employed to obtain first determinations of small-$x$ improved collinear PDFs~\cite{Ball:2017otu}.

A double-resummation approach that coherently incorporates both large-$x$ (threshold) and small-$x$ (BFKL) logarithms, matching these all-order contributions with fixed-order calculations taken at N$^3$LO accuracy~\cite{Ball:2013bra}, was applied to the study of inclusive Higgs production rates in gluon fusion~\cite{Bonvini:2018ixe}.
In that work it was argued that the double resummation has a 2\% impact on the Higgs cross section at the current LHC energies of 13 TeV. Then, its small-$x$ component makes the full-resummation weight grow up to around 5\%, and it definitely dominates on the large-$x$ one at nominal energies of the Future Circular Collider (FCC), $\sqrt{s} = 100\mbox{ TeV}$, where the resummation corrects the Higgs production rate by 10\%~\cite{Mangano:2016jyj}.
The analysis of Ref.~\cite{Bonvini:2018ixe} braced the message that electroweak physics at 100 TeV is \emph{de facto} small-$x$ physics. Although this estimate supports the statement that BFKL-resummation effects become more and more evident when energies increase, we remark that the inclusion of full NLL terms, once the NLO central-Higgs coefficient function will be available, could radically chance the scenario, possibly highlighting that high-energy logarithms need to be accounted for at lower energy values, closer to the current LHC ones.
More in general, NLO impact factors are core elements to reach the precision level in the description of high-energy differential distributions sensitive to Higgs-boson emissions.

In this paper we present the calculation of the NLO impact-factor correction for the production of a forward Higgs boson in hybrid high-energy and collinear factorization, where the gluon-Reggeon-Higgs vertex is calculated within the standard BFKL framework, and then convoluted with the incoming-gluon collinear PDF.
Our computation is carried out within perturbative QCD in the high-energy limit and relies upon the effective Higgs field theory valid in the large top-mass limit. It will cross-check 
a similar calculation~\cite{Hentschinski:2020tbi} based on the use of the Lipatov's high-energy QCD effective action~\cite{Lipatov:1995pn,Lipatov:1996ts}. Moreover, we will present our result both in momentum basis and and in the basis formed by the LO BFKL kernel eigenfunctions. The latter representation, in addition to being particularly useful for numeric applications, provides us with an unambiguous and straightforward procedure for the complete cancellation of all divergences. 

The paper is organized as follows. In Section~\ref{sec:theory} we recap generalities of the BFKL formalism and of the calculation of the LO Higgs impact factor, then we briefly illustrate the procedure followed for the NLO computation.
Feynman rules for the Higgs effective theory and general definitions are given in Appendix~\ref{AppendixA}.
Real and virtual corrections are presented and discussed in Sections~\ref{sec:real} and~\ref{sec:virtual}, respectively.
Useful integrals to calculate virtual corrections can be found in Appendix~\ref{AppendixB}.
In Section~\ref{sec:projection} we offer the final result for the NLO impact factor projected onto the LO eigenfunctions of the BFKL kernel.
Finally, we come out with conclusions and future perspectives (see Section~\ref{sec:conclusions}).

\section{Theoretical framework}
\label{sec:theory}

The aim of our computation is the next-to-leading order the impact factor (vertex) for the 
production of a forward\footnote{In this paper, if not otherwise stated, {\em forward/backward} will mean ``with (large) positive/negative rapidity''.} Higgs boson, in the limit of infinite top mass. It 
enters the theoretical predictions for the cross section, differential in the kinematics of the produced Higgs boson, of the inclusive hadroproduction of a forward Higgs in the limit of small Bjorken-$x$, in the framework of $\kappa_T$-factorization, and that of a forward Higgs plus a backward identified object, in the framework of the hybrid collinear/BFKL factorization.

\begin{figure}
\begin{center}
\includegraphics[scale=0.5]{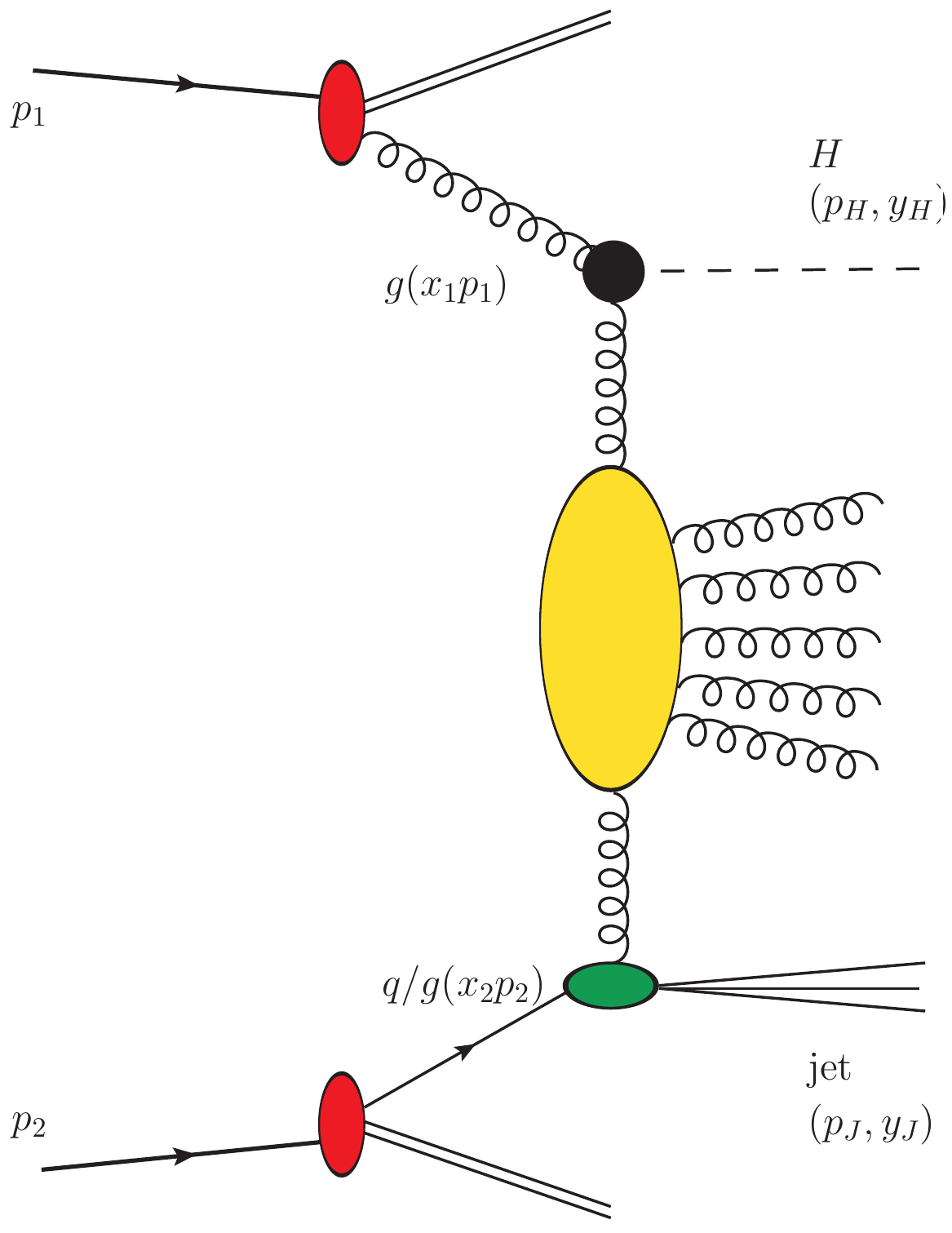} 
\caption{Schematic description of the process~(\ref{Process}) in the LL.}
\label{HiggsSemiHard}
\end{center}
\end{figure}

For a better understanding of the definition of this impact factor and for illustration of the main steps in the calculation, we make reference to the specific process proposed in Ref.~\cite{Celiberto:2020tmb}, {\it i.e.} the inclusive hadroproduction of a Higgs boson and a jet (see Fig. \ref{HiggsSemiHard}),
\begin{equation}
    {\rm{proton}} (p_1) + {\rm{proton}} (p_2) \rightarrow H(\vec{p}_H, y_H) + X + {\rm{jet}}(\vec{p}_J, y_J) \; ,
\label{Process}
\end{equation}
in the kinematic region where both the Higgs boson and the jet have large transverse momenta $\vec{p}_H^{\; 2}$, $\vec{p}_J^{\; 2}$ and the separation between their rapidities, $y_H$ and $y_J$ is large. This provides the hard scale, $Q^2 \sim p_{H,J}^{\; 2}$, which makes perturbative QCD applicable. Moreover, the squared center-of-mass energy of the proton collision,
$W^2 \equiv 2 p_1 \cdot p_2$, is assumed to be much larger than the hard scale, $W^2 \gg Q^2$.  \\
Considering the leading behaviour in the $1/Q$-expansion (leading-twist approximation), the cross section of this process can be written in the standard QCD collinear factorization as
\begin{equation}
\frac{d \sigma (W^2)}{ d x_H d x_J d^2 \vec{p}_H d^2 \vec{p}_J} = \sum_{A,B=q,g} \int_{0}^1 \int_{0}^1 d x_1 d x_2 f_A \left( x_1, \mu_F \right) f_B \left( x_2, \mu_F \right) \frac{d \hat{\sigma} (x_1 x_2 W^2,\mu_F)}{d x_H d x_J d^2 \vec{p}_H d^2 \vec{p}_J} \; ,
\label{CollFacSig}
\end{equation}
where the $A,B$ indices specify parton types, $A,B=q,\bar{q},g$, while $f_{A,B}(x_{1,2}, \mu_F)$ denotes the initial proton parton distribution function (PDF), $x_{1,2}$ are the longitudinal fractions of the proton momenta carried by the partons involved in the hard subprocess, $x_H$ and $x_J$ are the longitudinal fractions of the proton momenta carried by the Higgs and the jet, the factorization scale is denoted by $\mu_F$ and $d \hat{\sigma} (x_1 x_2 W^2, \mu_F)$ is the partonic cross section for the production of an Higgs boson and a jet. \\\
In the case of large interval of rapidity between the two identified object, the energy of the partonic subprocess is much larger than Higgs and jet transverse momenta, $s \equiv x_1 x_2 W^2 \gg \vec{p}_{H,J}^{\; 2}$. In this
region the perturbative partonic cross section receives at higher orders large contributions $\sim \alpha_s^n \ln^n (s/\vec{p}_{H,J}^{\;2})$, related with large energy logarithms.  The resummation of such enhanced contributions with NLL accuracy can be performed by using the BFKL approach.

\subsection{Generalities of the BFKL approach}
\label{ssec:BFKL_general}

We briefly recall here the generalities of the BFKL approach, starting our discussion from the fully inclusive parton-parton cross section $A(k_A)+B(k_B) \to {\rm all}$, which, through the optical theorem, can be related to the $s$-channel imaginary part of the 
elastic amplitude $A(k_A)+B(k_B) \to A(k_A)+B(k_B)$ at zero transferred momentum,
\begin{equation}
    \sigma_{AB} = \frac{\Im m_{s} \mathcal{A}}{s}\;, 
\end{equation}
with $s = ({k}_A + {k}_B)^2$.
The BFKL approach is introduced to describe this cross section in the limit $s\to \infty$.

We use for all vectors the Sudakov decomposition
\begin{equation}
p = \beta k_1 + \alpha k_2 + p_{\perp},\ \ \ \ \ \ \ \ p_{\perp}^2 =
- \vec p^{\:2}~,
\label{Sudakov1}
\end{equation}
the vectors $(k_1,\ k_2)$ being the light-cone basis of the initial
particle momenta plane $(k_A,\ k_B)$, so that we can put
\begin{equation}
k_A= k_1 +
\frac{m_A^2}{s}k_2~,\ \ \ \ \ \ \ k_B =k_2 +
\frac{m_B^2}{s} k_1~.
\label{Sudakov2}
\end{equation}
Here $m_A$ and $m_B$ are the masses of the colliding partons $A$ and $B$ (taken equal to zero) and the vector notation is used throughout this paper for the transverse
components of the momenta, since all vectors in the transverse subspace are 
evidently space-like.
\begin{figure}
\begin{center}
{\parbox[t]{8 cm}{\epsfysize 8 cm \epsffile{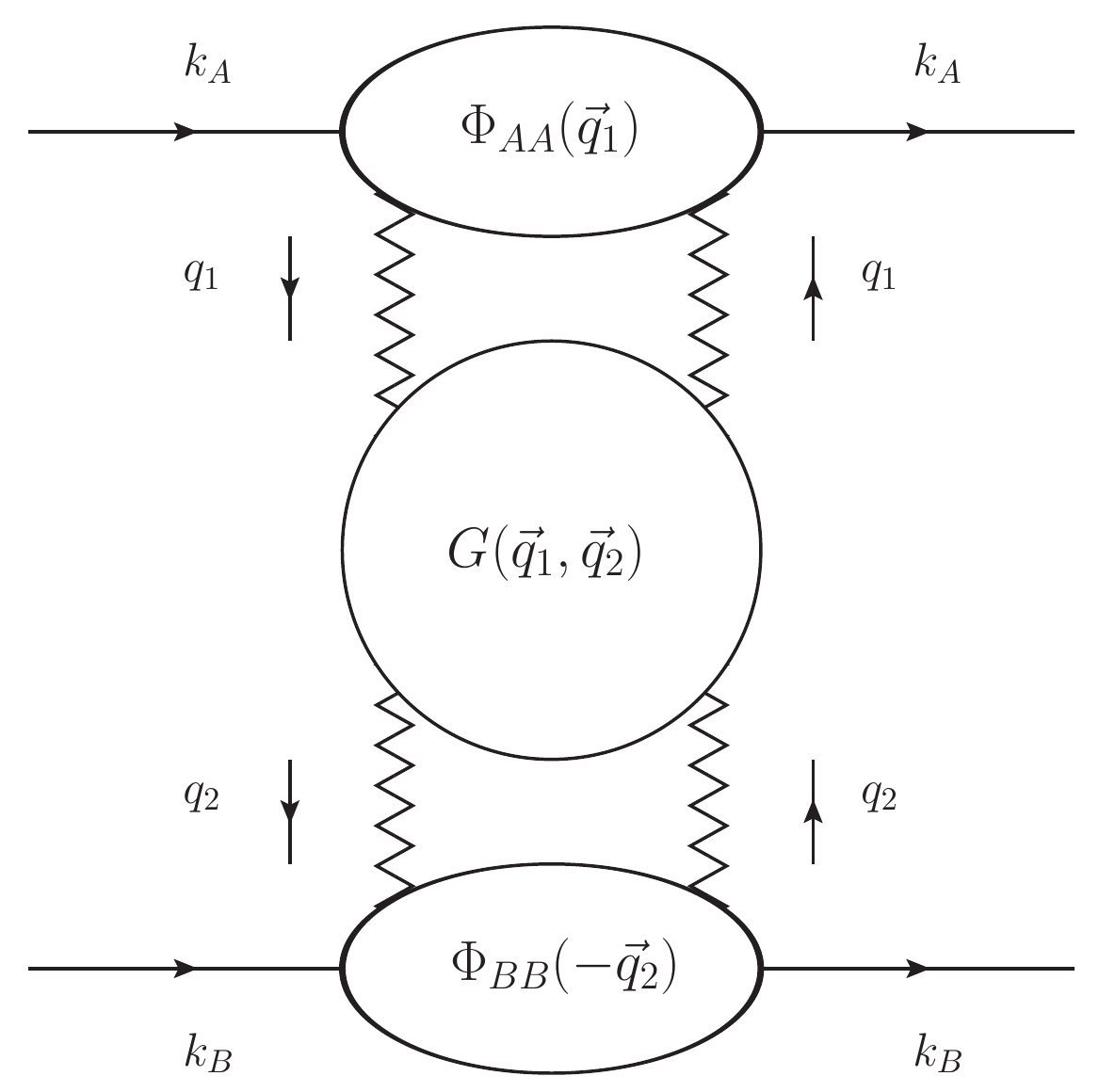}}}
\end{center}
\caption{Diagrammatic representation of the elastic scattering amplitude $A + B \rightarrow A + B$.}
\label{GreenCon}
\end{figure}
Within the NLL, the elastic scattering amplitude for $t=0$ can be written as
\[
\Im m_{s}\left(({\cal A})_{AB}^{AB}\right) = \frac{s}{\left( 2\pi \right)^{D-2}}
\int \frac{d^{D-2}q_1}{\vec{q}_{1}^{\:2}}
\int \frac{d^{D-2}q_2}{\vec{q}_{2}^{\:2}}
\]
\begin{equation}
\times \Phi _{AA}\left( \vec{q}_{1};s_{0}\right)\int_{\delta -i\infty}^{\delta+i\infty}
\frac{d\omega }{2\pi i}\left[ \left( \frac{s}{s_{0}}\right)^{\omega }
G_{\omega }\left( \vec{q}_{1},\vec{q}_{2}\right) 
\right] \Phi _{BB}\left( -\vec{q}_{2};s_{0}\right) \;,
\label{Ar}
\end{equation}
where momenta are defined in Fig.~\ref{GreenCon} and $D$ is the space-time dimension, which
will be taken equal to $4 - 2 \epsilon$ in order to regularize divergences. 
In the above equation $\Phi_{P P}$ are the impact factors and $G_{\omega}$ is the Mellin transform of the Green's function for the Reggeon-Reggeon 
scattering~\cite{Fadin:1998fv}. 
The parameter $s_0$ is an arbitrary energy scale introduced in order to define the partial wave expansion of the scattering amplitudes. The dependence 
on this parameter disappears in the full expressions 
for the amplitudes, within the NLL accuracy. The integration in the complex plane $\omega$ is performed along the line $\Re e(\omega)=\delta$ which is supposed to lie to the right of all
singularities in $\omega$ of $G_\omega$.

The Green's function obeys the BFKL equation
\begin{equation}
\omega G_{\omega }\left( \vec{q}_{1},\vec{q}_{2}\right) = \delta^{\left(D-2\right) }\left( \vec{q}_{1}-\vec{q}_{2}\right)
+\int d^{D-2} q_r \ {\cal K}\left( \vec{q}_{1},\vec{q}_r\right) G_{\omega }\left( \vec{q}_r,\vec{q}_{2}\right) \;,
\label{genBFKL}
\end{equation}
where ${\cal K}$ is the NLO kernel in the singlet color representation~\cite{Fadin:1998fv}. The definition of NLO impact factor can be found in Ref.~\cite{Fadin:1998fv} and for $t=0$ reads
$$
\Phi_{AA}(\vec q_1; s_0) = \left( \frac{s_0}
{\vec q_1^{\:2}} \right)^{\omega( - \vec q_1^{\:2})}
\sum_{\{f\}}\int\theta(s_{\Lambda} -
s_{AR})\frac{ds_{AR}}{2\pi}\ d\rho_f \ \Gamma_{\{f\}A}^c
\left( \Gamma_{\{f\}A}^{c^{\prime}} \right)^* 
\langle cc^{\prime} | \hat{\cal P}_0 | 0 \rangle
$$
\begin{equation}
-\frac{1}{2}\int d^{D-2}q_2\ \frac{\vec q_1^{\:2}}{\vec q_2^{\:2}}
\: \Phi_{AA}^{(0)}(\vec q_2)
\: {\cal K}^{(0)}_r (\vec q_2, \vec q_1)\:\ln\left(\frac{s_{\Lambda}^2}
{s_0(\vec q_2 - \vec q_1)^2} \right)~,
\label{ImpactUnpro}
\end{equation}
where $\omega(t)$ is the Reggeized gluon trajectory, which enters this expression at the leading order, given by
\begin{equation}
    \omega^{(1)}(t) = \frac{g^2 t}{(2 \pi)^{D-1}} \frac{N}{2} \int \frac{d^{D-2}k_{\perp}}{k_{\perp}^2 (q-k)_{\perp}^{2}} = - \frac{g^2 N \Gamma(1+\epsilon)(\vec{q}^{\; 2})^{-\epsilon}}{(4 \pi)^{2-\epsilon}} \frac{\Gamma^2(-\epsilon)}{\Gamma(-2\epsilon)} \; ,
    \label{ReggeTraj}
\end{equation}
with $t=q^2=-\vec q^{\:2}$ and $N$ the number of colors.
\begin{figure}
\begin{center}
\includegraphics[scale=0.50]{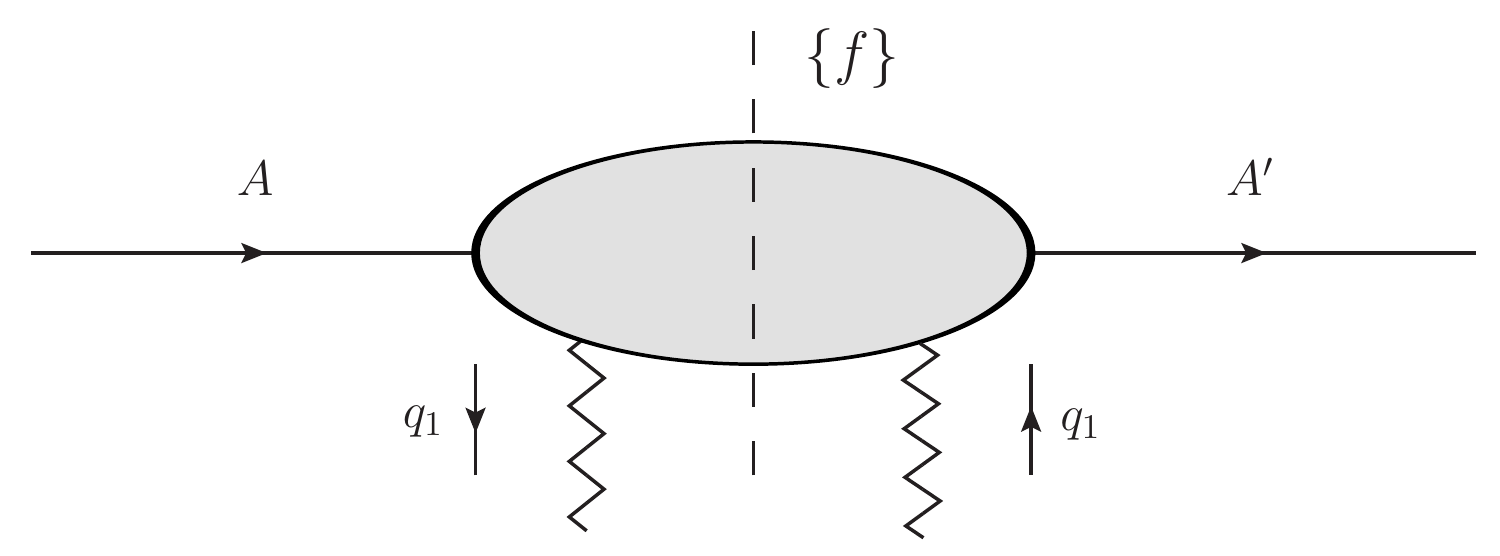}
\end{center}
\caption{Schematic description of an impact factor.}
\label{ImpactFactorRepres}
\end{figure}
$\Gamma_{\{f\}A}^c$ is the effective vertex for production of the
system ${\{f\}}$ (see Fig.~\ref{ImpactFactorRepres}) in the collision of the particle $A$
and the Reggeized gluon with color index $c$ and momentum $-q_1$, with
\begin{equation}
q_1 = \alpha k_2 + {q_1}_{\perp}~,\ \ \ \ \ \ \ \alpha 
\approx -\left( s_{AR} - m_A^2 + \vec q_1^{\:2} \right)/s \ll 1~, 
\end{equation}
and $s_{AR}$ is the particle-Reggeon squared invariant mass. In the fragmentation region of the particle $A$, 
where all transverse momenta as well as the invariant mass $\sqrt{s_{AR}}$ are not 
growing with $s$, we have $q_1^2 = - \vec q_1^{\:2}$.
The factor 
\begin{equation}
    \langle cc^{\prime} | \hat{\cal P}_{0} | 0 \rangle = \frac{\delta^{c c'}}{\sqrt{N^2-1}}
\end{equation}
is the projector on the singlet color state representation. Summation in
Eq.~(\ref{ImpactUnpro}) is carried out over all  systems $\{f\}$ which can be
produced in the NLL and the integration is performed over the phase
space volume of the produced system, which for a $n$-particle
system (if there are identical particles in this system, 
corresponding symmetry factors should also be introduced) reads
\begin{equation}
d\rho_f = (2\pi)^D\delta^{(D)}\biggl(p_A - q_1 - \sum_{m=1}^nk_m\biggr)
\prod_{m=1}^n\frac{d^{D-1}k_m}{2E_m(2\pi)^{D-1}}~,
\label{GenPhasSpa}
\end{equation}
as well as over the particle-Reggeon invariant mass. The average over initial-state color and spin degrees of freedom is implicitly assumed.
The parameter $s_{\Lambda}$, limiting the integration region over the invariant mass in the first term in the R.H.S. of Eq.~(\ref{ImpactUnpro}), is introduced for the separation of the contributions of multi-Regge and quasi-multi-Regge kinematics (MRK and QMRK) and should be considered as tending to infinity.
The dependence of the impact factors on this parameter disappears due to the cancellation between the first and the second term in
the R.H.S. of Eq.~(\ref{ImpactUnpro}). 
In the second term, usually called ``BFKL counterterm'', $\Phi_{AA}^{(0)}$ is the Born
contribution to the impact factor, which does not depend on $s_0$, while
${\cal K}^{(0)}_r$ is the part of the BFKL kernel in the Born approximation 
connected with real particle production:
\begin{equation}
{\cal K}^{(0)}_r(\vec q_1, \vec q_2) =
\frac{2 g^2 N}{(2\pi)^{D-1}} \frac{1}{(\vec q_1 - \vec q_2)^2}\;.
\label{BornKer}
\end{equation}

\subsection{The leading-order impact factor for forward Higgs production}
\label{ssec:LO_IF}

\begin{figure}
\begin{center}
\includegraphics[scale=0.50]{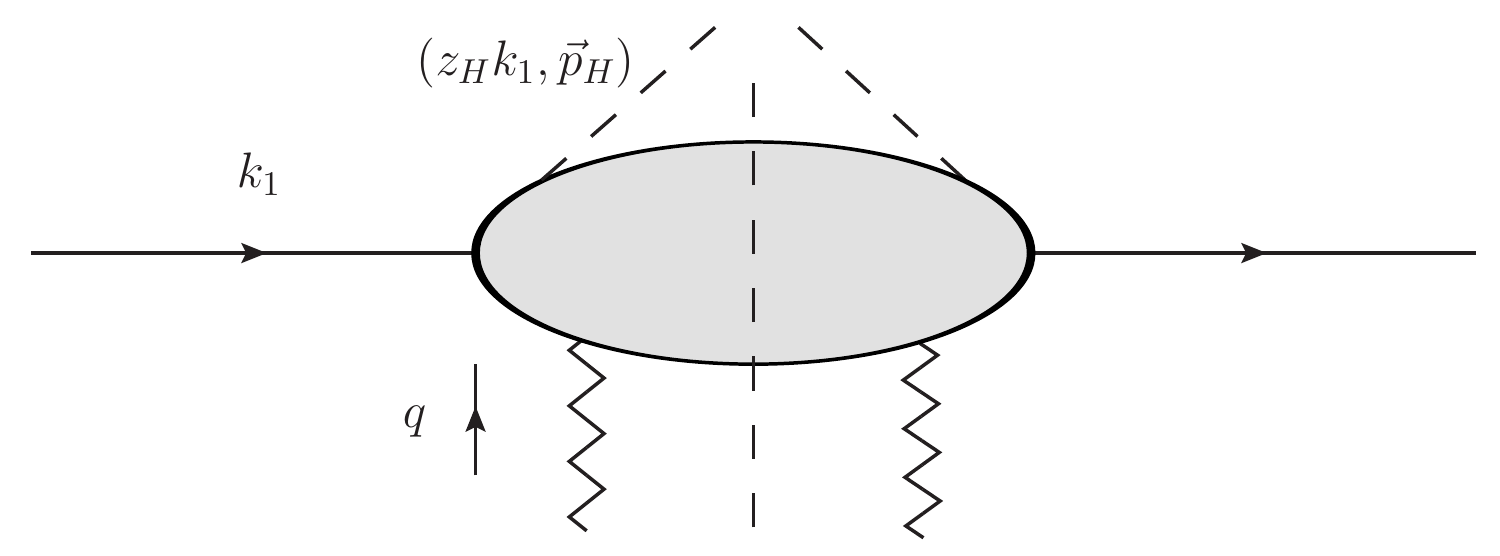}
\end{center}
\caption{Schematic description of the Higgs impact factor: $z_H$ is the fraction of longitudinal momentum of the initial parton carried by the outgoing Higgs and $\vec{p}_H$ is its transverse momentum.}
\label{HiggsImpactFactorRepres}
\end{figure}
Since we are interested in the description of the inclusive production of
a Higgs in proton-proton collisions, say in the forward region, the setup sketched 
in the previous subsection must be suitably modified. First of all, we need to consider 
processes in which the parton $A$ can be a gluon or any of the active quarks and 
must then take the convolution the partonic impact factor with the corresponding PDF. Secondly, the integration over the intermediate states $\{f\}$ in~(\ref{ImpactUnpro}) must exclude the state of the Higgs, so that the resulting impact factor will be differential in the Higgs kinematics
 (see Fig.~\ref{HiggsImpactFactorRepres}).

Let us illustrate the procedure in the simplest case of the leading-order (LO) contribution.
In the infinite top-mass approximation, the Higgs field couples to QCD {\it via} the 
effective Lagrangian~\cite{Ellis:1975ap,Shifman:1979eb},
\begin{equation}
\mathcal{L}_{ggH} = - \frac{g_H}{4} F_{\mu \nu}^{a} F^{\mu \nu,a} H \; ,
\label{EffLagrangia}
\end{equation}
where $H$ is the Higgs field, $F_{\mu \nu}^a = \partial_{\mu} A_{\nu}^a - \partial_{\nu} A_{\mu}^a  + g f^{abc} A_{\mu}^b A_{\nu}^c$ is the field strength tensor, 
\begin{equation}
g_H = \frac{\alpha_s}{3 \pi v} \left( 1 + \frac{11}{4} \frac{\alpha_s}{\pi} \right) + {\cal O} (\alpha_s^3)
\label{gH}
\end{equation}
is the effective coupling~\cite{Dawson:1990zj,Ravindran:2002} and $v^2 = 1/(G_F \sqrt{2})$ with $G_F$ the Fermi constant. Feynman rules deriving from this theory can be found in Appendix~\ref{AppendixA}.
We introduce the Sudakov decomposition for the Higgs and Reggeon momenta:
\begin{equation}
  p_H = z_H k_1 + \frac{m_H^2+ \vec{p}_H^{\; 2}}{z_H s} k_2 + p_{H \perp}\; , \hspace{0.5 cm} p_H^2 = m_H^2 \; ,
 \end{equation}
 \begin{equation}
 -q_1 \equiv q = - \alpha_q k_2 + q_{\perp} \; , \hspace{0.5 cm} q^2 = -\vec{q}^{\; 2} \; .
  \end{equation}
We notice that we have redefined the Reggeon momentum $\vec q_1$ to $-\vec q$. 
At LO, the impact factor takes contribution only from the case when the initial-state parton (particle $A$) is a gluon, for which $k_A=k_1$. For the polarization vector of all gluons in the external lines involved in the calculation we will use the light-cone gauge,
\begin{equation}
\varepsilon (k) \cdot k_2 = 0 \;,
\label{ExternalGluonGauge}
\end{equation}
which leads to the following Sudakov decomposition:
\begin{equation}
\varepsilon(k)=-\frac{(k_\perp \cdot \varepsilon_\perp(k))}{(k\cdot k_2)}k_2+\varepsilon_\perp(k)\;,
\end{equation}
so that for the initial-state gluon we have $\varepsilon (k_1) = \varepsilon_{\perp}(k_1)$.
The transverse polarization vectors have the properties
$$
\left( \varepsilon_{\perp}^*(k_1, \lambda_1) \cdot \varepsilon_{\perp}(k_1, \lambda_2) \right) =
\left( \varepsilon^*(k_1, \lambda_1) \cdot \varepsilon(k_1, \lambda_2) \right) = - \delta_{\lambda_1,\,
\lambda_2},
$$
\begin{equation}\label{217}
\sum_{\lambda} \varepsilon_{\perp}^{*\mu}(k, \lambda) \varepsilon_{\perp}^{\nu}(k, \lambda) =
- g_{\perp\perp}^{\mu\nu}~,
\end{equation}
where the index $\lambda$ enumerates the independent polarizations of gluon (sometimes it will be omitted for simplicity, but it should be always understood),
$g^{\mu\nu}$ is the metric tensor in the full space and 
$g_{\perp\perp}^{\mu\nu}$ the one in the transverse subspace, 
\begin{equation}
g_{\perp\perp}^{\mu\nu} = g^{\mu\nu} - 
\frac{k_1^{\mu} k_2^{\nu} + k_2^{\mu}k_1^{\nu}}
{(k_1 \cdot k_2)}~.
\label{metricTensorDec}
\end{equation}

\begin{figure}
\begin{center}
\includegraphics[scale=0.50]{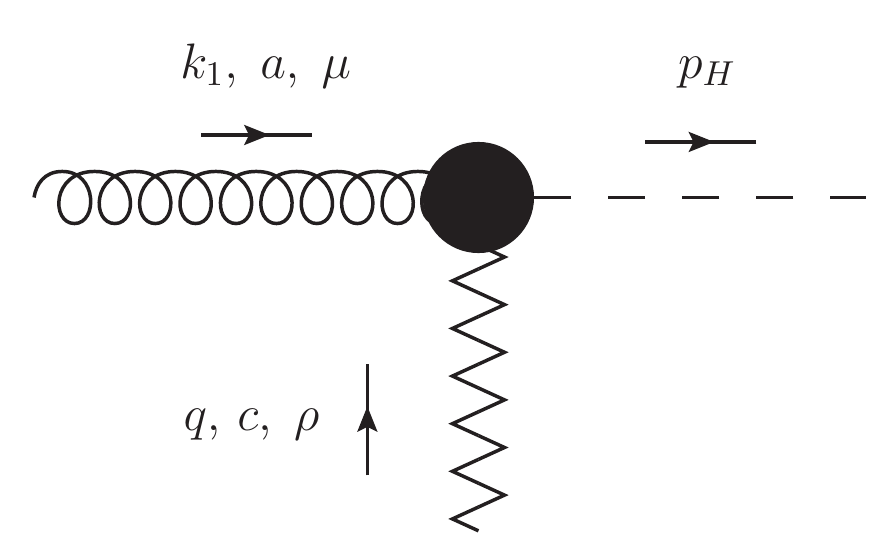} \hspace{1.5 cm} \includegraphics[scale=0.50]{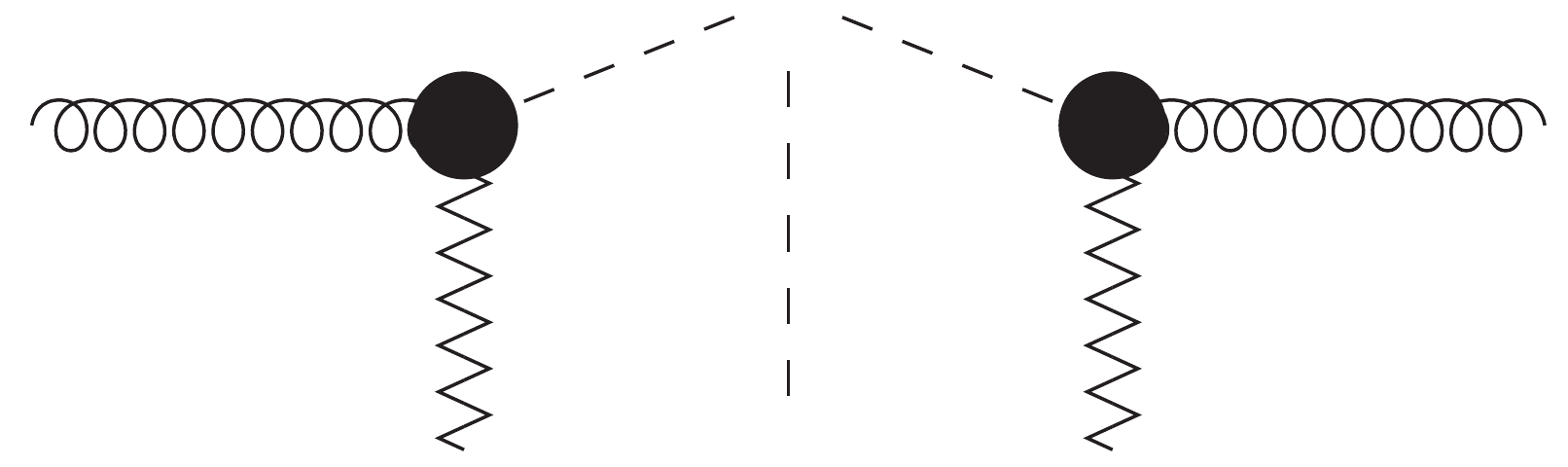}
\end{center}
\caption{Born gluon-Reggeon-Higgs effective vertex (left) and schematic description of the single contribution to the impact factor LO (right). We draw the Higgs boson above the cut to emphasize that we do not integrate over its kinematic variables.}
\label{BornVertex}
\end{figure}

The LO impact factor before differentiation in the kinematic variables of the Higgs
is given by 
\begin{equation}
    \Phi_{gg}^{\{ H \}(0)} (\vec{q} \; ) = \frac{\langle
cc^{\prime} | \hat{\cal P}_{0} | 0 \rangle}{2(N^2-1)} \sum_{a, \lambda } \int \frac{d s_{gR}}{2 \pi} d \rho_H  \Gamma_{\{ H \} g}^{ac(0)} (q) \left( \Gamma_{\{ H \} g}^{ac'(0)} (q) \right)^{*} \; ,
\label{Higgsssec:LO_IF1}
\end{equation}
where the only state contributing to the intermediate state $\{f\}$ is the Higgs boson
and the particle $A$ is identified with an on-shell gluon. Here, 
\begin{equation}
    \Gamma_{\{ H \} g}^{ac(0)} (q) = \frac{g_H}{2} \delta^{ac} \ \bigl(q_{ \perp}\cdot  \varepsilon_{\perp} (k_1)\bigr) \; ,
    \label{gRHBorn}
\end{equation}
is the high-energy gluon-Reggeon-Higgs (gRH) Born vertex (see Fig.~\ref{BornVertex}). The
overall factor $1/2(N^2-1)$ comes from the average over the polarization and color states of the incoming gluon. The vertex in Eq.~(\ref{gRHBorn}) is obtained from the Higgs effective theory Feynman rules, taking for the Reggeon in the $t$-channel the so-called ``nonsense'' polarization $(-k_2^{\rho}/s)$. This effective polarization arises from the fact that, for gluons in the $t$-channel that connect strongly separated regions in rapidity, the substitution
\begin{equation}
g^{\rho \nu} = g_{\perp\perp}^{\rho \nu} +
2 \frac{k_1^{\rho} k_2^{\nu} + k_2^{\rho} k_1^{\nu}}{s} \longrightarrow 2 \frac{k_2^{\rho} k_1^{\nu}}{s} = 2 s \left(\frac{-k_2^{\rho}}{s} \right) \left( \frac{-k_1^{\nu}}{s} \right) ~ \; ,
\label{Gribov}
\end{equation}
known as Gribov trick, can be used in the kinematic region relevant for the ``upper''
impact factor. Since the integration over the invariant mass $s_{PR}$ and over the phase space is completely trivial, we simply obtain
\begin{equation}
    \Phi_{gg}^{\{ H \}(0)} (\vec{q} \; ) = \frac{g_H^2}{8 \sqrt{N^2-1}} \vec{q}^{\; 2}\;.
\end{equation}
Hence, it is straightforward to construct the differential (in the kinematic variables of the Higgs) LO order impact factor, which in $D=4-2 \epsilon$ dimensions, reads\footnote{Not to burden our formulas, we write $\delta^{(2)}$ instead of $\delta^{(D-2)}$ in the R.H.S. and 
$d^2\vec p_H$ instead of $d^{D-2}\vec p_H$ in the L.H.S.; the latter convention will apply
also in the following. Moreover, if $g_H$ is meant to have dimension of an inverse mass, then in $D$-dimension it should be replaced by $g_H \mu^{-\epsilon}$, where $\mu$ is an arbitrary mass scale.} 
    \begin{equation}
    \frac{d\Phi_{gg}^{\{ H \}(0)} (\vec{q} \; )}{d z_H d^2 \vec{p}_H} = \frac{g_H^2}{8 (1-\epsilon) \sqrt{N^2-1}} \vec{q}^{\; 2} \delta (1-z_H) \delta^{(2)} (\vec{q}-\vec{p}_H) \; .
    \label{LOHiggsImp2}
\end{equation}
After convolution with the gluon PDF, we obtain the proton-initiated LO impact factor,
  \begin{equation}
    \frac{d \Phi_{PP}^{ \{ H \}(0)} (x_H, \vec{p}_H, \vec{q})}{d x_H d^2 \vec{p}_H} = \int_{x_H}^1 \frac{d z_H}{z_H} f_g \left( \frac{x_H}{z_H} \right) \frac{d \Phi_{gg}^{ \{H \}(0)} (z_H, \vec{p}_H, \vec{q})}{d z_H d^2 \vec{p}_H} = \frac{g_H^2 \vec{q}^{\; 2} f_g (x_H) \delta^{(2)} ( \vec{q} - \vec{p}_H) }{8 (1-\epsilon) \sqrt{N^2-1}} \; .
    \label{Factorization}
\end{equation}

\subsection{NLO computation in a nutshell}
\label{ssec:NLO_nutshell}

\begin{figure}
  \begin{center}
  \includegraphics[scale=0.40]{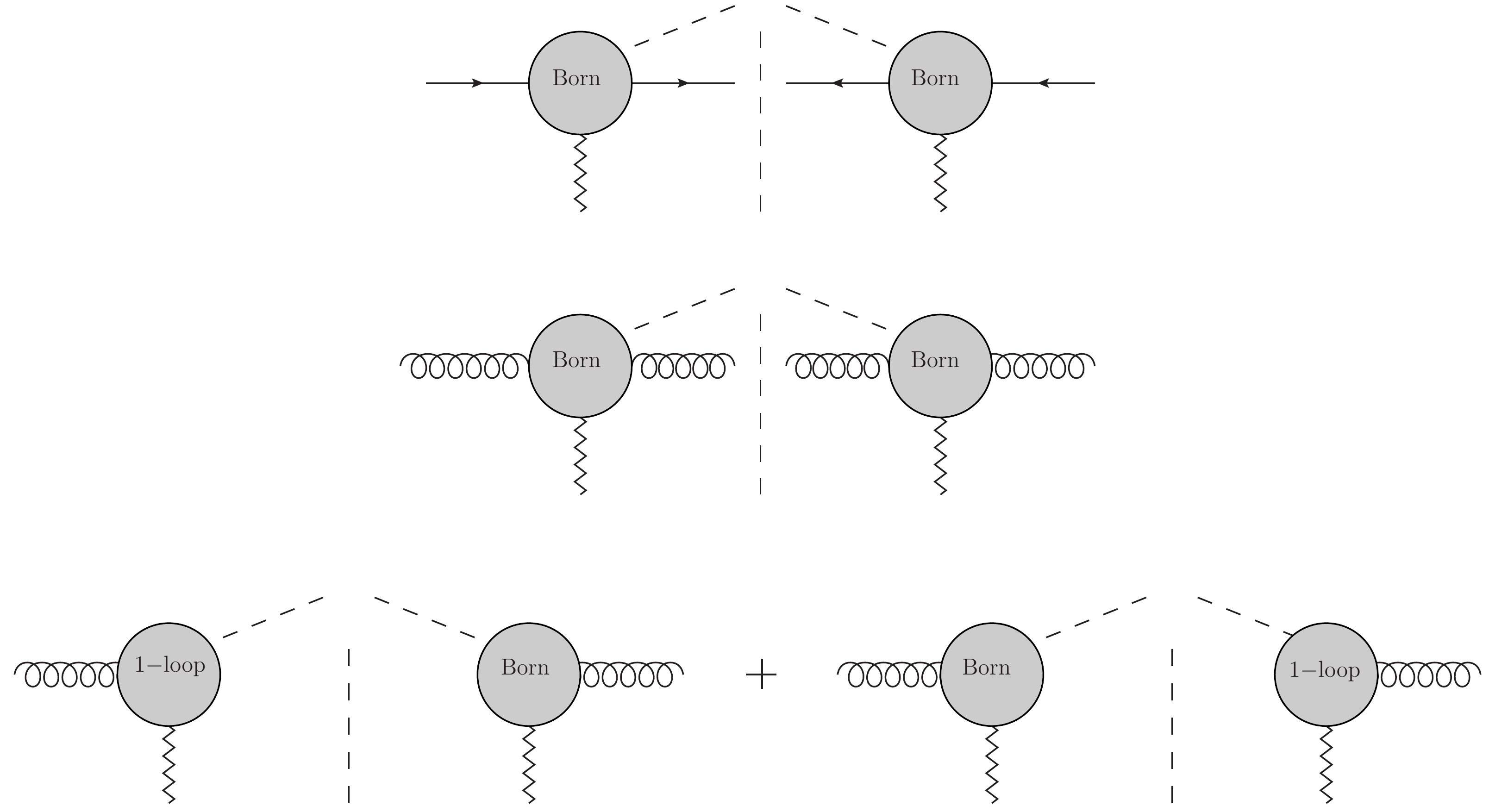}
  \end{center}
  \caption{Schematic description of the calculation of the impact factor at the NLO. In the first figure from above the hard process is initiated by a quark, and an additional quark is produced through the cut. Note that this quark ``crosses'' the cut to indicate that we will integrate on its kinematic variables.}
  \label{ProgramHiggsNLO}
\end{figure}
The program for calculating the NLO Higgs impact factor is the following: 
\begin{itemize}
    \item \textbf{Real corrections} \\
    At NLO the process can be initiated either by a gluon or by a quark extracted from the proton. Moreover, the Higgs must be accompanied by an additional parton (see the first two diagrams from the top of Fig.~\ref{ProgramHiggsNLO}). These corrections will be calculated in section~\ref{sec:real}, separating the quark-initiated case (subsection~\ref{ssec:quark}) and the gluon-initiated one (subsection~\ref{ssec:gluon}).
    \item \textbf{Virtual corrections} \\
    Another NLO order correction is obtained when we take one of the two effective vertices $\Gamma$ that appear in the definition of the impact factor at one loop and the other at Born level (see diagrams in the last line of Fig.~\ref{ProgramHiggsNLO}). This correction is trivial to compute once the 1-loop correction to the vertex~(\ref{gRHBorn}) has been extracted. This problem will be addressed in the section~\ref{sec:virtual}. 
    \item \textbf{Projection onto the eigenfunctions of the LO BFKL kernel and cancellation of divergences} \\
    In section~\ref{sec:projection} we perform the convolution with the PDFs for the previously calculated contributions and show that, after (i) carrying out the UV renormalization of the strong coupling, (ii) introducing the counterterms associated with the PDFs, (iii) performing the projection on the eigenfunctions of the LO BFKL kernel (more details in section~\ref{sec:projection}), the final result is free from any kind of divergence.
    
\end{itemize}

\section{NLO impact factor: Real corrections}
\label{sec:real}

In this section, we compute real corrections to the Higgs impact factor. At NLO both initial-state gluon and quark can contribute: if the process is initiated by a quark (gluon), then the intermediate state $\{f\}$ will contain the Higgs and a quark (gluon).
The Sudakov decomposition of the momentum of the produced parton reads
\begin{equation}
    p_p = z_p k_1 + \frac{\vec{p}_p^{\; 2}}{z_p s} k_2 + p_{p \perp} \; , \hspace{0.5 cm} p_p^2 = 0 \; , 
\end{equation}
where the subscript $p$ is equal to $q$ ($g$) in the quark (gluon) case. We have then
\begin{equation}
     s_{pR} = (p_p+p_H)^2 = \frac{z_p (z_H+z_p) m_H^2 + (z_p \vec{p}_H - z_H \vec{p}_p)^2}{z_H z_p} \; ,
     \label{SpR}
\end{equation}
\begin{equation}
    \frac{d s_{pR}}{2 \pi} d \rho_{pH} = \delta (1 - z_p - z_H) \delta^{(2)} (\vec{p}_p + \vec{p}_H - \vec{q} \; ) \frac{d z_p d z_H}{z_p z_H} \frac{d^{D-2} p_p d^{D-2} p_H}{2 (2 \pi)^{D-1}} \; .
\end{equation}
The integration over the kinematic variables of the produced parton is trivial due to the
presence of the delta functions and results in the constraints
\begin{equation}
    z_p = 1 - z_H \; , \hspace{0.6 cm} \vec{p}_p \equiv \vec{q} - \vec{p}_H \; .
\end{equation}
The integration over $z_H$ and $\vec p_H$ is not performed, since our target is an impact factor differential in the kinematic variables of the Higgs.

In the gluon case we will use the Sudakov decompositions of the polarization vectors
of the initial-state gluon with momentum $k_1$ and of the produced gluon with momentum $p_g$,
\begin{equation}
   \varepsilon (k_1) =  \varepsilon_{\perp} (k_1) \;, \hspace{0.6 cm}
   \varepsilon (p_g) = - \frac{(p_{g , \perp} \cdot \varepsilon_{\perp} (p_g))}{(p_g \cdot k_2)} k_2 + \varepsilon_{\perp} (p_g) \; ,
   \label{Pol12}
  \end{equation}
which encode the gauge condition $\varepsilon\cdot k_2=0$ for both gluons. 

We observe that, when calculating NLO real corrections, the $s_0$-dependent factor in the definition~(\ref{ImpactUnpro}) can be taken at the lowest order in the perturbative expansion and therefore put equal to one. Moreover, in the quark case there is no rapidity divergence for $s_{pR}\to \infty$, therefore the regulator $s_\Lambda$ can be sent to infinity in the argument of the theta function and the second term in Eq.~(\ref{ImpactUnpro}) (the so-called ``BFKL counterterm'') can be omitted.

\subsection{Quark-initiated contribution}
\label{ssec:quark}

\begin{figure}
\begin{center}
{\parbox[t]{6 cm}{\epsfysize 4cm \epsffile{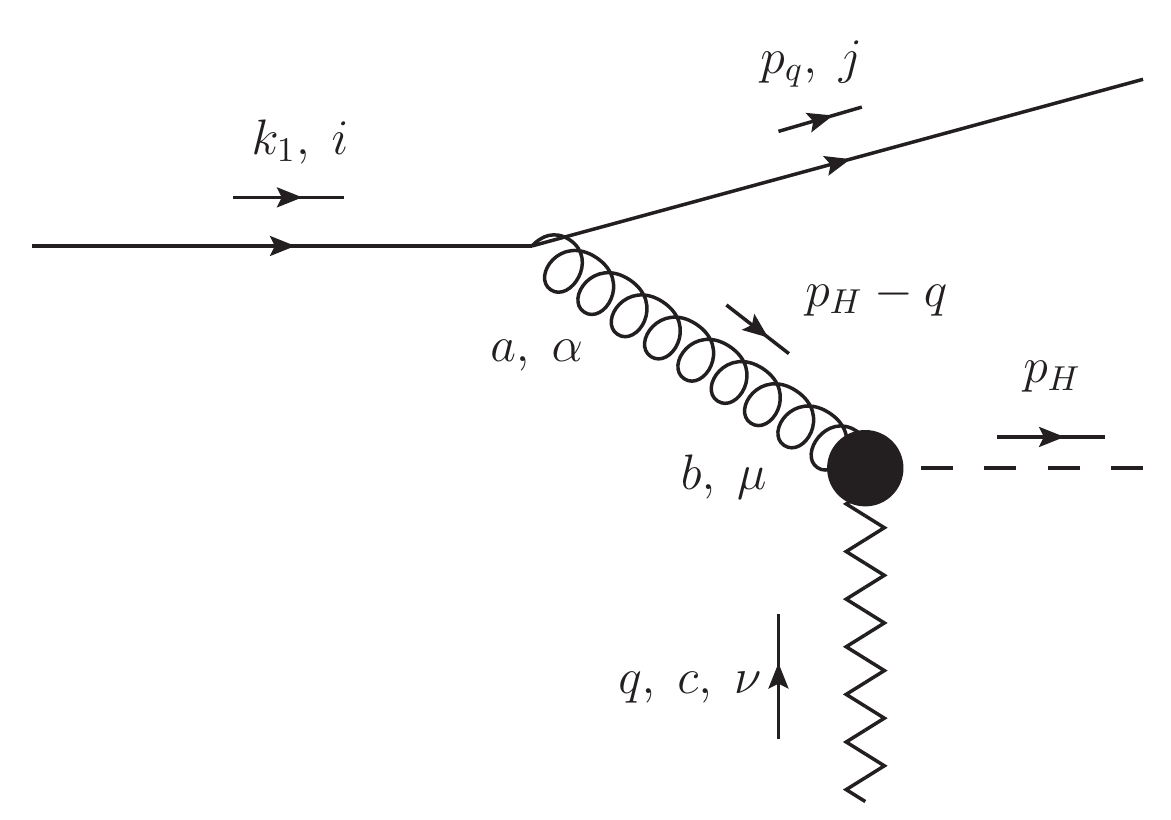}}}
\end{center}
\caption{Feynman diagram contributing to the $q \;R \rightarrow q \; H$ amplitude.}
\label{QuarkDiagr}
\end{figure}

In the case of incoming quark, the contribution to the impact factor reads
\begin{equation}
    d \Phi_{q q}^{\{H q \}} (\vec{q}) = \langle cc^{\prime} | \hat{\cal P}_{0} | 0 \rangle \frac{1}{2 N} \sum_{\substack{i, j \\ \lambda,\lambda'}}\int \frac{d s_{q R}}{2 \pi} d \rho_{\{H q \}} \Gamma_{\{ H q \} q}^{c(0)} (q) \left( \Gamma_{\{ H q \} q}^{c'(0)} (q) \right)^{*}  \; .
    \label{QuarkImpDef}
\end{equation}
where color and spin states have been averaged for the initial-state quark and summed over for 
the produced one. There is only one Feynman diagram to consider, shown in Fig.~\ref{QuarkDiagr}, leading to
\begin{equation}
  \Gamma_{\{ H q \} q}^{c(0)} = - g g_H t_{i j}^c \frac{1-z_H}{(\vec{q}-\vec{p}_H)^2} \bar{u} (p_q) \left( \frac{\slashed{k}_2}{s} (p_H-q) \cdot q - \slashed{q} \left( (p_H-q) \cdot \frac{k_2}{s} \right) \right) u(k_1) \; .
  \label{QuarkAmpBorn}
\end{equation}
Substituting Eq.~(\ref{QuarkAmpBorn}) into Eq.~(\ref{QuarkImpDef}), one easily gets the quark initiated contribution to the impact factor,
\begin{equation*}
    \frac{d \Phi_{q q}^{\{H q \}} (z_H, \vec{p}_H, \vec{q})}{d z_H d^2 \vec{p}_H} = \frac{\sqrt{N^2-1}}{16 N (2 \pi)^{D-1}} \frac{g^2 g_H^2}{[(\vec{q}-\vec{p}_H)^2]^2}
    \end{equation*}
    \begin{equation}
    \times \left[ \frac{4 (1-z_H) \left[(\vec{q}-\vec{p}_H) \cdot \vec{q} \; \right]^2 + z_H^2 \vec{q}^{\; 2} (\vec{q} - \vec{p}_H)^2}{z_H} \right] \; .
    \label{QuarkConImpacFin}
\end{equation}
We observe that this contribution to the impact factor vanishes in the limit $\vec q \to 0$, as required by gauge invariance.

\subsection{Gluon-initiated contribution}
\label{ssec:gluon}

\begin{figure}
  \begin{center}
  \includegraphics[scale=0.50]{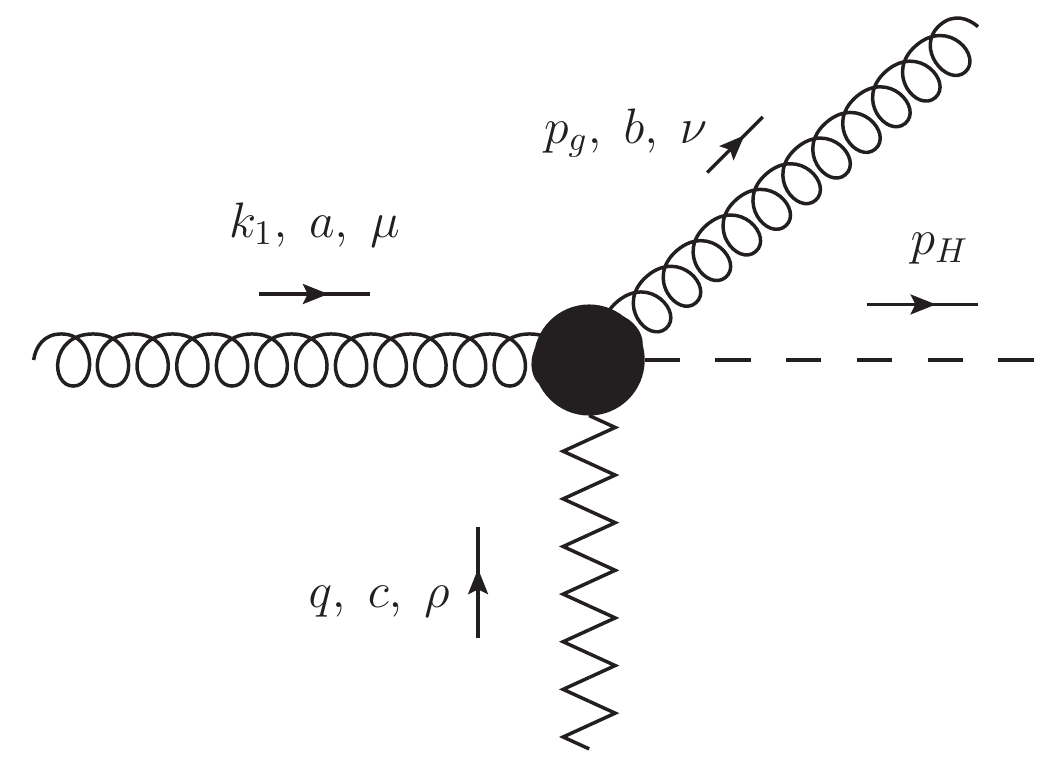} \hspace{1.5 cm}
  \includegraphics[scale=0.50]{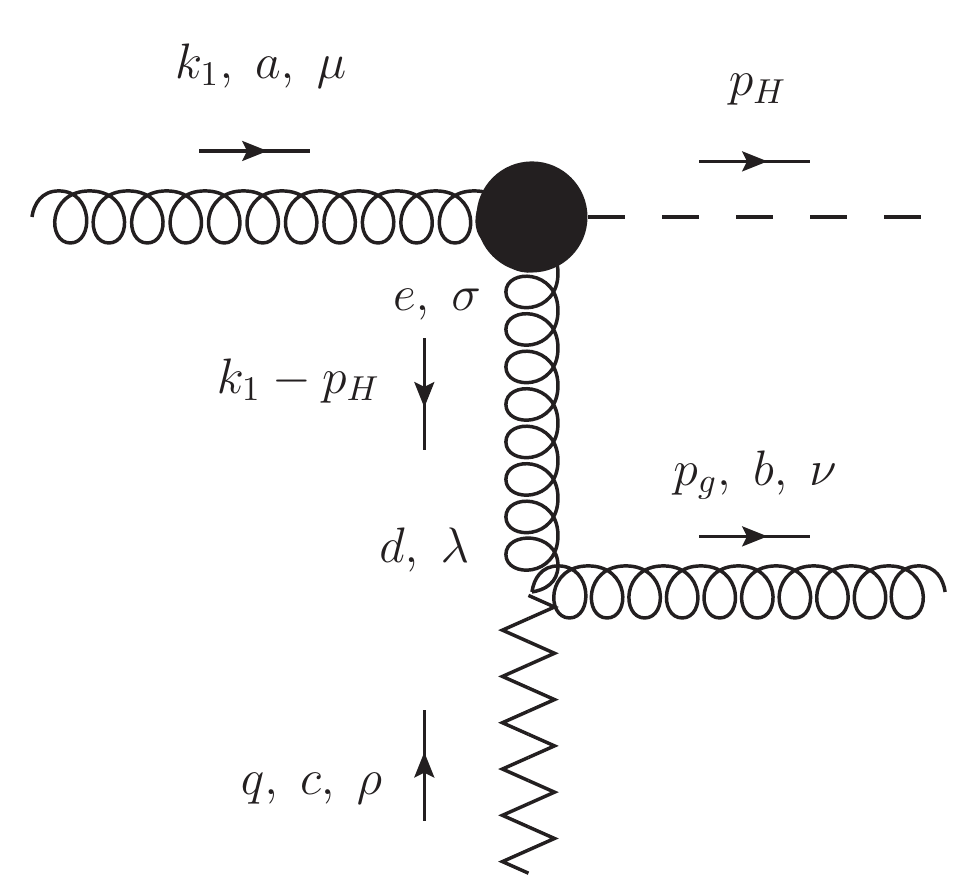} \vspace{1.0 cm} \\
  \includegraphics[scale=0.55]{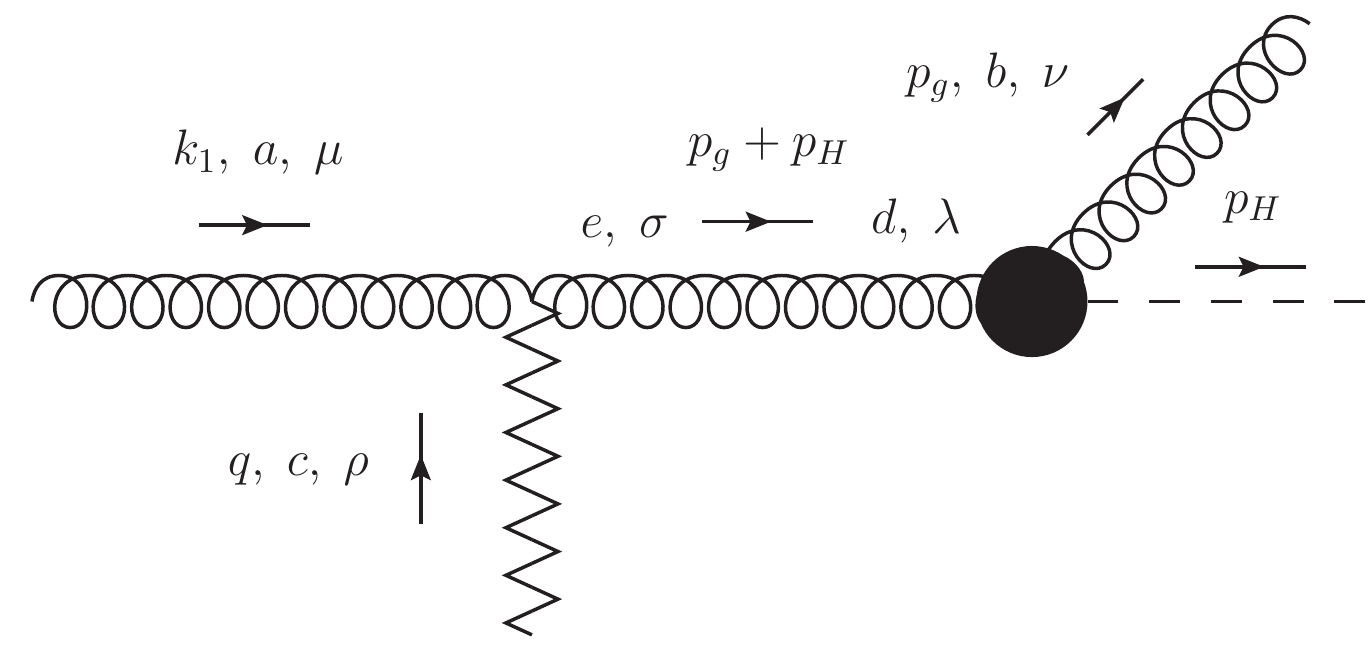} \hspace{0.3 cm}
  \includegraphics[scale=0.55]{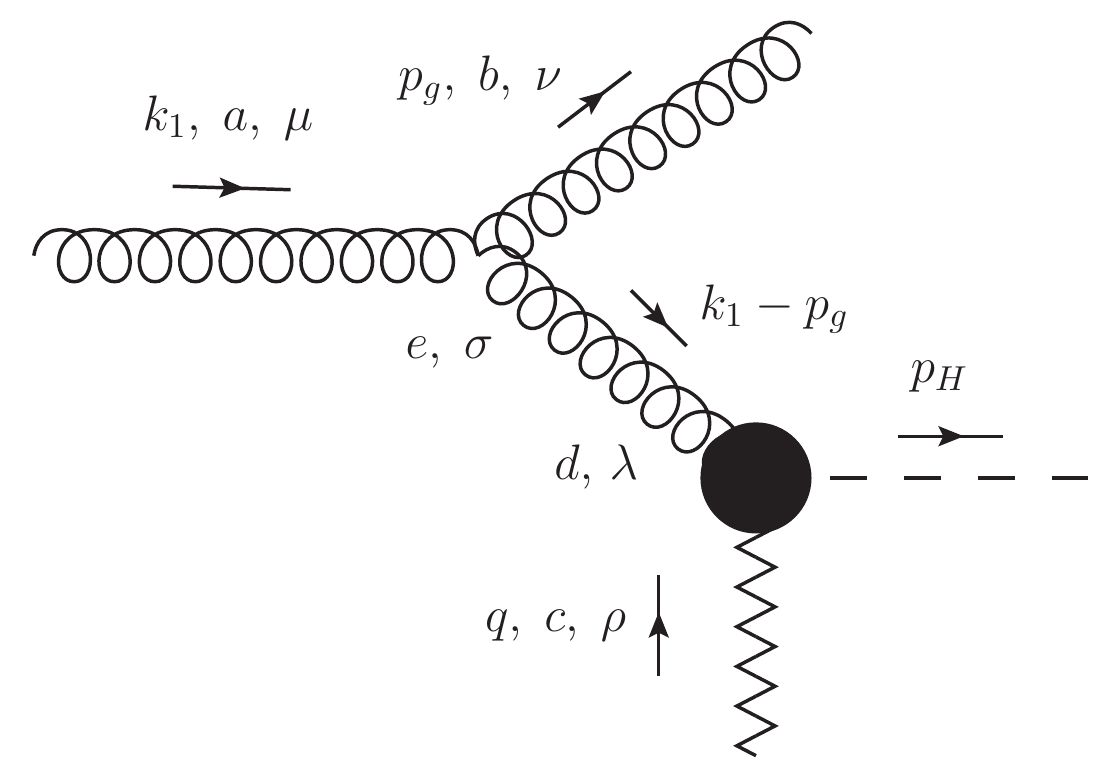}
  \end{center}
  \caption{Feynman diagrams contributing to the $g R \rightarrow g H$ amplitude. In the text we label them from (1) to (4) starting from the top left.}
  \label{GluonDiagrams}
\end{figure}

In the case of gluon in the initial state, the NLO real corrections, which are given, up to the BFKL counterterm, by the first term in Eq.~(\ref{ImpactUnpro}), take the following form
\begin{equation}
    d \Phi_{g g}^{\{H g \}} (\vec{q}) =  \frac{\braket{c c'|\mathcal{P}|0}}{2(1-\epsilon) (N^2-1)} \sum_{\substack{a, b \\ \lambda,\lambda'}}\int \frac{d s_{g R}}{2 \pi} d \rho_{\{H g \}}  \Gamma_{\{ H g \} g}^{a b c(0)} (q) \left( \Gamma_{\{ H g \} g}^{a b c' (0)} (q) \right)^{*} \theta (s_{\Lambda}-s_{gR})\; ,
    \label{GluonIniImp}
\end{equation}
where color and polarization states have been averaged for the initial-state gluon and summed over for the produced one.
The Feynman diagrams contributing to the amplitude $\Gamma_{\{ H g \} g}^{a b c(B)}$ of the $g R \rightarrow g H$ process are shown in Fig.~\ref{GluonDiagrams}, and give
\begin{equation}
\Gamma_1 = - i g g_H f^{abc} \varepsilon_{\mu} (k_1) \varepsilon_{\nu}^{*} (p_g)  \frac{2-z_H}{2}  g^{\mu \nu} \; ,
\end{equation}
\begin{equation}
\Gamma_2 = \frac{-i g g_H f^{abc} \varepsilon_{\mu} (k_1) \varepsilon_{\nu}^{*} (p_g) }{2[(z_H-1)m_H^2 - \vec{p}_H^{\; 2}]} \left[ (1-z_H) (m_H^2 + \vec{p}_H^{\; 2})g^{\mu \nu} - 2 z_H p_H^{\mu} (p_H^{\nu} - z_H k_1^{\nu}) \right] \; , 
\end{equation}
\begin{equation*}
\Gamma_3 = \frac{i g g_H f^{abc}\varepsilon_{\mu} (k_1) \varepsilon_{\nu}^{*} (p_g)}{2 \left[(1-z_H) m_H^2 + (\vec{p}_H-z_H \vec{q} \; )^2 \right]}
\end{equation*}
\begin{equation}
    \times \left[ g^{\mu \nu} \left( (1-z_H)^2 m_H^2 + (\vec{p}_H-z_H \vec{q} \; )^2 \right) +2 z_H (1-z_H)^2 p_H^{\mu} p_H^{\nu} -2 z_H^2 (1-z_H) p_g^{\mu} p_H^{\nu} )\right] \; ,
\end{equation}
\begin{equation}
\Gamma_4 = -i g g_H f^{abc} \varepsilon_{\mu} (k_1) \varepsilon_{\nu}^{*} (p_g) \frac{(1-z_H)}{(\vec{q}-\vec{p}_H)^2} \left[ - g^{\mu \nu} (\vec{q} - \vec{p}_H) \cdot \vec{q} - z_H (p_H^{\mu} k_1^{\nu} + p_g^{\mu} p_H^{\nu}) \right] \; ,
\end{equation}
which sum up to
\begin{equation*}
\Gamma_{\{ H g \} g}^{a b c(0)} = i g g_H f^{abc} \varepsilon_{\mu} (k_1) \varepsilon_{\nu}^{*} (p_g) \left[ \frac{\left[ (1-z_H) (m_H^2 + \vec{p}_H^{\; 2})g^{\mu \nu} - 2 z_H p_H^{\mu} (p_H^{\nu} - z_H k_1^{\nu}) \right] }{2[(1-z_H) m_H^2 + \vec{p}_H^{\; 2}]} \right.
\end{equation*}
\begin{equation*}
    + \left. \frac{\left[ g^{\mu \nu} \left( (1-z_H)^2 m_H^2 + (\vec{p}_H-z_H \vec{q} \; )^2 \right) +2 z_H (1-z_H)^2 p_H^{\mu} p_H^{\nu} -2 z_H^2 (1-z_H) p_g^{\mu} p_H^{\nu} )\right]}{2 \left[(1-z_H) m_H^2 + (\vec{p}_H-z_H \vec{q} \; )^2 \right]} \right. 
\end{equation*}
\begin{equation}
    \left. -\frac{2-z_H}{2}  g^{\mu \nu} + \frac{(1-z_H)\left[ g^{\mu \nu} (\vec{q} - \vec{p}_H) \cdot \vec{q} + z_H (p_H^{\mu} k_1^{\nu} + p_g^{\mu} p_H^{\nu}) \right]}{(\vec{q}-\vec{p}_H)^2}  \right] \; .
\end{equation}
Using Eqs.~(\ref{Pol12}), we can decouple longitudinal and transverse degrees of freedom and get
\begin{equation*}
\Gamma_{\{ H g \} g}^{a b c(0)} = -i g g_H f^{abc} \varepsilon_{\mu \perp} (k_1) \varepsilon_{\nu \perp}^{*} (p_g) \left[ \frac{2 z_H p_{H \perp}^{\mu} p_{H \perp}^{\nu} - z_H (1-z_H) m_H^2 g^{\mu \nu}}{2 \left[ (1-z_H) m_H^2 + \vec{p}_H^{\;2} \right]} \right.
\end{equation*}
\begin{equation}
    \left. - \frac{2 z_H \Delta_{\perp}^{\mu} \Delta_{\perp}^{\nu} - z_H (1-z_H) m_H^2 g^{\mu \nu}}{2 \left[ (1-z_H) m_H^2 + \vec{\Delta}^2 \right]}  + \frac{z_H (q_{\perp}^{\mu} r_{\perp}^{\nu} - (1-z_H) r_{\perp}^{\mu} q_{\perp}^{\nu})-(1-z_H) \vec{r} \cdot \vec{q} \; g^{\mu \nu}}{\vec{r}^{\; 2}} \right] \; ,
    \label{FinFormGluAmp}
\end{equation}
where we have defined, similarly to~\cite{Hentschinski:2020tbi}, 
\begin{equation}
    \Delta_{\perp}^{\mu} \equiv  p_{H,\perp}^{\mu} - z_H q_{\perp}^{\mu} \; , \hspace{1 cm} r_{\perp} \equiv q_{\perp}^{\mu} - p_{H,\perp}^{\mu} \; .
\end{equation}
We can finally plug this expression into Eq.~(\ref{GluonIniImp}) and get the gluon-initiated contribution to the impact factor,
\begin{equation*}
    \frac{d \Phi_{g g}^{\{H g \}} (z_H, \vec{p}_H, \vec{q})}{d z_H d^{2} p_H} = \frac{g^2 g_H^2 C_A}{8 (2 \pi)^{D-1}(1-\epsilon) \sqrt{N^2-1}} \left \{ \frac{2}{z_H (1-z_H)} \right.
\end{equation*}
\begin{equation*}
     \left. \left[ 2 z_H^2 + \frac{(1-z_H)z_H m_H^2 (\vec{q} \cdot \vec{r}) [z_H^2 - 2 (1-z_H) \epsilon]+2 z_H^3 (\vec{p}_H \cdot \vec{r}) (\vec{p}_H \cdot \vec{q})}{\vec{r}^{\; 2} \left[ (1-z_H) m_H^2 + \vec{p}_H^{\; 2} \right]} - \frac{2 z_H^2 (1-z_H) m_H^2}{\left[ (1-z_H) m_H^2 + \vec{p}_H^{\; 2}  \right]}  \right. \right. 
\end{equation*}
\begin{equation*}
    -\frac{(1-z_H)z_H m_H^2 (\vec{q} \cdot \vec{r}) [z_H^2 - 2 (1-z_H) \epsilon]+2 z_H^3 (\vec{\Delta} \cdot \vec{r}) (\vec{\Delta} \cdot \vec{q})}{\vec{r}^{\; 2} \left[ (1-z_H) m_H^2 + \vec{\Delta}^{ 2} \right]} - \frac{2 z_H^2 (1-z_H) m_H^2}{\left[ (1-z_H) m_H^2 + \vec{\Delta}^{2}  \right]}
\end{equation*}
\begin{equation*}
   \left. + \frac{(1-\epsilon) z_H^2 (1-z_H)^2 m_H^4}{2} \left( \frac{1}{\left[ (1-z_H) m_H^2 + \Delta^{2}  \right]} + \frac{1}{\left[ (1-z_H) m_H^2 + \vec{p}_H^{\; 2}  \right]}  \right)^2 \right.
\end{equation*}
\begin{equation*}
   \left. - \frac{2 z_H^2 (\vec{p}_H \cdot \vec{\Delta})^2 - 2 \epsilon (1-z_H)^2 z_H^2 m_H^4}{\left[ (1-z_H) m_H^2 + \vec{p}_H^{\; 2}  \right] \left[ (1-z_H) m_H^2 + \Delta^{2}  \right]} \right]
\end{equation*}
\begin{equation}
     \left. + \frac{2 \vec{q}^{\; 2}}{\vec{r}^{\; 2}} \left[ \frac{z_H}{1-z_H} + z_H (1-z_H) + 2 (1-\epsilon) \frac{(1-z_H)}{z_H} \frac{(\vec{q} \cdot \vec{r})^2}{\vec{q}^{\; 2} \vec{r}^{\; 2}} \right] \right \} \theta \left( s_{\Lambda} - \frac{(1-z_H) m_H^2 + \vec{\Delta}^2}{z_H (1-z_H)} \right) \; .
     \label{GluonImp}
\end{equation}
We notice that this expression is compatible with gauge invariance, since it vanishes for $\vec q \to 0$. We have presented our result for $\Phi_{g g}^{\{H g \}}$ in a form similar to that given in Eq.~(46) of Ref.~\cite{Hentschinski:2020tbi} to facilitate the comparison. One can see that the expression in Ref.~\cite{Hentschinski:2020tbi} agrees with ours\footnote{We stress that in Ref.~\cite{Hentschinski:2020tbi} authors work in $D=4+2 \epsilon$, while we work in $D=4-2 \epsilon$}, except for:
\begin{itemize}
    \item two terms which are proportional to $z_H (1-z_H)^2$ instead of $z_H^2 (1-z_H)$; 
    \item two terms, proportional to $z_H^3$, which have a sign different from ours.
\end{itemize}
These little discrepancies are due to misprints in Ref.~\cite{Hentschinski:2020tbi}, as privately communicated to us by the authors of that paper.

\section{NLO impact factor: Virtual corrections}
\label{sec:virtual}

In this section, we compute the contribution to the impact factor coming from virtual corrections. The basic ingredient we need is the 1-loop correction to the high-energy vertex in Eq.~(\ref{gRHBorn}), that we indicate as\footnote{We remind that this is the effective vertex for the coupling of the initial-state gluon with the final-state Higgs and a $t$-channel Reggeized gluon, which is an object in the octet color representation with negative signature. Up to the NLO, the effective vertex takes contribution from the exchange in the $t$-channel of either one or two gluons; in the latter case, the projection onto the octet color representation with negative signature should be taken; in our calculation this projection is automatic in presence of an initial-state gluon, since the Higgs is a color singlet state.}
\begin{equation}
    \Gamma_{\{ H\} g}^{ac(1)} (q) =  \Gamma_{\{ H\} g}^{ac(0)} (q) \left[ 1 + \delta_{{\rm{NLO}}} \right] \; .
\end{equation}
Thereafter, the virtual contribution to the impact factor can be calculated as
\begin{equation*}
    \Phi_{gg}^{\{ H \}(1)} (\vec{q}; s_0 ) = \frac{\langle
cc^{\prime} | \hat{\cal P}_{0} | 0 \rangle}{2 (1-\epsilon) (N^2-1)}
\end{equation*}
\begin{equation}
   \times \sum_{a, \lambda } \int \frac{d s_{gR}}{2 \pi} d \rho_H  \Gamma_{\{ H \} g}^{ac(0)} (q) \left( \Gamma_{\{ H \} g}^{ac'(0)} (q) \right)^{*} \left[ \omega^{(1)}(-\vec{q}^{\; 2}) \ln \left( \frac{s_0}{\vec{q}^{\; 2}} \right) + \delta_{{\rm{NLO}}} + \delta_{{\rm{NLO}}}^{*} \right] \; ,
\label{HiggsVirtImp}
\end{equation}
where the $s_0$-dependent term comes from the expansion of $(s_0/\vec q^{\,2})^{\omega^{(1)}(-\vec q^{\,2})}$ in~(\ref{ImpactUnpro}). 

\begin{figure}
  \begin{center}
  \includegraphics[scale=0.50]{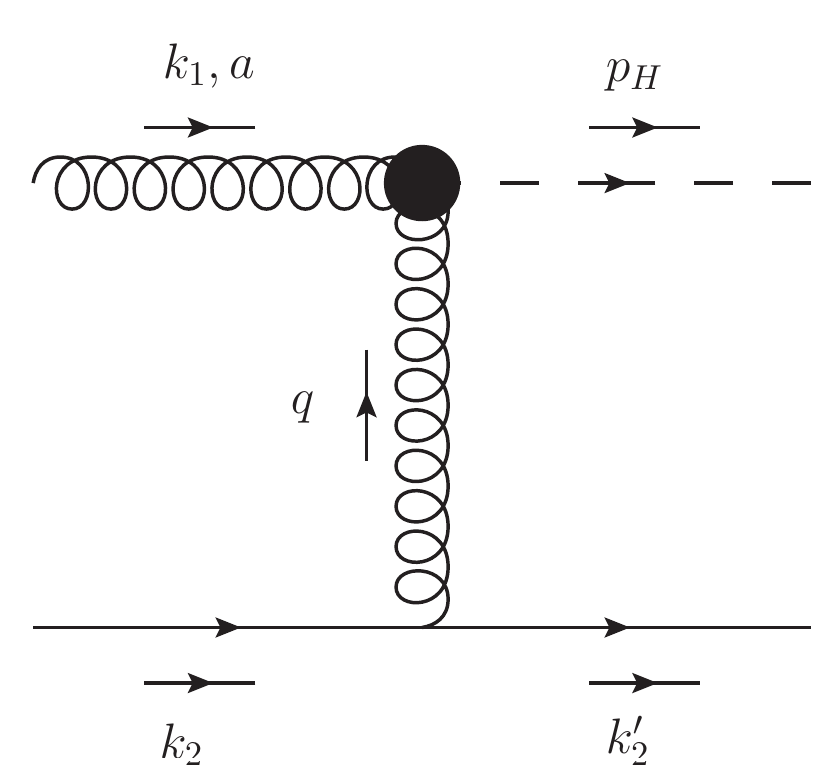} \hspace{1.5 cm}
  \end{center}
  \caption{$g q \rightarrow H q$ amplitude.}
  \label{LOgqHqAmp}
\end{figure}

The general strategy to find a NLO high-energy effective vertex is to compute, in the 
NLO and in the high-energy limit, a suitable amplitude and to compare it with the expected Regge form, written in terms of the needed effective vertex and of another known one. For our
purposes, we consider the diffusion of a gluon off a quark to produce a Higgs plus a quark, whose amplitude ${\cal A}_{g q \rightarrow H q}$, for the octet colour state and the negative signature in the $t$-channel has the following Reggeized form (see Fig.~\ref{LOgqHqAmp}):
$$
{\cal A}_{g q \rightarrow H q}^{(8,-)} = \Gamma_{ \{H\} g}^{ac} \frac{s}{t}\left[ \left( \frac{s}{-t} \right)^{\omega(t)} + \left(
\frac{-s}{-t} \right)^{\omega(t)} \right]\Gamma_{qq}^{c} \approx
\Gamma_{\{H\} g}^{ac(0)} \frac{2s}{t} \Gamma_{qq}^{c(0)}
$$
\begin{equation}
+ \Gamma_{\{H\} g}^{ac(0)} \frac{s}{t}\omega^{(1)}(t)
\left[ \ln\left( \frac{s}{-t} \right) + \ln\left(
\frac{-s}{-t} \right) \right]\Gamma_{q q}^{c(0)} + \Gamma_{\{H\} g}^{ac(0)} \frac{2s}{t}\Gamma_{q q}^{c(1)} +
\Gamma_{\{H\} g}^{ac(1)}\frac{2s}{t} \Gamma_{qq}^{c(0)} \; ,
\label{ReggeFormEx1}
\end{equation}
where 
\begin{equation}
    \Gamma_{qq}^{c(0)} = g t_{ji}^c \bar{u} (k_2-q) \frac{\slashed{k}_1}{s} u(k_2)
\end{equation} 
is the LO quark-quark-Reggeon effective vertex and $\Gamma_{qq}^{c(1)}$ its 1-loop correction,
known since long~\cite{Fadin:1993qb,Fadin:1995km}, and $t=q^2=-\vec q^{\,2}$.

Let us now split the 1-loop contributions to this amplitude into three pieces, related to
2-gluon (2g) or 1-gluon (1g) exchange or self-energy (se) diagrams in the $t$-channel:
$$
{\cal A}_{g q \rightarrow H q}^{(2g)(8,-)(1)} + {\cal A}
_{g q \rightarrow H q}^{(se)(1)} + {\cal A}_{g q \rightarrow H q}^{(1g)(1)} = \left\{ \Gamma_{\{H\} g}^{(2g)ac(1)}\frac{2s}{t}\Gamma_{qq}^{c(0)} + \Gamma_{\{H\} g}^{ac(0)}
\frac{2s}{t}\Gamma_{q q}^{(2g)c(1)} \right.
$$
$$
\left. + \Gamma_{\{H\} g}^{c(0)}\frac{s}{t}\omega^{(1)}(t)
\left[ \ln\left( \frac{s}{-t} \right) + \ln\left( \frac{-s}{-t}
\right) \right]\Gamma_{q q}^{c(0)} \right\}
$$
\begin{equation}
+ \left\{ \Gamma_{\{H\} g}^{(se)ac(1)}\frac{2s}{t}\Gamma
_{qq}^{c(0)} +  \Gamma_{\{H\} g}^{ac(0)}\frac{2s}{t}
\Gamma_{qq}^{(se)c(1)} \right\} +  \left\{ \Gamma_{\{H\} g}
^{(1g)ac(1)}\frac{2s}{t}\Gamma_{qq}^{c(0)} + \Gamma_{\{H\} g}
^{c(0)}\frac{2s}{t}\Gamma_{qq}^{(1g)c(1)} \right\},
\label{ReggeFormEx2}
\end{equation}
where the self-energy diagrams and 1-gluon exchange diagrams are automatically in the $8^-$
color representation.

We stress that we will use a unique regulator $\epsilon \equiv \epsilon_{\rm UV} \equiv \epsilon_{\rm IR}$ for both ultraviolet (UV) and infrared (IR) divergences. As a consequences, scaleless integrals such as
\begin{equation}
   \int \frac{d^D k}{i (2 \pi)^D} \frac{1}{k^2 (k+k_1)^2} 
\end{equation}
are equal to zero due to the exact cancellation between IR- and UV-divergence. In our case of massless partons, this implies that the contribution from the renormalization of the external quark and gluon lines is absent in dimensional regularization.

The 1-gluon-exchange contribution $\Gamma_{g H}^{(1g)ac(1)}$ can be calculated in a straightforward way by taking the radiative corrections to the amplitude of Higgs production in the collision of an on-shell gluon with an off-shell one (the gluon in the $t$-channel) having momentum $q$, colour index $c$ and ``nonsense'' polarization vector $-k_2/s$. In a similar manner one could calculate the 1-gluon-exchange contribution to the quark-quark-Reggeon effective vertex; in this case, the $t$-channel off-shell gluon must be taken with ``nonsense'' polarization vector $-k_1/s$. This is the consequence of the Gribov trick on the $t$-channel gluon propagator, valid in the high-energy limit.
The 1-gluon-exchange contribution does not include self-energy corrections to the $t$-channel gluon, which must be calculated separately and assigned with weight equal to one half to both $\Gamma_{\{H\}g}^{(se)ac(1)}$ and $\Gamma_{qq}^{(se)c(1)}$. 

For the 2-gluon exchange contributions we have the relation
\begin{equation*}
\Gamma_{\{H\} g}^{(2g)ac(1)}\frac{2s}{t}\Gamma_{qq}^{c(0)} +
\Gamma_{\{H\} g}^{ac(0)}\frac{2s}{t}\Gamma_{qq}^{(2g)c(1)} 
\end{equation*}
\begin{equation}
={\cal A}_{g q \rightarrow H q}^{(2g)(8,-)(1)} - \Gamma
_{\{H\} g}^{ac(0)}\frac{s}{t}\omega^{(1)}(t)\left[ \ln\left(
\frac{s}{-t} \right) + \ln\left( \frac{-s}{-t}
\right) \right]\Gamma_{qq}^{c(0)},
\label{FromExt2Gluon}
\end{equation}
which shows that we need to know the correction
$\Gamma_{qq}^{(2g)c(1)}$. This correction has been obtained in Ref.~\cite{Fadin:2001ap} and reads
\begin{equation}\label{311} \Gamma_{qq}^{(2g)c(1)} =\delta_{qq}^{(2g)}(t)
\Gamma_{qq}^{c(0)}\;, 
\end{equation}
with
$$
\delta_{qq}^{(2g)}(t) = \frac{1}{2}\omega^{(1)}(t)\left[ -\frac{1}{\epsilon}
+ \psi(1) + \psi(1+\epsilon) - 2\psi(1-\epsilon) \right]
$$
\begin{equation}
= g^2N\frac{\Gamma(2+\epsilon)}{(4\pi)^{2-\epsilon}}\frac{1}
{\epsilon}\left( -t \right)^{-\epsilon}\left( -\frac{1}{\epsilon} + 1 -
\epsilon + 4\epsilon\psi^\prime(1) \right) + {\cal O}(\epsilon).
\end{equation}

\subsection{The 1-loop correction: 1-gluon exchange and self-energy diagrams}
\label{ssec:virtual1Gluon}

\begin{figure}
  \begin{center}
  \includegraphics[scale=0.50]{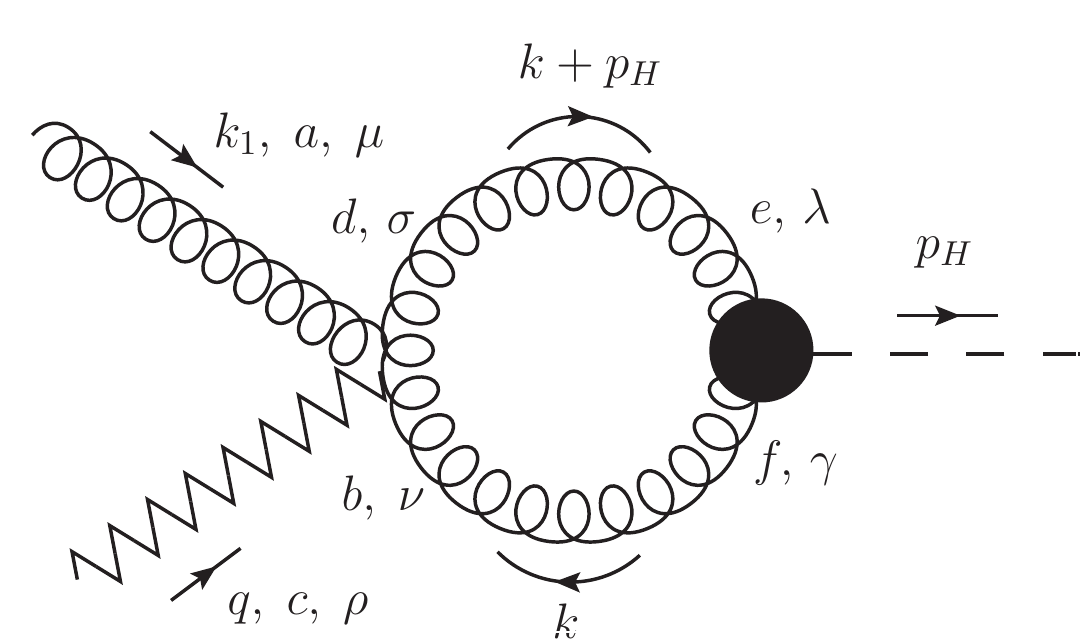} \hspace{1.5 cm}
  \includegraphics[scale=0.50]{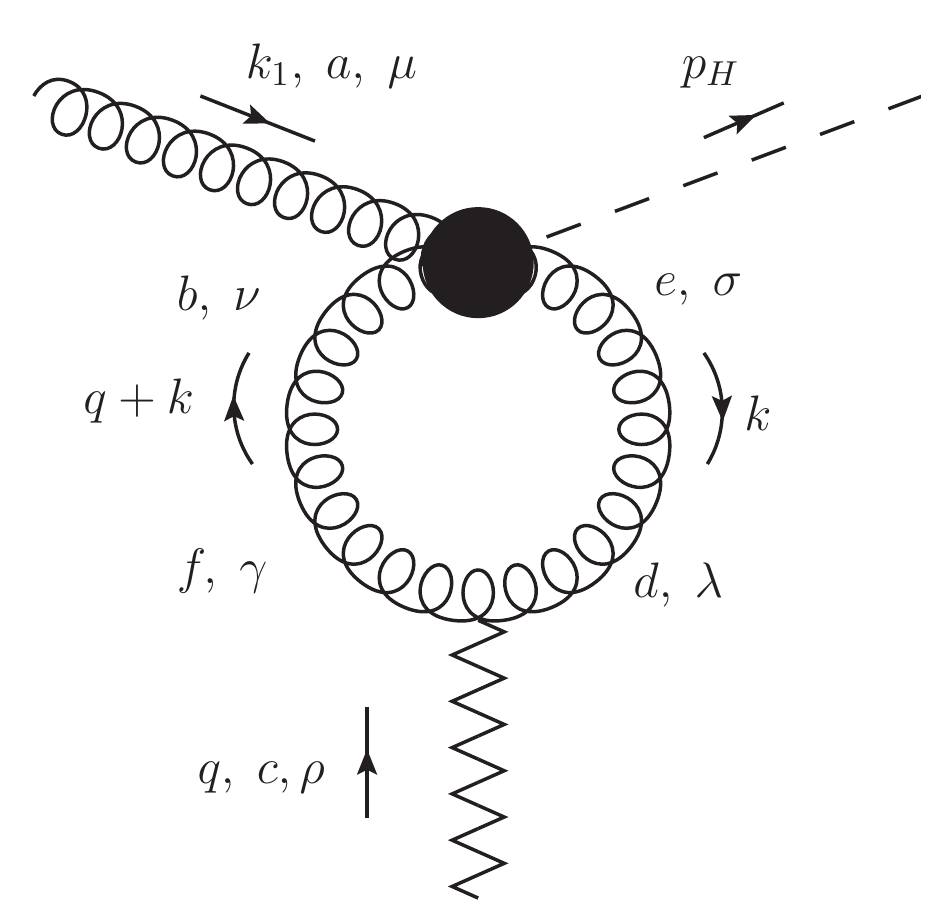} \vspace{1.0 cm} \\ \hspace{1.2 cm}
  \includegraphics[scale=0.53]{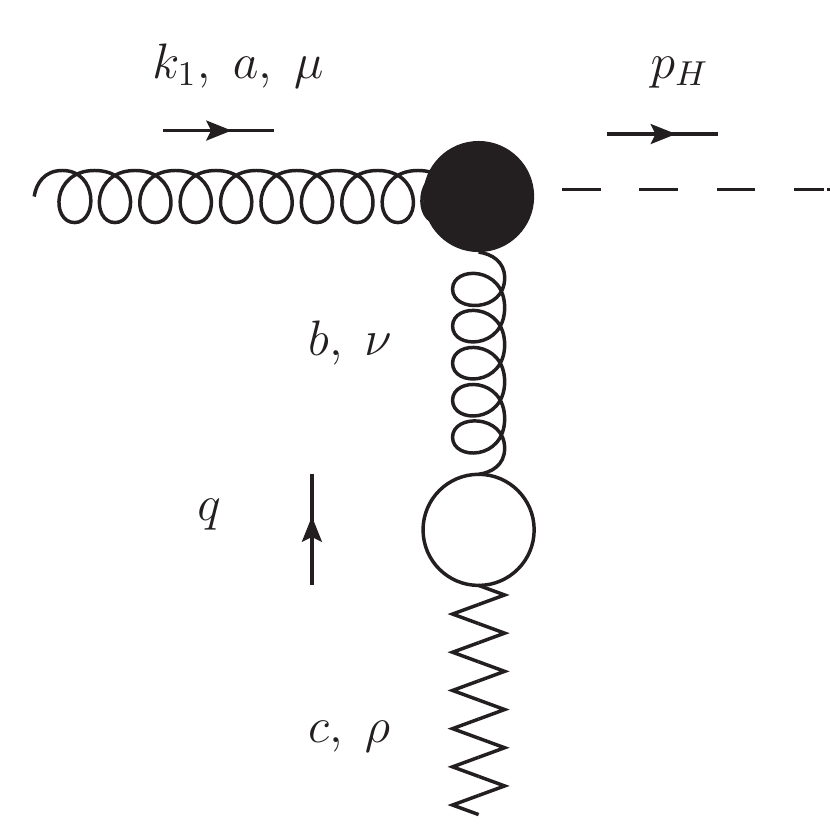} \hspace{1.0 cm} 
  \includegraphics[scale=0.50]{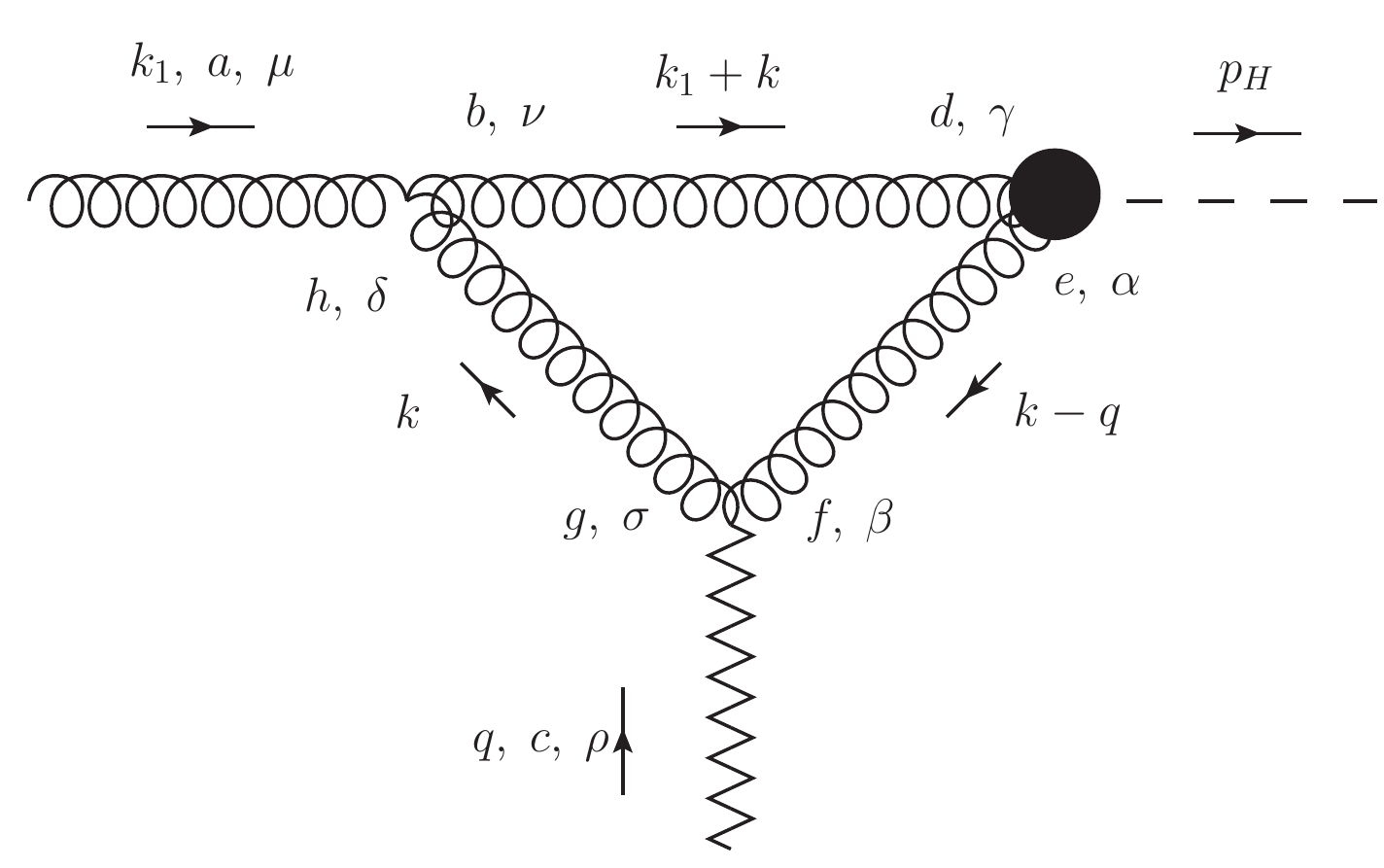} \\
  \end{center}
  \caption{1-gluon in the $t$-channel exchange diagrams. In the text we label them from (1) to (4) starting from the top left.}
  \label{VirtualDiagrams}
\end{figure}
There are four non-vanishing diagrams with 1-gluon exchange in the $t$-channel, including the one with self-energy corrections to the exchanged gluon; they are shown in Fig.~\ref{VirtualDiagrams}, where, the blob in the third diagram means the sum of gluon, ghost and quark loop contributions. 

Let's start by considering the first diagram; it gives
\begin{equation}
    \Gamma_{1,V}^{(1g)} = \frac{1}{2} i g^2 g_H \epsilon_{\mu} (k_1) \frac{k_{2,\rho}}{s} \int \frac{d^D k}{(2 \pi)^D} \frac{\Gamma_{abcb}^{\mu \rho \nu \sigma} H_{\sigma \nu} (-k-p_H,k)}{k^2 (k+p_H)^2} \; ,
\end{equation}
where 
\begin{equation*}
     \Gamma_{abcd}^{\mu \rho \nu \sigma} = \left[ (T^g)_{ab} (T^g)_{cd} (g^{\mu \rho} g^{\nu \sigma} - g^{\mu \sigma} g^{\nu \rho}) + (T^g)_{ac} (T^g)_{bd} (g^{\mu \nu} g^{\rho \sigma} - g^{\mu \sigma} g^{\nu \rho} ) \right. 
\end{equation*}
\begin{equation}
   \left. + (T^g)_{ad} (T^g)_{cb} (g^{\mu \rho} g^{\nu \sigma} - g^{\mu \nu} g^{\sigma \rho} ) \right]\;, \;\;\;\;\;  (T^a)_{bc}=-i f_{abc} \; ,
\end{equation}
and $H_{\sigma\nu}$ is defined in~(\ref{ggH}).  
After a trivial computation, we obtain
\begin{equation}
    \Gamma_{1,V}^{(1g)} =\Gamma_{\{H\} g}^{ac(0)} N g^2 \frac{(D-2)}{4(D-1)}  B_0(m_H^2) \equiv \Gamma_{\{H\} g}^{ac(0)} \delta_{1,V}^{(1g)} \; .
    \label{DeltaBubble}
\end{equation}
The definition and the value of the integral $B_0(m_H^2)$, together with all the other integrals that appear in the calculation of the virtual corrections, can be found in the Appendix~\ref{AppendixB}.

The second diagram gives the following contribution
\begin{equation}
    \Gamma_{2,V}^{(1g)} = \frac{-i}{2} \epsilon_{\mu} (k_1) \frac{k_{2,\rho}}{s} N \delta^{ac} g^2 g_H \int \frac{d^D k}{(2 \pi)^D} \frac{V^{\mu \nu \sigma}(-k_1,-k-q,k) A^{\rho}_{\; \sigma \nu}(-k,k+q)}{k^2 (q+k)^2} \; ,
\end{equation}
where
\begin{equation}
      A^{\mu \rho \nu} (p,q) = g^{\nu \rho} (q-p)^{\mu} + g^{\rho \mu} (2p+q)^{\nu} -g^{\mu \nu} (p+2q)^{\rho}
  \end{equation}
and $V^{\mu\nu\sigma}$ is defined in~(\ref{gggH}). We find
\begin{equation}
       \Gamma_{2,V}^{(1g)} = \Gamma_{\{H\} g}^{ac(0)} \left[ -\frac{3}{2} N g^2 B_0 (-\vec{q}^{\; 2}) \right] \equiv \Gamma_{\{H\} g}^{ac(0)} \delta_{2,V}^{(1g)} \; .
       \label{DeltaFish}
  \end{equation}
  
The correction due to the Reggeon self-energy is universal and reads (see, {\it e.g.}, Ref.~\cite{Fadin:2001dc}) 
\begin{equation}
    \Gamma_{3,V}^{(1g)} = \Gamma_{\{H\} g}^{ac(0)} \left[ \omega^{(1)}(t) \frac{(5-3\epsilon)N-2(1-\epsilon)n_f}{4(1-2\epsilon)(3-2\epsilon)N} \right] \equiv \Gamma_{\{H\} g}^{ac(0)} \delta_{3,V}^{(1g)} \; .
    \label{DeltaSelf}
\end{equation}

The last correction is the one associated with the ``triangular'' diagram, 
\begin{equation*}
    \Gamma_{4,V}^{(1g)} = g^2 N \epsilon_{\mu} (k_1) \frac{k_{2,\rho}}{s} \delta^{ac} g_H
\end{equation*}
\begin{equation}
    \times \int \frac{d^D k}{i (2 \pi)^D} \frac{A_{\:\: \sigma}^{\mu \; \nu}(-k,k+k_1) A^{\; \sigma \rho \beta}(-q,q-k) H_{\nu \beta}(-k_1-k,k-q)}{k^2 (k+k_1)^2 (k-q)^2} \; .
\end{equation}
This correction is less trivial and we made use of the \textsc{FeynCalc}~\cite{Mertig:1991ca,Shtabovenko:2016cp} package to perform some analytic steps; up to terms power-suppressed in $s$, we end up with
\begin{equation*}
   \Gamma_{4,V}^{(1g)}  \equiv \Gamma_{H g}^{ac(0)} \delta_{4,V}^{(1g)} = \Gamma_{H g}^{ac(0)} \frac{N g^2}{4 (D-2) (1-D)
   \left(m_H^2+\vec{q}^{\; 2}\right)^2}  
   \end{equation*}
   \begin{equation*}
   \times \Bigg \{ 2 (D-1) \left[ \left(-5 (D-2) m_H^4-4(3 D-8) m_H^2
   \vec{q}^{\; 2}+(16-7 D) (\vec{q}_2^{\; 2})^2 \right) B_0(-\vec{q}^{\; 2}) \right. 
   \end{equation*}
   \begin{equation*}
   \left. +2 (D-2) m_H^2 \left(2
   m_H^4 +3 m_H^2 \vec{q}^{\; 2} + (\vec{q}^{\; 2})^2\right)
   C_0\left(0,-\vec{q}^{\; 2},m_H^2\right)\right] + \left( -D(D-2) (\vec{q}^{\; 2})^2 \right.
   \end{equation*}
   \begin{equation}
   + \left. (D (D (4 D-35)+92)-60)
   m_H^4-2 (D (-2 (D-9) D-41)+24) m_H^2 \vec{q}^{\; 2}\right)
   B_0\left(m_H^2\right)\Bigg\} \; .
   \label{DeltaTri}
\end{equation}

\subsection{The 1-loop correction: 2-gluon exchange diagrams}
\label{ssec:virtual2Gluon}

\begin{figure}
  \begin{center}
  \includegraphics[scale=0.60]{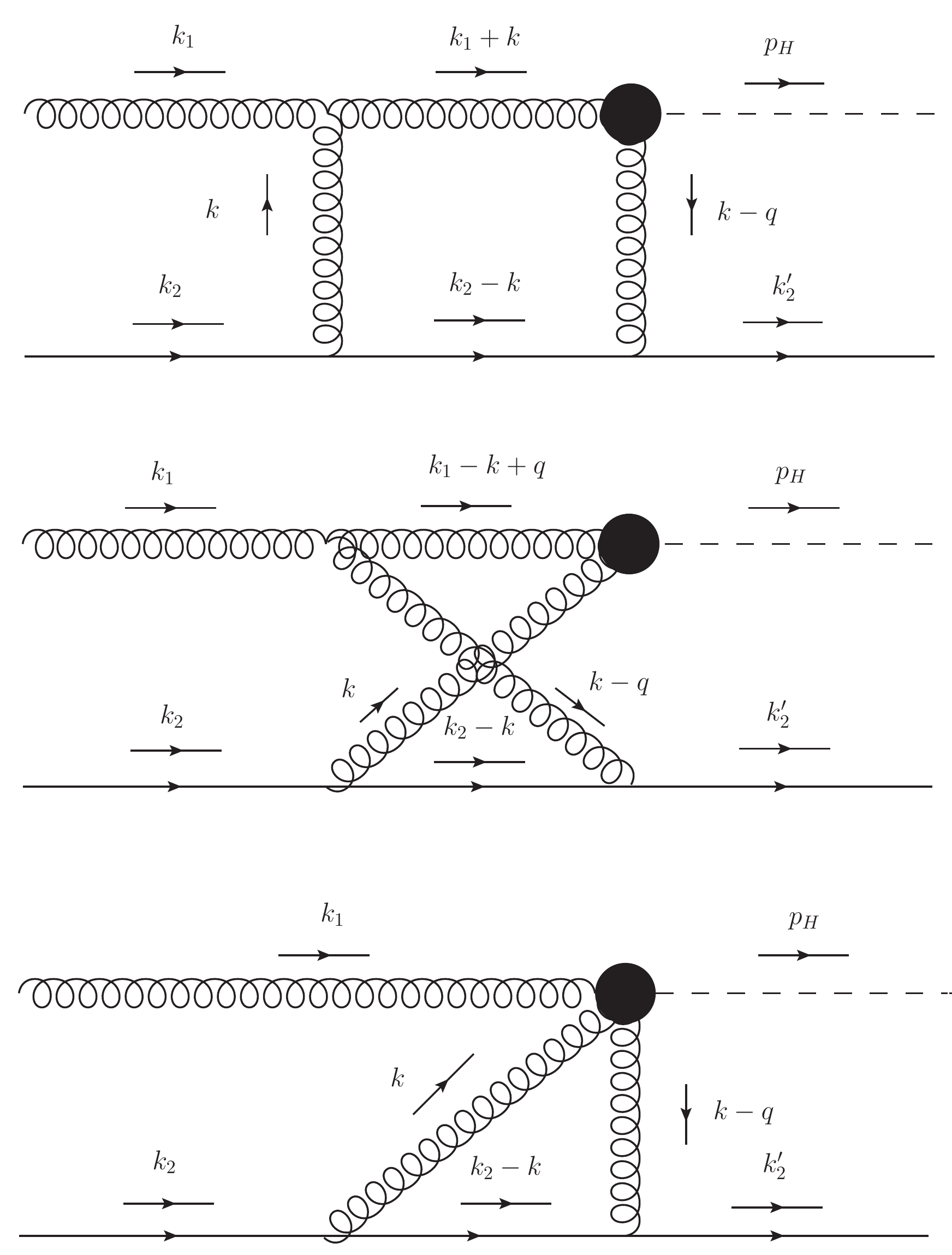} \hspace{1.5 cm}
  \end{center}
  \caption{Feynman diagrams contributing to the $g R \rightarrow g H$ amplitude.}
  \label{2GluonTchannel}
\end{figure}
As already mentioned, to extract this contribution we need to compute ${\cal A}_{g q \rightarrow H q}^{(2g)(8,-)(1)}$ and use Eq.~(\ref{FromExt2Gluon}). There are three diagrams contributing to this amplitude (see Fig.~\ref{2GluonTchannel}). If all vertices involved in
these diagrams were from QCD, one could use the Gribov trick~(\ref{Gribov}) on both gluons in the $t$-channel and adopt the powerful technique explained in Refs.~\cite{Fadin:1999qc,Fadin:2001dc} to simplify the calculation. In presence of an effective gluon-Higgs vertex, the Gribov trick cannot be applied to the $t$-channel gluon(s) entering that vertex\footnote{Detailed explanations about this failure and possible extensions of the above technique will be
given in a separate paper~\cite{GribovNext}.}. In particular, the third diagram in Fig.~\ref{2GluonTchannel}, which would be vanishing if the Gribov trick were applied, gives a non-zero contribution.

We begin by computing the first diagram; the second one can be obtained from the first by analytic continuation from the $s$-channel to the $u$-channel, namely
\begin{equation}
    {\cal A}_{g q \rightarrow H q , \;2 }^{(2g)(8,-)(1)} (s) = - {\cal A}_{g q \rightarrow H q , \;1 }^{(2g)(8,-)(1)} (-s) \; .
\end{equation}
The amplitude relative to the first diagram is
\begin{equation*}
    {\cal A}_{g q \rightarrow H q , \;1 }^{(2g)(8,-)(1)} = \epsilon_{\mu}(k_1) g^3 g_H f^{abc} (t^b t^c)_{ji} \int \frac{d^D k}{(2 \pi)^{D}} \frac{A^{\mu \rho \nu}(-k,k+k_1) H_{\nu}^{\; \sigma} (-k-k_1, k-q)}{k^2(k+k_1)^2(k_2-k)^2 (k-q)^2}
\end{equation*}
\begin{equation}
    \times g_{\sigma \xi} g_{\rho \gamma} \bar{u} (k_2-q) \gamma^{\xi} (\slashed{k}_2 - \slashed{k}) \gamma^{\gamma} u(k_2) \; .
\end{equation}
Observing that 
\begin{equation*}
    f^{a b c} (t^b t^c)_{ji} = \frac{1}{2} f^{a b c} (t^b t^c - t^c t^b)_{ji} = \frac{i}{2} C_A t_{ji}^a = \frac{i}{2} C_A \delta^{ac} t_{ji}^c 
\end{equation*}
and then using \textsc{FeynCalc}~\cite{Mertig:1991ca,Shtabovenko:2016cp}, we obtain
\begin{equation*}
    {\cal A}_{g q \rightarrow \{H\} q , \;1 }^{(2g)(8,-)(1)} + {\cal A}_{g q \rightarrow H q , \;2 }^{(2g)(8,-)(1)} = \Gamma_{H g}^{ac(0)}  \left[ \frac{g^2 N}{(-4)} \left( \frac{2 \vec{q}^{\; 2} \left((5 D-12) m_H^2+(D-2)
   \vec{q}^{\; 2}\right) B_0\left(m_H^2\right)}{(D-2)
   \left(m_H^2+\vec{q}^{\; 2}\right)^2} \right. \right. 
   \end{equation*}
   \begin{equation*}
   \left. 
   +\frac{4 \left((D-2) m_H^4 + (D-2) m_H^2
   \vec{q}^{\; 2}+(2 D-5) (\vec{q}^{\; 2})^2 \right) B_0(-\vec{q}^{\; 2})}{(D-2)
   \left(m_H^2+\vec{q}^{\; 2}\right)^2}+2 s C_0\left(m_H^2,0,-s\right) \right. 
   \end{equation*}
   \begin{equation*}
   \left. -2 s
   C_0\left(m_H^2,0,s\right)-\frac{4 \left(m_H^4-m_H^2
   \vec{q}^{\; 2}-\vec{q}^{\; 2}\right)
   C_0\left(m_H^2,0,-\vec{q}^{\; 2}\right)}{m_H^2+\vec{q}^{\; 2}} +4 \vec{q}^{\; 2}
   C_0(0,0,-\vec{q}^{\; 2}) \right. \vspace{0.1cm}
   \end{equation*}
   \begin{equation*}
   \left. -2 s C_0(0,0,-s)+2 s C_0(0,0,s) +2 \vec{q}^{\; 2} s D_0\left(m_H^2,0,0,0;-\vec{q}^{\; 2},-s\right) \right. 
   \end{equation*}
   \begin{equation}
    -2 \vec{q}^{\; 2} s
   D_0\left(m_H^2,0,0,0;-\vec{q}^{\; 2},s\right) \Bigg ) \Bigg ] \frac{2s}{t} \Gamma_{qq}^{c(0)} \; .
\end{equation}

The third diagram gives a much simpler contribution:
\begin{equation}
   {\cal A}_{g q \rightarrow H q , \;3 }^{(2g)(8,-)(1)} = \Gamma_{\{H\} g}^{ac(0)} \left[ N g^2 B_0(-\vec{q}^{\; 2};0,0) \right] \frac{2 s}{t} \Gamma_{q q}^{c(0)} \; .
\end{equation}

At this point, using Eq.~(\ref{FromExt2Gluon}), we are able to extract the contribution to the gRH vertex from the exchange of two gluons in the $t$-channel:
\begin{equation}
    \delta_{Hg}^{(2g)} = \frac{{\cal A}_{g q \rightarrow H q , \;1 }^{(2g)(8,-)(1)}+{\cal A}_{g q \rightarrow H q , \;2 }^{(2g)(8,-)(1)}+{\cal A}_{g q \rightarrow H q , \;3 }^{(2g)(8,-)(1)}}{\Gamma_{\{H\} g}^{ac(0)}\frac{2s}{t}\Gamma_{qq}^{c(0)}} -\delta_{qq}^{(2g)} - \frac{\omega^{(1)}(t)}{2} \left[ \ln\left(
\frac{s}{-t} \right) + \ln\left( \frac{-s}{-t}
\right) \right] \; .
\end{equation}

\subsection{Full 1-loop correction to the impact factor}
\label{ssec:virtual1full}

Adding $\delta_{Hg}^{(2g)}$ to $\delta_{1,V}^{(1g)}$, $\delta_{2,V}^{(1g)}$, $\delta_{3,V}^{(1g)}$, and $\delta_{4,V}^{(1g)}$ as defined in Eqs.~(\ref{DeltaBubble}), (\ref{DeltaFish}), (\ref{DeltaSelf}), and (\ref{DeltaTri}), respectively, we obtain
\begin{equation}
    \delta_{\rm NLO} \simeq \frac{\bar{\alpha}_s}{4 \pi} \left( \frac{\vec{q}^{\; 2}}{\mu^2} \right)^{-\epsilon} \left \{  - \frac{C_A}{\epsilon^2} + \frac{11 C_A -2 n_f}{6 \epsilon} - \frac{5 n_f}{9} + C_A \left( 2 {\rm{Li}}_2 \left( 1 + \frac{m_H^2}{\vec{q}^{\; 2}} \right) + \frac{\pi^2}{3} + \frac{67}{18} \right) \right \} \: , 
    \label{FullvertexCorrect}
\end{equation}
where we have kept only terms non-vanishing in the limit $\epsilon \to 0$ and have used
\begin{equation}
    \bar{\alpha}_s = \frac{g^2 \Gamma (1+\epsilon) \mu^{-2 \epsilon}}{(4 \pi)^{1-\epsilon}} \; , \hspace{0.5 cm} C_A = N \; .
\end{equation}
Substituting~(\ref{FullvertexCorrect}) into Eq.~(\ref{HiggsVirtImp}), we finally get the virtual correction to the Higgs impact factor:
\begin{equation*}
\frac{d \Phi_{gg}^{\{ H \}(1)} (z_H, \vec{p}_H, \vec{q}; s_0 )}{d z_H d^2 \vec{p}_H} = \frac{d \Phi_{gg}^{\{ H \}(0)} (z_H, \vec{p}_H,\vec{q})}{d z_H d^2 \vec{p}_H} \; \frac{\bar{\alpha}_s}{2 \pi} \left( \frac{\vec{q}^{\; 2}}{\mu^2}  \right)^{- \epsilon}  \left[  - \frac{C_A}{\epsilon^2} \right.
\end{equation*}
\begin{equation}
      \left. + \frac{11 C_A - 2n_f}{6 \epsilon} - \frac{C_A}{\epsilon} \ln \left( \frac{\vec{q}^{\; 2}}{s_0} \right) - \frac{5 n_f}{9} + C_A \left( 2\;\Re e \left( {\rm{Li}}_2 \left( 1 + \frac{m_H^2}{\vec{q}^{\; 2}} \right)\right) + \frac{\pi^2}{3} + \frac{67}{18} \right) + 11 \right] \; .
      \label{VirtualPartIMF}
\end{equation}
where $g_H$ has to be taken at leading order, while the last term equal to $11$ takes into account its next-to-leading contribution (see Eq.~(\ref{gH})). The result in Eq.~(\ref{VirtualPartIMF}) can be also obtained from Ref.~\cite{Schmidt:1997wr}, by taking the high-energy limit and using crossing symmetry. We used this alternative strategy as an independent check, finding perfect agreement.~\footnote{To perform the comparison, we first used crossing symmetry to obtain from Ref.~\cite{Schmidt:1997wr} the NLO amplitude for the  Higgs plus quark production in the collision of a gluon with a quark, then we took the high-energy limit of this amplitude and confronted with the BFKL-factorized form of the same amplitude, taking advantage of the known expression for the NLO quark-quark-Reggeon vertex.}
The comparison with Ref.~\cite{Hentschinski:2020tbi}, whose virtual part contribution is based on Ref.~\cite{Nefedov:2019mrg}, is not immediate, since their calculation is based on the Lipatov effective action and on a different definition of the impact factor. This implies, for instance,
that in their purely virtual result, Eq.~(43) of Ref.~\cite{Hentschinski:2020tbi}, the rapidity regulator is still present, which cancels when one combines the impact factor with their definition of the unintegrated gluon distribution\footnote{We recall that they considered the single-forward Higgs boson production.}.

\section{Projection onto the eigenfunctions of the BFKL kernel and cancellation of divergences}
\label{sec:projection}

In this section we will carry out the projection of the Higgs impact factor onto the eigenfunctions of the LO BFKL kernel. There are two main reasons for performing this procedure. 

First, being our impact factor differential in the Higgs kinematic variables,
IR divergences do not appear in the real corrections and, therefore, it is not possible to observe their cancellation in the final result for the impact factor, as it would happen for a fully inclusive {\em hadronic} one (see Ref.~\cite{Fadin:1999qc}). The cancellation must of course be observed at the level of the cross section: the projection onto the LO BFKL eigenfunctions is an effective way to anticipate the integrations needed to get the cross section and, therefore, the projected impact factor, when all counterterms are taken into account, turns to be IR- and UV-finite. This procedure has been already successfully applied in 
Refs.~\cite{Ivanov:2012ms,Ivanov:2012iv}. The second reason is more practical: moving from the momentum representation to the representation in terms of the LO BFKL kernel eigenfunctions makes the numeric implementation somewhat simpler.

To understand the idea behind the projection, it is enough to consider the {\em partonic} amplitude~(\ref{Ar}) in the LL,
\[
\Im m_{s}\left(({\cal A}^{(0)})_{AB}^{AB}\right) = \frac{s}{\left( 2\pi \right)^{D-2}}
\int \frac{d^{D-2}q_1}{\vec{q}_{1}^{\:2}}
\int \frac{d^{D-2}q_2}{\vec{q}_{2}^{\:2}}
\]
\begin{equation*}
\times \Phi _{A A}^{\left(0\right) }
\left( \vec{q}_{1};s_{0}\right)\int_{\delta -i\infty}^{\delta+i\infty}
\frac{d\omega }{2\pi i}\left[ \left( \frac{s}{s_{0}}\right)^{\omega }
G_{\omega }^{\left( 0\right) }\left( \vec{q}_{1},\vec{q}_{2}\right) 
\right] \Phi _{B B}^{\left( 0 \right) }\left( -\vec{q}_{2};s_{0}\right) \;,
\end{equation*}
and to use the following spectral representation for the BFKL Green's function:
\begin{equation}
    G_{\omega }^{\left( 0\right) }\left( \vec{q}_{1},\vec{q}_{2}\right) = \sum_{n=-\infty}^\infty \int_{-\infty}^{+\infty} d \nu \frac{\phi_{\nu}^{n} (\vec{q}_1^{\; 2}) \phi_{\nu}^{n *}(\vec{q}_2^{\; 2})}{\omega - \frac{\alpha_s C_A}{\pi} \chi(n, \nu)} \; ,
\end{equation}
where $\phi_{\nu}^{n} (\vec{q}^{\; 2})$ are the LO BFKL kernel eigenfunctions and $(\alpha_s C_A/\pi) \chi(n, \nu)$ the corresponding eigenvalues, with
\begin{equation}
    \phi_{\nu}^{n} (\vec{q}^{\; 2}) = \frac{1}{\pi \sqrt{2}} (\vec{q}^{\; 2})^{i \nu - \frac{1}{2}} e^{i n \phi} \; , \hspace{0.5 cm} \chi(n, \nu) = 2 \psi (1) - \psi \left( \frac{n}{2} + \frac{1}{2} + i \nu \right) - \psi \left( \frac{n}{2} + \frac{1}{2} - i \nu \right) \; .
\end{equation}
Here $\phi$ is the azimuthal angle of the vector $\vec q$ counted from some fixed direction in the transverse space.
Then, integrations over transverse momenta decouple and each impact factor can be separately projected onto the eigenfunctions of the BFKL kernel, so that
\begin{equation}
     \int \frac{d^{2-2\epsilon} q}{\pi \sqrt{2}} (\vec{q}^{\; 2})^{i \nu - \frac{3}{2}} e^{i n \phi} \Phi _{A A}^{\left(0\right)} (\vec{q} \;) \equiv 
     \Phi^{(0)}_{A A}(n,\nu)\;,
\label{ProjectionDef}
\end{equation}
and similarly for the $\Phi_{BB}$. The extension of the procedure to the NLL is straightforward (see, {\it e.g.}, Ref.~\cite{Caporale:2015uva}). Since we work in $D = 4 - 2\epsilon$, we introduce the ``continuation'' of the LO BFKL kernel eigenfunctions to non-integer dimensions,
\[
(\vec q^{\:2})^\gamma e^{i n \phi} \ \rightarrow (\vec q^{\:2})^{\gamma -\frac{n}{2}} (\vec q \cdot \vec l)^n\;, \;\;\;\;\; \gamma\equiv i\nu-
\frac{1}{2}\;, 
\]
and the vector $\vec l$ lies only in the first two of the $2 -\epsilon$ transverse space dimensions, {\it i.e.} $\vec l = (1,i)$, so that $\vec l^{\:2}=0$.

As already discussed in subsection~\ref{ssec:LO_IF}, to get the Higgs impact factor for a proton-initiated process, we must take the convolution with the initial-state PDFs. This 
brings along initial-state collinear singularities, which must be canceled by suitable counterterms. In the rest of this section we will construct the $(n,\nu)$-projection
of all contributions to the hadronic Higgs impact factor, including the PDF counterterms, and will check the explicit cancellation of all IR divergences. The residual UV divergence will
be taken care of by the renormalization of the QCD coupling. To perform the $(n,\nu)$-projection, according to Eq.~(\ref{ProjectionDef}), we will make use of the integrals computed in Appendix~\ref{AppendixC}.

\subsection{Projection of the LO impact factor}
\label{ssec:projection_LO}

Recalling the expression of the LO impact factor,
\begin{equation}
    \frac{d \Phi_{PP}^{ \{ H \}(0)} (x_H, \vec{p}_H, \vec{q})}{d x_H d^2 \vec{p}_H} = \int_{x_H}^1 \frac{d z_H}{z_H} f_g \left( \frac{x_H}{z_H} \right) \frac{d \Phi_{gg}^{ \{H \}(0)} (z_H, \vec{p}_H, \vec{q})}{d z_H d^2 \vec{p}_H} = \frac{g_H^2 \vec{q}^{\; 2} f_g (x_H) \delta^{(2)} ( \vec{q} - \vec{p}_H) }{8 (1-\epsilon) \sqrt{N^2-1}} \: ,
    \tag{\ref{Factorization}}
\end{equation}
we get immediately from Eq.~(\ref{ProjectionDef}) its projected counterpart,
\begin{equation}
    \frac{d \Phi_{ PP }^{\{H \}(0)} (x_H, \vec{p}_H, n, \nu)}{d x_H d^2 \vec{p}_H}  = \frac{g_H^2}{8 (1-\epsilon) \sqrt{N^2-1}} \frac{(\vec{p}_H^{\; 2})^{i \nu - \frac{1}{2}} e^{i n \phi_H}}{\pi \sqrt{2}} f_g (x_H)  \; .
\end{equation}
Here $\phi_H$ is the azimuthal angle of the vector $\vec p_H$ counted from the fixed direction in the transverse space. The projected LO impact factor is the starting point for the 
calculation of the NLO contribution to the projected impact factor from the gluon PDF and QCD coupling counterterms. From now on, we will omit the apex $^{(1)}$, since all contributions
to the projected impact factor are NLO.

\subsection{Projection of gluon PDF and coupling counterterms}
\label{ssec:projection_gluon_PDF}

Taking into account the running of $\alpha_s$,
\begin{equation}
    \alpha_s (\mu^2) = \alpha_s (\mu_R^2) \left[ 1 + \frac{\alpha_s (\mu_R^2)}{4 \pi} \left( \frac{11 C_A}{3} - \frac{2 n_f}{3} \right) \left( - \frac{1}{\epsilon} - \ln (4 \pi e^{-\gamma_E}) + \ln \left( \frac{\mu_R^2}{\mu^2} \right) \right) \right] \; ,
\end{equation}
and the running of the gluon PDF,
\begin{equation*}
    f_g (x, \mu) = f_g (x, \mu_F) - \frac{\alpha_s(\mu_F)}{2 \pi} \left( - \frac{1}{\epsilon} - \ln (4 \pi e^{- \gamma_E}) + \ln \left( \frac{\mu_F^2}{\mu^2} \right) \right) 
\end{equation*}
\begin{equation}
    \times \int_x^1 \frac{dz}{z} \left[ P_{gq} (z) \sum_{a=q \bar{q}} f_a \left( \frac{x}{z} , \mu_F \right) + P_{gg} (z) f_g \left( \frac{x}{z} , \mu_F \right) \right] \; ,
\end{equation}
where
\begin{equation}
    P_{gq} (z) = C_F \frac{1+(1-z)^2}{z} \; ,
\end{equation}
\begin{equation}
    P_{gg} (z) = 2 C_A \left( \frac{z}{(1-z)_{+}} + \frac{(1-z)}{z} + z (1-z) \right) + \frac{11 C_A - 2 n_f}{6} \delta (1-z) \; ,
\end{equation}
with plus prescription defined as
\begin{equation}
    \int_a^1 dx \frac{F(x)}{(1-x)_+} = \int_a^1 dx \frac{F(x)-F(1)}{(1-x)} - \int_0^a dx \frac{F(1)}{1-x} \; ,
\end{equation}
we obtain the following projected counterterms,
 \begin{equation*}
    \frac{d \Phi_{PP}^{ \{H \}} (x_H, \vec{p}_H, n, \nu)}{d x_H d^2 \vec{p}_H} \bigg|_{{\rm{coupling \; c.t.}}} = \frac{d \Phi_{PP}^{ \{H \}(0)} (x_H, \vec{p}_H, n, \nu)}{d x_H d^2 \vec{p}_H} \frac{\bar{\alpha}_s}{2 \pi} \left( \frac{\vec{p}_H^{\; 2}}{\mu^2} \right)^{-\epsilon}  
 \end{equation*}
 \begin{equation*}
    \times \left( \frac{11 C_A}{3} - \frac{2 n_f}{3} \right) \left( - \frac{1}{\epsilon} + \ln \left( \frac{\mu_R^2}{\vec{p}_H^{\; 2}}\right) \right) 
 \end{equation*}
 \begin{equation}
     \equiv \frac{d \Phi_{PP, {\rm{div}}}^{ \{H \}} (x_H, \vec{p}_H, n, \nu)}{d x_H d^2 \vec{p}_H} \bigg|_{{\rm{coupling \; c.t.}}} + \frac{d \Phi_{PP, {\rm{fin}}}^{ \{H \}} (x_H, \vec{p}_H, n, \nu)}{d x_H d^2 \vec{p}_H} \bigg|_{{\rm{coupling \; c.t.}}} \; ,
     \label{CoupliforDiv}
 \end{equation}
 \begin{equation*}
    \frac{d \Phi_{PP}^{ \{H \}}(x_H, \vec{p}_H, n, \nu)}{d x_H d^2 \vec{p}_H}\bigg|_{{\rm{P_{qg} \; c.t.}}} = - \frac{1}{f_g (x_H)} \frac{d \Phi_{PP}^{ \{H \}(0)} (x_H, \vec{p}_H, n, \nu)}{d x_H d^2 \vec{p}_H} \frac{\bar{\alpha}_s}{2 \pi} \left( \frac{\vec{p}_H^{\; 2}}{\mu^2} \right)^{-\epsilon} 
\end{equation*}
\begin{equation*}
    \times \left( - \frac{1}{\epsilon} + \ln \left( \frac{\mu_F^2}{\vec{p}_H^{\; 2}}\right) \right) \int_{x_H}^1 \frac{dz_H}{z_H} \left[ P_{gq} (z_H) \sum_{a=q \bar{q}} f_a \left( \frac{x_H}{z_H} , \mu_F \right) \right] 
\end{equation*}
\begin{equation}
    \equiv \frac{d \Phi_{PP, {\rm{div}}}^{ \{H \}}(x_H, \vec{p}_H, n, \nu)}{d x_H d^2 \vec{p}_H}\bigg|_{{\rm{P_{qg} \; c.t.}}} + \frac{d \Phi_{PP,{\rm{fin}}}^{ \{H \}}(x_H, \vec{p}_H, n, \nu)}{d x_H d^2 \vec{p}_H}\bigg|_{{\rm{P_{qg} \; c.t.}}} 
\label{PqgforDiv}
\end{equation}
and
\begin{equation*}
    \frac{d \Phi_{PP}^{ \{H \}}(x_H, \vec{p}_H, n, \nu)}{d x_H d^2 \vec{p}_H}\bigg|_{{\rm{P_{gg} \; c.t.}}} = - \frac{1}{f_g (x_H)} \frac{d \Phi_{PP}^{ \{H \}(0)} (x_H, \vec{p}_H, n, \nu)}{d x_H d^2 \vec{p}_H} \frac{\bar{\alpha}_s}{2 \pi} \left( \frac{\vec{p}_H^{\; 2}}{\mu^2} \right)^{-\epsilon} 
\end{equation*}
\begin{equation*}
    \times \left( - \frac{1}{\epsilon} + \ln \left( \frac{\mu_F^2}{\vec{p}_H^{\; 2}}\right) \right) \int_{x_H}^1 \frac{dz_H}{z_H} \left[ P_{gg} (z_H) f_g \left( \frac{x_H}{z_H} , \mu_F \right) \right] 
\end{equation*}
\begin{equation}
    \equiv \frac{d \Phi_{PP, {\rm{div}}}^{ \{H \}}(x_H, \vec{p}_H, n, \nu)}{d x_H d^2 \vec{p}_H}\bigg|_{{\rm{P_{gg} \; c.t.}}} + \frac{d \Phi_{PP,{\rm{fin}}}^{ \{H \}}(x_H, \vec{p}_H, n, \nu)}{d x_H d^2 \vec{p}_H}\bigg|_{{\rm{P_{gg} \; c.t.}}} \; .
    \label{PggforDiv}
\end{equation}
With obvious notation, we have implicitly defined divergent (``div'') and finite (``fin'') part of each contribution. We will keep adopting this notation in the following.

\subsection{Projection of high-rapidity real gluon contribution and BFKL counterterm}
\label{ssec:projection_BFKL}

The contribution from a real gluon emission has a divergence for $z_H \to 1$ or $ (z_g \to 0)$ at any value of the gluon momenta $\vec{q}-\vec{p}_H$. This divergence is regulated by the parameter $s_{\Lambda}$ and, in the final result, cancelled by the BFKL counterterm appearing in the definition of the NLO impact factor. In this section we make the cancellation of $s_{\Lambda}$ explicit and give the $(n,\nu)$-projection of the high-rapidity part of the real gluon production impact factor combined with the BFKL counterterm.

First of all, let's take the convolution of the impact factor for real gluon production, given in~(\ref{GluonImp}), with the gluon PDF and rewrite it in an equivalent form, by adding and subtracting three terms, for later convenience:
\[
\frac{d \Phi^{\{Hg\}}_{PP}(x_H, \vec{p}_H,\vec q;s_0)}{d x_H d^2 p_H}
= \frac{d \tilde{\Phi}^{\{Hg\}}_{PP}(x_H, \vec{p}_H,n,\vec q;s_0)}{d x_H d^2 p_H}
\]
\[
+ \frac{d \Phi^{\{Hg\} (1-x_H)}_{PP}(x_H, \vec{p}_H,\vec q;s_0)}{d x_H d^2 p_H}
+ \frac{d \Phi^{\{Hg\}{\rm plus}}_{PP}(x_H, \vec{p}_H,\vec q;s_0)}{d x_H d^2 p_H}
\]
\begin{equation}
+ \int_{x_H}^1 d z_H f_g (x_H)  \frac{d \Phi_{gg}^{ \{H g \}} (z_H, \vec{p}_H, \vec{q} ; s_0)}{d z_H d^2 \vec{p}_H} \bigg|_{z_H \rightarrow 1} \; ,
\label{subtractions}
\end{equation}
where
\[
    \frac{d \tilde{\Phi}^{\{Hg\}}_{PP}(x_H, \vec{p}_H,\vec q;s_0)}{d x_H d^2 p_H}
    =\int_{x_H}^1 \frac{d z_H}{z_H} f_g \left( \frac{x_H}{z_H} \right) \frac{d \Phi_{gg}^{ \{ H g \}}(z_H, \vec{p}_H, \vec{q};s_0)}{d z_H d^2 \vec{p}_H} 
\]
\begin{equation}
    - \int_{x_H}^1 d z_H f_g \left( \frac{x_H}{z_H} \right) \left[ \frac{d \Phi_{gg}^{ \{ H g \}}(z_H, \vec{p}_H, \vec{q};s_0)}{d z_H d^2 \vec{p}_H} \right]_{z_H=1} \;,
\end{equation}
\begin{equation}
\frac{d \Phi^{\{Hg\} (1-x_H)}_{PP}(x_H, \vec{p}_H,\vec q;s_0)}{d x_H d^2 p_H}
=\int_{0}^{x_H} d z_H f_g (x_H) \frac{d \Phi_{gg}^{ \{ H g \}}(z_H, \vec{p}_H, \vec{q};s_0)}{d z_H d^2 \vec{p}_H} \Bigg |_{z_H=1} 
\end{equation}
and
\[
\frac{d \Phi^{\{Hg\}{\rm plus}}_{PP}(x_H, \vec{p}_H,\vec q;s_0)}{d x_H d^2 p_H}
= - \int_{0}^{x_H} d z_H f_g (x_H) \frac{d \Phi_{gg}^{ \{ H g \}}(z_H, \vec{p}_H, \vec{q};s_0)}{d z_H d^2 \vec{p}_H} \Bigg |_{z_H=1}
\]
\begin{equation}
    + \int_{x_H}^1 d z_H \left( f_g \left( \frac{x_H}{z_H} \right) - f_g (x_H) \right) \left[\frac{d \Phi_{gg}^{ \{ H g \}}(z_H, \vec{p}_H, \vec{q};s_0)}{d z_H d^2 \vec{p}_H} \right]_{z_H=1} \; .
\end{equation}
The pieces $d\tilde{\Phi}$, $d\Phi^{\{Hg\}(1-x_H)}$, and $d\Phi^{\{Hg\}{\rm plus}}$ are free from the divergence for $z_H\to 1$ and therefore in their expressions the limit $s_{\Lambda} \rightarrow \infty$ can be safely taken, which means that $\theta(s_{\Lambda} - s_{PR})$ can be set to one.
The projection of these terms will be considered in subsection~\ref{ssec:projection_gluon}.
The last term in Eq.~(\ref{subtractions}) can be easily calculated, since
\begin{equation*}
    \frac{d \Phi_{gg}^{ \{H g \}} (z_H, \vec{p}_H, \vec{q} ; s_0)}{d z_H d^2 \vec{p}_H} \bigg|_{z_H \rightarrow 1} = \frac{\braket{cc'|\hat{\mathcal{P}}|0}}{2(1-\epsilon)(N^2-1)} 
\end{equation*}
\begin{equation*}
   \times \left[ \sum_{\{ f \}} \int \frac{d s_{PR} d\rho_f}{2 \pi} \Gamma^c_{P \{ f \}} \left( \Gamma^c_{P \{ f \}} \right)^{*} \theta \left( s_{\Lambda} - s_{PR} \right) \right]_{z_H \rightarrow 1}
\end{equation*}
\begin{equation}
    = \frac{g^2 g_H^2 C_A}{4 (1-\epsilon) \sqrt{N^2-1} (2 \pi)^{D-1} } \frac{\vec{q}^{\; 2}}{(\vec{q}- \vec{p}_H)^2} \frac{1}{(1-z_H)} \theta \left( s_{\Lambda} - \frac{(\vec{q}-\vec{p}_H)^2}{(1-z_H)} \right) \; .
\end{equation}
We get
\begin{equation*}
    \int_{x_H}^1 d z_H f_g (x_H)  \frac{d \Phi_{gg}^{ \{H g \}} (z_H, \vec{p}_H, \vec{q} ; s_0)}{d z_H d^2 \vec{p}_H} \bigg|_{z_H \rightarrow 1}
\end{equation*}
\begin{equation*}
=    \frac{g^2 g_H^2 C_A}{4 (1-\epsilon) \sqrt{N^2-1} (2 \pi)^{D-1} } \frac{\vec{q}^{\; 2}}{(\vec{q}- \vec{p}_H)^2} \int_{x_H}^1 d z_H \frac{1}{(1-z_H)} f_g (x_H) \theta \left( s_{\Lambda} - \frac{(\vec{q}-\vec{p}_H)^2}{(1-z_H)} \right) 
\end{equation*}
\begin{equation}
    = \frac{g^2 g_H^2 C_A}{4 (1-\epsilon) \sqrt{N^2-1} (2 \pi)^{D-1} } \frac{\vec{q}^{\; 2}}{(\vec{q}- \vec{p}_H)^2} f_g (x_H) \left[ \ln (1-x_H) - \frac{1}{2} \ln \left( \frac{\left[(\vec{q}-\vec{p}_H)^2 \right]^2}{s_{\Lambda}^2} \right) \right] \; .
\label{Phi_g_rap}
\end{equation}

Let's consider now the BFKL counterterm
\begin{equation}
   \frac{d \Phi^{\rm{BFKL \ c.t.}}_{gg}(z_H, \vec{p}_H, \vec q;s_0)}{d z_H d^2 p_H} = - \frac{1}{2} \int d^{D-2} q' \frac{ \vec{q}^{\; 2}}{\vec{q}^{\; '2}}  \frac{d \Phi_{gg}^{ \{H \} (0)} (\vec{q} \; ' )}{d z_H d^2 p_H}  \mathcal{K}^{(0)}_r (\vec{q} \; ', \vec{q} \; ) \ln \left( \frac{s_{\Lambda}^2}{(\vec{q} \; ' - \vec{q} \; )^2 s_0} \right) \; .
\end{equation}
Using Eq.~(\ref{LOHiggsImp2}) and Eq.~(\ref{BornKer}), we find
\begin{equation}
    \frac{d \Phi^{\rm{BFKL \ c.t.}}_{gg}(z_H, \vec{p}_H, \vec q;s_0)}{d z_H d^2 p_H} \!=\! \frac{-g^2 g_H^2 C_A}{8 (2 \pi)^{D-1} (1-\epsilon) \sqrt{N^2-1}} \frac{\vec{q}^{\; 2}}{(\vec{q}-\vec{p}_H)^2} \ln \left( \frac{s_{\Lambda}^2}{(\vec{q} - \vec{p}_H )^2 s_0}  \right) \delta (1-z_H)
\end{equation}
and, after convolution with the gluon PDF, we get
\[
 \frac{d \Phi^{{\rm{BFKL\ c.t.}}}_{PP}(x_H,\vec p_H,\vec q;s_0)}{d x_H d^2 p_H} = \int_{x_H}^1 \frac{d z_H}{z_H} f_g \left( \frac{x_H}{z_H} \right) \frac{d \Phi_{gg}^{{\rm{BFKL\ c.t.}}} (z_H, \vec{p}_H, \vec{q})}{d z_H d^2 \vec{p}_H} 
 \]
 \begin{equation}
= - \frac{g^2 g_H^2 C_A}{8 (2 \pi)^{D-1} \sqrt{N^2-1}} \frac{\vec{q}^{\; 2}}{(\vec{q}-\vec{p}_H)^2} \frac{f_g (x_H)}{(1-\epsilon)} \ln \left( \frac{s_{\Lambda}^2}{(\vec{q} - \vec{p}_H )^2 s_0}  \right) \; .
 \label{BFKLct}
\end{equation}
When we combine the last term of Eq.~(\ref{subtractions}), given in~(\ref{Phi_g_rap}), with the BFKL counterterm, given in~(\ref{BFKLct}), we obtain  
\begin{equation}
 \frac{d \Phi^{{\rm{BFKL}}}_{PP}(x_H, \vec{p}_H, \vec q;s_0)}{d x_H d^2 p_H} \equiv \frac{g^2 g_H^2 C_A}{4 (2 \pi)^{D-1} (1-\epsilon) \sqrt{N^2-1}} \frac{\vec{q}^{\; 2}}{(\vec{q}-\vec{p}_H)^2} f_g (x_H) \ln \left( \frac{(1-x_H) \sqrt{s_0}}{|\vec{q} - \vec{p}_H|}  \right) \,.
 \label{CountFin}
\end{equation}
Note that this term is finite as far as the high-energy divergence is concerned. The remaining divergences can be isolated after the projection, 
\begin{equation*}
   \frac{d \Phi^{{\rm{BFKL}}}_{PP}(x_H, \vec{p}_H,n,\nu;s_0)}{d x_H d^2 p_H} = \frac{d \Phi_{PP}^{\{ H \}(0)} (x_H, \vec{p}_H, n, \nu)}{d x_H d^2 \vec{p}_H} \frac{\bar{\alpha}_s}{2 \pi} \left( \frac{\vec{p}_H^{\; 2}}{\mu^2} \right)^{-\epsilon} \left \{ \frac{C_A}{\epsilon^2} + \frac{C_A}{\epsilon} \ln \left( \frac{\vec{p}_H^{\; 2}}{s_0} \right)  \right.
\end{equation*}
\begin{equation*}
    \left. - 2 \frac{C_A}{\epsilon} \ln (1-x_H) + C_A \left[ \ln \left( \frac{\vec{p}_H^{\; 2}}{s_0 (1-x_H)^2} \right) \left( 2  \gamma_E + \psi \left(\frac{1}{2} + \frac{n}{2} - i \nu \right) \right. \right. \right.
\end{equation*}
\begin{equation*}
    \left. + \psi \left( \frac{1}{2} + \frac{n}{2} + i \nu \right) \right) - 2 \gamma_E^2 - \zeta (2) - \frac{1}{2}\left(\psi' \left(\frac{1}{2} + \frac{n}{2} - i \nu\right) - \psi' \left(\frac{1}{2} + \frac{n}{2} + i \nu\right)\right) 
\end{equation*}
\begin{equation*}
    - 2 \gamma_E \left( \psi \left(\frac{1}{2} + \frac{n}{2} - i \nu\right) + \psi \left(\frac{1}{2} + \frac{n}{2} + i \nu\right) \right)
\end{equation*}
\begin{equation*}
- \left.\left. \frac{1}{2}  \left(\psi \left(\frac{1}{2} + \frac{n}{2} - i \nu \right) + \psi \left( \frac{1}{2} + \frac{n}{2} + i \nu \right) \right)^2 \right] \right \}
\end{equation*}
\begin{equation}
    \equiv \frac{d \Phi^{{\rm{BFKL}}}_{PP, {\rm{div}}}(x_H, \vec{p}_H,n,\nu;s_0)}{d x_H d^2 p_H} + \frac{d \Phi^{{\rm{BFKL}}}_{PP, {\rm{fin}}}(x_H, \vec{p}_H,n,\nu;s_0)}{d x_H d^2 p_H} \; .
    \label{BFKLforDiv}
\end{equation}

\subsection{Projection of virtual and real quark contributions}
\label{ssec:projection_quark}

Since the virtual contribution is proportional to the LO impact factor, the convolution with the gluon PDF and the $(n,\nu)$-projection are trivial and give
\begin{equation*}
\frac{d \Phi_{PP}^{\{ H \}(1)} (x_H, \vec{p}_H, n, \nu; s_0 )}{d x_H d^2 \vec{p}_H} = \frac{d \Phi_{PP}^{\{ H \}(0)} (x_H, \vec{p}_H,n, \nu)}{d x_H d^2 \vec{p}_H} \; \frac{\bar{\alpha}_s}{2 \pi} \left( \frac{\vec{p}_H^{\; 2}}{\mu^2}  \right)^{- \epsilon}  \left[  - \frac{C_A}{\epsilon^2} \right.
\end{equation*}
\begin{equation*}
      \left. + \frac{11 C_A - 2n_f}{6 \epsilon} - \frac{C_A}{\epsilon} \ln \left( \frac{\vec{p}_H^{\; 2}}{s_0} \right) - \frac{5 n_f}{9} + C_A \left( 2\;\Re e \left({\rm{Li}}_2 \left( 1 + \frac{m_H^2}{\vec{p}_H^{\; 2}} \right)\right) + \frac{\pi^2}{3} + \frac{67}{18} \right) + 11 \right] 
\end{equation*}
\begin{equation}
    \equiv \frac{d \Phi_{PP,{\rm{div}}}^{\{ H \}(1)} (x_H, \vec{p}_H, n, \nu; s_0 )}{d x_H d^2 \vec{p}_H} + \frac{d \Phi_{PP,{\rm{fin}}}^{\{ H \}(1)} (x_H, \vec{p}_H, n, \nu)}{d x_H d^2 \vec{p}_H} \; . \vspace{0.3 cm}
    \label{VirforDiv}
\end{equation}

We recall that the real quark contribution is
\begin{equation}
    \frac{d \Phi_{q q}^{\{H q \}} (z_H, \vec{p}_H, \vec{q})}{d z_H d^2 \vec{p}_H} = \frac{\sqrt{N^2-1}}{16 N (2 \pi)^{D-1}} \frac{g^2 g_H^2}{[(\vec{q}-\vec{p}_H)^2]^2}
\end{equation}
\begin{equation}
\times    \left[ \frac{4 (1-{z}_H) \left[(\vec{q}-\vec{p}_H) \cdot \vec{q} \; \right]^2 + z_H^2 \vec{q}^{\; 2} (\vec{q} - \vec{p}_H)^2}{z_H} \right] \; .
\tag{\ref{QuarkConImpacFin}}
\end{equation}
Using
\[
(\vec{q}-\vec{p}_H) \cdot \vec{q} = \frac{1}{2}\bigl[\vec q^{\:2} + (\vec{q}-\vec{p}_H)^2 - \vec p_{H}^{\:2} \bigl]\;,
\]
this contribution can be projected using the integral $I_1(\gamma_1, \gamma_2, n, \nu)$, defined in~(\ref{FirstMasterIntegral}), and suitably choosing $\gamma_1$ and $\gamma_2$ for the different terms. The projected quark contribution gives 
    \begin{equation*}
    \frac{d \Phi_{q q}^{\{H q \}} (z_H, \vec{p}_H, n, \nu)}{d z_H d^2 \vec{p}_H} = \frac{\sqrt{N^2-1}}{16 N (2 \pi)^{D-1}} g^2 g_H^2 \left \{ \left(z_H +2\, \frac{1-z_H}{z_H}\right) I_1(-1 , 1 , n , \nu) \right.
    \end{equation*}
\begin{equation*}    
    + \frac{1-z_H}{z_H} \bigg [ (\vec p_H^{\;2})^2 I_1(0, 2, n , \nu)  \bigg.
\end{equation*}
\begin{equation*}
     \left. \left. -2 \vec{p}_H^{\; 2} \bigg( I_1(0,1, n, \nu) + I_1(-1 , 2, n, \nu)\bigg)
    +I_1(-2,2, n, \nu)  \right] \right \} \; .
\end{equation*}
If we replace $I_1(\gamma_1, \gamma_2, n, \nu)$ by its explicit expression~(\ref{FirstMasterIntegral}), perform a partial $\epsilon$-expansion and take the convolution with the quark PDFs, we obtain
\begin{equation*}
    \frac{d \Phi_{P P}^{\{H q \}} (x_H, \vec{p}_H, n, \nu)}{d x_H d^2 \vec{p}_H} = \frac{1}{f_g(x_H)} \frac{d \Phi_{PP}^{\{H\}(0)}(x_H, \vec{p}_H, n , \nu)}{d x_H d^2 p_H} \frac{\bar{\alpha}_s}{2 \pi} \left( \frac{\vec{p}_H^{\; 2}}{\mu^2} \right)^{-\epsilon}     
\end{equation*}
\begin{equation*}
   \times \int_{x_H}^1 \frac{d z_H}{z_H} \sum_{a=q \bar{q}} f_a \left( \frac{x_H}{z_H}, \mu_F \right) \left \{ - \frac{1}{\epsilon} C_F \left( \frac{1+(1-z_H)^2}{z_H} \right) + (1-\gamma_E) C_F \left( \frac{1+(1-z_H)^2}{z_H} \right)  \right.
\end{equation*}
\begin{equation*}
    \left. \left. + \frac{C_F}{z_H} \bigg [ 4 (z_H-1)  - \frac{(1+n)(z_H-1)}{(\frac{1}{2} + \frac{n}{2} - i \nu)} - \frac{(1+(1-z_H)^2)}{(-\frac{3}{2} + \frac{n}{2} + i \nu)} - \frac{3+n(z_H-1)+z_H(z_H-3)}{(-\frac{1}{2} + \frac{n}{2} + i \nu)} \right. \right.
\end{equation*}
\begin{equation*}
    \left. \left.- (1 + (1-z_H)^2) \left( H_{-1/2+n/2-i \nu} + \psi (-\frac{3}{2}+\frac{n}{2}+i \nu) \right) \right] \right \} 
\end{equation*}
\begin{equation}
    \equiv \frac{d \Phi_{P P, {\rm{div}}}^{\{H q \}} (x_H, \vec{p}_H, n, \nu)}{d x_H d^2 \vec{p}_H} + \frac{d \Phi_{P P, {\rm{fin}}}^{\{H q \}} (x_H, \vec{p}_H, n, \nu)}{d x_H d^2 \vec{p}_H} \; .
    \label{QuarkforDiv}
\end{equation}
We observe that the $\epsilon$-singularity is cancelled when we combine this object with the counterterm containing $P_{g q} (z_H)$, which appears in~(\ref{PqgforDiv}). We emphasize that the limits $n \to 1$ or $n\to 3$, $\nu \rightarrow 0$ are safe from divergences.

\subsection{Projection of the real gluon contribution}
\label{ssec:projection_gluon}

In this subsection we discuss the projection of all terms appearing in~(\ref{subtractions}), but the last, which was already treated in subsection~\ref{ssec:projection_BFKL}.

We start with the terms labeled ``plus'', which, after $(n,\nu)$-projection, gives
\begin{equation*}
     \frac{d \Phi_{P P}^{\{H g\} {\rm{plus}}} (x_H, \vec{p}_H, n, \nu; s_0)}{d x_H d^2 \vec{p}_H} = - \frac{1}{f_g(x_H)} \frac{d \Phi_{P P}^{\{H \}(0)} (x_H, \vec{p}_H, n, \nu)}{d x_H d^2 \vec{p}_H} \frac{\bar{\alpha}_s}{2 \pi} \left( \frac{\vec{p}_H^{\; 2}}{\mu^2} \right)^{-\epsilon} 
\end{equation*}
\begin{equation*}
    \times \int_{x_H}^1 \frac{d z_H}{z_H} f_g \left( \frac{x_H}{z_H} \right) 2 C_A \frac{z_H}{(1-z_H)_{+}}  \left[ \frac{1}{\epsilon}  + \left( H_{-1/2+n/2 - i \nu} + H_{-1/2+n/2+i \nu} \right) \right]
\end{equation*}
\begin{equation}
    \equiv \frac{d \Phi_{P P, {\rm{div}}}^{\{H g\} {\rm{plus}}} (x_H, \vec{p}_H, n, \nu)}{d x_H d^2 \vec{p}_H} + \frac{d \Phi_{P P, {\rm{fin}}}^{\{H g\} {\rm{plus}}} (x_H, \vec{p}_H, n, \nu)}{d x_H d^2 \vec{p}_H} \; .
    \label{PlusforDiv}
\end{equation}
The divergence in this term is cancelled by the term containing the plus prescription in $P_{gg}(z_H)$, which appears in~(\ref{PggforDiv}).

Then, we project the term labeled ``$(1-x_H)$'' and find
\begin{equation*}
    \frac{d \Phi_{P P}^{\{H g\} (1-x_H)} (x_H, \vec{p}_H, n, \nu)}{d x_H d^2 \vec{p}_H} = \frac{d \Phi_{P P}^{\{H \}(0)} (x_H, \vec{p}_H, n, \nu)}{d x_H d^2 \vec{p}_H} \frac{\bar{\alpha}_s}{2 \pi} \left( \frac{\vec{p}_H^{\; 2}}{\mu^2} \right)^{-\epsilon} 
\end{equation*}
\begin{equation*}
     \times 2 C_A \ln (1-x_H) \left[ \frac{1}{\epsilon} + \left( H_{-1/2 + n/2 - i \nu} + H_{-1/2 + n/2 + i \nu} \right) \right] 
\end{equation*}
\begin{equation}
    \equiv \frac{d \Phi_{P P, \rm{div}}^{\{H g\} (1-x_H)} (x_H, \vec{p}_H, n, \nu)}{d x_H d^2 \vec{p}_H} + \frac{d \Phi_{P P, \rm{fin}}^{\{H g\} (1-x_H)} (x_H, \vec{p}_H, n, \nu)}{d x_H d^2 \vec{p}_H} \; .
    \label{(1-xH)forDiv}
\end{equation}
The divergence cancels with an analogous term present in $d \Phi_{PP}^{{\rm{BFKL}}}$, given in~(\ref{BFKLforDiv}).

We are left with the first term in Eq.~(\ref{subtractions}), {\it i.e.} 
\begin{equation}
\frac{d \tilde{\Phi}^{\{Hg\}}_{PP}(x_H, \vec{p}_H,\vec q;s_0)}{d x_H d^2 p_H}
\end{equation}
\[
=\int_{x_H}^1 \frac{d z_H}{z_H} f_g \left( \frac{x_H}{z_H} \right) \left[ \frac{d \Phi_{gg}^{ \{ H g \}}(z_H, \vec{p}_H, \vec{q})}{d z_H d^2 \vec{p}_H} - z_H \frac{d \Phi_{gg}^{ \{ H g \}}(z_H, \vec{p}_H, \vec{q})}{d z_H d^2 \vec{p}_H} \Bigg |_{z_H=1} \right] \; .
\]
The effect of this subtraction is to remove the first term of the last line in Eq.~(\ref{GluonImp}), which in fact represents the only true divergence for $z_H \rightarrow 1$, and we obtain
\begin{equation*}
    \frac{d \Phi_{gg}^{ \{ H g \}}(z_H, \vec{p}_H, \vec{q})}{d z_H d^2 \vec{p}_H} - z_H \frac{d \Phi_{gg}^{ \{ H g \}}(z_H, \vec{p}_H, \vec{q};s_0)}{d z_H d^2 \vec{p}_H} \Bigg |_{z_H=1}
    = \frac{g^2 g_H^2 C_A}{8 (2 \pi)^{D-1}(1-\epsilon) \sqrt{N^2-1}}
\end{equation*}
\begin{equation*}
    \times \left \{ \frac{2}{z_H (1-z_H)} \left. \left[ 2 z_H^2 + \frac{(1-z_H)z_H m_H^2 (\vec{q} \cdot \vec{r}) [z_H^2 - 2 (1-z_H) \epsilon]+2 z_H^3 (\vec{p}_H \cdot \vec{r}) (\vec{p}_H \cdot \vec{q})}{\vec{r}^{\; 2} \left[ (1-z_H) m_H^2 + \vec{p}_H^{\; 2} \right]} \right. \right.  \right.
\end{equation*}
\begin{equation*}
    - \frac{2 z_H^2 (1-z_H) m_H^2}{\left[ (1-z_H) m_H^2 + \vec{p}_H^{\; 2}  \right]}  -\frac{(1-z_H)z_H m_H^2 (\vec{q} \cdot \vec{r}) [z_H^2 - 2 (1-z_H) \epsilon]+2 z_H^3 (\vec{\Delta} \cdot \vec{r}) (\vec{\Delta} \cdot \vec{q})}{\vec{r}^{\; 2} \left[ (1-z_H) m_H^2 + \vec{\Delta}^{ 2} \right]} 
\end{equation*}
\begin{equation*}
   \left. - \frac{2 z_H^2 (1-z_H) m_H^2}{\left[ (1-z_H) m_H^2 + \vec{\Delta}^{2}  \right]} + \frac{(1-\epsilon) z_H^2 (1-z_H)^2 m_H^4}{2} \left( \frac{1}{\left[ (1-z_H) m_H^2 + \Delta^{2}  \right]} \right. \right.
\end{equation*}
\begin{equation*}
   \left. \left. + \frac{1}{\left[ (1-z_H) m_H^2 + \vec{p}_H^{\; 2}  \right]}  \right)^2 - \frac{2 z_H^2 (\vec{p}_H \cdot \vec{\Delta})^2 - 2 \epsilon (1-z_H)^2 z_H^2 m_H^4}{\left[ (1-z_H) m_H^2 + \vec{p}_H^{\; 2}  \right] \left[ (1-z_H) m_H^2 + \Delta^{2}  \right]} \right]
\end{equation*}
\begin{equation*}
     \left. + \frac{2 \vec{q}^{\; 2}}{\vec{r}^{\; 2}} \left[ z_H (1-z_H) + 2 (1-\epsilon) \frac{(1-z_H)}{z_H} \frac{(\vec{q} \cdot \vec{r})^2}{\vec{q}^{\; 2} \vec{r}^{\; 2}} \right] \right \}
\end{equation*}
\begin{equation}
    \equiv \frac{d \Phi_{gg}^{ \{ H g \}{\rm{coll}}}(z_H, \vec{p}_H, \vec{q})}{d z_H d^2 \vec{p}_H} + \frac{d \Phi_{gg}^{ \{ H g \} (1-z_H)}(z_H, \vec{p}_H, \vec{q})}{d z_H d^2 \vec{p}_H} + \frac{d \Phi_{gg}^{ \{ H g \}{\rm{rest}}}(z_H, \vec{p}_H, \vec{q})}{d z_H d^2 \vec{p}_H} \; .
\end{equation}
In the last equality, we split this term in three contributions: 1) $d \Phi_{g g}^{\{H g \}{\rm coll}}$ contains the pure collinear divergence remained, 2) $d \Phi_{g g}^{\{H g \}(1-z_H)}$ contains terms that taken alone are singular for $z_H \rightarrow 1$, but when combined are safe from divergences, 3) $d \Phi_{g g}^{\{H g \}{\rm rest}}$ is the rest. 

\vspace{0.2 cm}

\textit{Collinear term}

\begin{equation*}
\frac{d \Phi_{g g}^{\{H g \}{\rm{coll}}} (z_H, \vec{p}_H, \vec{q})}{d z_H d^2 \vec{p}_H} = \frac{g^2 g_H^2 C_A}{8 (2 \pi)^{D-1}(1-\epsilon)  \sqrt{N^2-1}} \frac{2 \vec{q}^{\; 2}}{\vec{r}^{\; 2}} 
\end{equation*}
\begin{equation}
\times \left[ z_H (1-z_H) + 2 (1-\epsilon) \frac{(1-z_H)}{z_H} \frac{(\vec{q} \cdot \vec{r})^2}{\vec{q}^{\; 2} \vec{r}^{\; 2}} \right] \; .
\end{equation}
This term can be projected and taken in convolution with the gluon PDF in a way quite analogous to the quark case; we find
\begin{equation*}
    \frac{d \Phi_{P P}^{\{H g \} {\rm{coll}}} (x_H, \vec{p}_H, n, \nu)}{d x_H d^2 \vec{p}_H} \equiv \frac{1}{f_g(x_H)} \frac{d \Phi_{P P}^{\{ H \}(0)}(x_H, \vec{p}_H, n , \nu)}{d x_H d^2 \vec{p}_H} \frac{\bar{\alpha}_s}{2 \pi} \left( \frac{\vec{p}_H^{\; 2}}{\mu^2} \right)^{-\epsilon} \int_{x_H}^1 \frac{d z_H}{z_H} f_g \left( \frac{x_H}{z_H} \right)
\end{equation*}
\begin{equation*}
 \times \left \{ - \frac{1}{\epsilon} 2 \; C_A \left( z_H (1-z_H) + \frac{(1-z_H)}{z_H} \right) - 2 \gamma_E  C_A \left( z_H (1-z_H) + \frac{(1-z_H)}{z_H} \right) \right.
\end{equation*}
\begin{equation*}
    \left. - \frac{2 C_A (1-z_H)}{z_H} \left[ 1 + \gamma_E + \gamma_E z_H^2 -  \frac{1+n}{2 (\frac{1}{2} + \frac{n}{2} - i \nu)} + \frac{1+z_H^2}{\left( -\frac{3}{2} + \frac{n}{2} + i \nu \right)} + \frac{3-n+2 z_H^2}{2(-\frac{1}{2} + \frac{n}{2} + i \nu)} \right. \right.
\end{equation*}
\begin{equation*}
    \left. \left. + (1 + z_H^2) \left( \psi \left(\frac{1}{2}+\frac{n}{2}-i \nu \right)  + \psi \left( -\frac{3}{2}+\frac{n}{2}+i \nu \right) \right) \right] \right \} 
\end{equation*}
\begin{equation}
   \equiv \frac{d \Phi_{P P, {\rm{div}}}^{\{H g \} {\rm{coll}}} (x_H, \vec{p}_H, n, \nu)}{d x_H d^2 \vec{p}_H} + \frac{d \Phi_{P P, {\rm{fin}}}^{\{H g \}{\rm{coll}}} (x_H, \vec{p}_H, n, \nu)}{d x_H d^2 \vec{p}_H} \; .
   \label{CollforDiv}
\end{equation}
We observe that the $\epsilon$-singularity is cancelled by a terms contained in the counterterm with $P_{g g} (z_H)$, which is given in~(\ref{PggforDiv}). As in the quark case, the above expression is safe from divergences in $n=1$ or $n=3$, $\nu=0$. 

\vspace{0.2 cm}

$(1-z_H)$-\textit{term}

\begin{equation*}
 \frac{d \Phi_{g g}^{\{H g \}(1-z_H)} (z_H, \vec{p}_H, \vec{q})}{d z_H d^2 \vec{p}_H} = \frac{g^2 g_H^2 C_A}{4 (2 \pi)^{D-1}(1-\epsilon)  \sqrt{N^2-1}}  \frac{1}{z_H (1-z_H)}\Biggl[ 2 z_H^2 \Biggr.
\end{equation*}
\begin{equation}
    \left. + \frac{2 z_H^3 (\vec{p}_H \cdot \vec{r}) (\vec{p}_H \cdot \vec{q})}{\vec{r}^{\; 2} \left[ (1-z_H) m_H^2 + \vec{p}_H^{\; 2} \right]} 
    -\frac{2 z_H^3 (\vec{\Delta} \cdot \vec{r}) (\vec{\Delta} \cdot \vec{q})}{\vec{r}^{\; 2} \left[ (1-z_H) m_H^2 + \vec{\Delta}^{ 2} \right]} 
    \right.
\end{equation}
\begin{equation*}
    \left. - \frac{2 z_H^2 (\vec{p}_H \cdot \vec{\Delta})^2}{\left[ (1-z_H) m_H^2 + \vec{p}_H^{\; 2}  \right] \left[ (1-z_H) m_H^2 + \Delta^{2}  \right]} \right] \; . 
\end{equation*}
Here, and in the rest of the calculation, we will use the following formula, 
\begin{equation}
    \vec{q} \cdot \vec{p}_H  = (\vec{q}^{\; 2})^{\frac{1}{2}} (\vec{p}_H^{\; 2})^{\frac{1}{2}} \cos( \phi - \phi_H ) = (\vec{q}^{\; 2})^{\frac{1}{2}} (\vec{p}_H^{\; 2})^{\frac{1}{2}} \left( \frac{e^{i(\phi-\phi_H)} + e^{-i(\phi-\phi_H)}}{2} \right)
    \label{angle trick} \; .
\end{equation}
Let us first observe that the term in square bracket gives zero in the limit $z_H \rightarrow 1$, so that there is no singularity in this limit. Now, using~(\ref{angle trick}), we perform the projection and convolution with the gluon PDF and obtain, up to vanishing terms in the $\epsilon\to 0$ limit 
\begin{equation*}
    \frac{d \Phi_{P P}^{\{H g \}(1-z_H)} (x_H, \vec{p}_H, n, \nu)}{d x_H d^2 \vec{p}_H} = \frac{1}{f_g(x_H)} \frac{d \Phi_{PP}^{ \{ H \}(0)} (x_H, \vec{p}_H, n, \nu)}{d x_H d^2 \vec{p}_H} \frac{2 \sqrt{2} C_A}{(\vec{p}_H^{\; 2})^{i \nu - \frac{1}{2}} e^{i n \phi_H}}  \frac{\alpha_s}{2 \pi}  
\end{equation*}
\begin{equation*}
    \times \int_{x_H}^1 \frac{d z_H}{z_H} f_g \left( \frac{x_H}{z_H} \right)  \frac{1}{z_H (1-z_H)} \left \{ \frac{2 z_H^3}{\left[ (1-z_H) m_H^2 + \vec{p}_H^{\; 2} \right]} \left[ \frac{\vec{p}_H^{\; 2}}{4} \left( e^{-2 i \phi_H} I_1(-1,1,n+2, \nu) \right. \right. \right.
\end{equation*}
\begin{equation*}
    \left. \left. \left. + e^{2 i \phi_H} I_1(-1,1,n-2, \nu) + 2 I_1(-1,1,n, \nu) \right) - \frac{(\vec{p}_H^{\; 2})^{\frac{3}{2}}}{2} \left( e^{- i \phi_H} I_1 \left(-\frac{1}{2},1,n+1, \nu \right) \right. \right. \right. 
\end{equation*}
\begin{equation*}
    \left. \left. \left.  + e^{ i \phi_H} I_1 \left( -\frac{1}{2},1,n-1, \nu \right) \right) \right] - \frac{2 z_H^2}{\left[ (1-z_H) m_H^2 + \vec{p}_H^{\; 2} \right]} \bigg[ (\vec{p}_H^{\; 2})^{2} I_3 (0,1,n,\nu)  \right.
\end{equation*}
\begin{equation*}
    \left. + \frac{z_H^2 \vec{p}_H^{\; 2}}{4} \left( e^{-2 i \phi_H} I_3(-1,1,n+2, \nu) + e^{2 i \phi_H} I_3(-1,1,n-2, \nu) + 2 I_3(-1,1,n, \nu) \right) \right. 
\end{equation*}
\begin{equation*}
    \left. - z_H (\vec{p}^{\;2}_H)^{\frac{3}{2}} \left( e^{-i \phi_H} I_3 \left(- \frac{1}{2}, 1, n+1, \nu \right) + e^{i \phi_H} I_3 \left(- \frac{1}{2}, 1, n-1, \nu \right) \right) \right] 
\end{equation*}
\begin{equation*}
    - 2 z_H^3 \left[ \frac{(\vec{p}_H^{\; 2})^{\frac{1}{2}}}{2} \left( e^{-i \phi_H} I_3 \left( -\frac{1}{2}, 1, n+1, \nu \right) + e^{i \phi_H} I_3 \left( -\frac{1}{2}, 1, n-1, \nu \right) \right) - z_H I_3 (-1,1,n, \nu) \right]
\end{equation*}
\begin{equation*}
    -2 z_H^3 (1-z_H) \left[ \frac{(\vec{p}^{\;2}_H)^{\frac{1}{2}}}{2} (1+z_H) \left( e^{-i \phi_H} I_2 \left( -\frac{3}{2},n+1,\nu \right) + e^{i \phi_H} I_2 \left( -\frac{3}{2},n-1,\nu \right) \right) -z_H \right.  
\end{equation*}
\begin{equation*}
   \left. \left. \times I_2 (-2,n,\nu) - \frac{\vec{p}_H^{\; 2}}{4} \left( e^{-2 i\phi_H} I_2 ( -1,n+2,\nu ) + e^{2 i \phi_H} I_2 ( -1,n-2,\nu ) + 2 I_2 (-1, n, \nu) \right) \right] \right \} \; .
   \label{(1-z_H)ForFin1}
\end{equation*}
Setting $\epsilon = 0$ and using the limits~(\ref{First Limit}), (\ref{Second Limit}), it is easy to show that the result is safe from $z_H \rightarrow 1$ divergences. In this result, there are combinations of integrals, which are safe from $\epsilon$-divergences, even if single integrals, taken alone, are divergent. In particular, we note that if we use Eq.~(\ref{FirstMasterIntegral}) for the integrals of the type $I_1$, the first five terms vanish up to ${\cal O}(\epsilon)$. Moreover, if we apply the replacement~(\ref{TrickAsy}), we see that the ``asymptotic'' counterparts of the last six integrals cancel completely and therefore any $I_2$ integral in the previous expression can be replaced by $I_{2, {\rm{reg}}}$. The final form for this contribution is
\begin{equation*}
    \frac{d \Phi_{P P}^{\{H g \}(1-z_H)} (x_H, \vec{p}_H, n, \nu)}{d x_H d^2 \vec{p}_H} = \frac{1}{f_g(x_H)} \frac{d \Phi_{P P}^{ \{ H \}} (x_H, \vec{p}_H, n, \nu)}{d x_H d^2 \vec{p}_H} \frac{2 \sqrt{2} C_A}{(\vec{p}_H^{\; 2})^{i \nu - \frac{1}{2}} e^{i n \phi_H}}  \frac{\alpha_s}{2 \pi}  
\end{equation*}
\begin{equation*}
    \times \int_{x_H}^1 \frac{d z_H}{z_H} f_g \left( \frac{x_H}{z_H} \right)  \frac{1}{z_H (1-z_H)} \left \{ - \frac{2 z_H^2}{\left[ (1-z_H) m_H^2 + \vec{p}_H^{\; 2} \right]} \bigg[ (\vec{p}_H^{\; 2})^{2} I_3 (0,1,n,\nu)  \right.
\end{equation*}
\begin{equation*}
    \left. + \frac{z_H^2 \vec{p}_H^{\; 2}}{4} \left( e^{-2 i \phi_H} I_3(-1,1,n+2, \nu) + e^{2 i \phi_H} I_3(-1,1,n-2, \nu) + 2 I_3(-1,1,n, \nu) \right) \right. 
\end{equation*}
\begin{equation*}
    \left. - z_H (\vec{p}^{\;2}_H)^{\frac{3}{2}} \left( e^{-i \phi_H} I_3 \left(- \frac{1}{2}, 1, n+1, \nu \right) + e^{i \phi_H} I_3 \left(- \frac{1}{2}, 1, n-1, \nu \right) \right) \right] 
\end{equation*}
\begin{equation*}
    - 2 z_H^3 \left[ \frac{(\vec{p}_H^{\; 2})^{\frac{1}{2}}}{2} \left( e^{-i \phi_H} I_3 \left( -\frac{1}{2}, 1, n+1, \nu \right) + e^{i \phi_H} I_3 \left( -\frac{1}{2}, 1, n-1, \nu \right) \right) - z_H I_3 (-1,1,n, \nu) \right]
\end{equation*}
\begin{equation*}
    -2 z_H^3 (1-z_H) \left[ \frac{(\vec{p}^{\;2}_H)^{\frac{1}{2}}}{2} (1+z_H) \left( e^{-i \phi_H} I_{2,\rm{reg}} \left( -\frac{3}{2},n+1,\nu \right) + e^{i \phi_H} I_{2,\rm{reg}} \left( -\frac{3}{2},n-1,\nu \right) \right) \right.  
\end{equation*}
\begin{equation}
\begin{split}
   \left. \left. -z_H I_{2,\rm{reg}} (-2,\nu, n) - \frac{\vec{p}_H^{\; 2}}{4} \left( e^{-2i  \phi_H} I_{2,\rm{reg}} ( -1,n+2,\nu ) \right. \right. \right. \\
   \left. \left. \left. + e^{2i \phi_H} I_{2,\rm{reg}} ( -1,n-2,\nu ) + 2 I_{2,\rm{reg}} (-1, n, \nu) \right) \right. \Bigg] \right. \Bigg \} \;. 
   \label{(1-z_H)ForFin}
\end{split}
\end{equation}

\vspace{0.5 cm} 

\textit{Rest term}

\begin{equation*}
    \frac{d \Phi_{g g}^{\{H g \}{\rm{rest}}} (z_H, \vec{p}_H, \vec{q})}{d z_H d^2 \vec{p}_H} \equiv \frac{g^2 g_H^2 C_A}{4 (2 \pi)^{D-1}(1-\epsilon) \sqrt{N^2-1}} \left \{ \frac{ m_H^2 (\vec{q} \cdot \vec{r}) [z_H^2 - 2 (1-z_H) \epsilon]}{\vec{r}^{\; 2} \left[ (1-z_H) m_H^2 + \vec{p}_H^{\; 2} \right]} \right.
\end{equation*}
\begin{equation*}
    \left.  + \frac{(1-\epsilon) z_H (1-z_H) m_H^4}{2} \left( \frac{1}{\left[ (1-z_H) m_H^2 + \Delta^{2}  \right]} + \frac{1}{\left[ (1-z_H) m_H^2 + \vec{p}_H^{\; 2}  \right]}  \right)^2 \right.
\end{equation*}
\begin{equation*}
  + \frac{ 2 \epsilon (1-z_H) z_H m_H^4}{\left[ (1-z_H) m_H^2 + \vec{p}_H^{\; 2}  \right] \left[ (1-z_H) m_H^2 + \Delta^{2}  \right]} - 2 z_H m_H^2 \left( \frac{1}{\left[ (1-z_H) m_H^2 + \vec{p}_H^{\; 2}  \right]} \right.
\end{equation*}
\begin{equation}
    \left. \left.  + \frac{1}{\left[ (1-z_H) m_H^2 + \vec{\Delta}^{2}  \right]} \right) -\frac{m_H^2 (\vec{q} \cdot \vec{r}) [z_H^2 - 2 (1-z_H) \epsilon]}{\vec{r}^{\; 2} \left[ (1-z_H) m_H^2 + \vec{\Delta}^{ 2} \right]} \right \} \; .
\end{equation}
We perform the projection and convolution with gluon PDF and, finally, we find, up to terms ${\cal O}(\epsilon)$,
\begin{equation*}
    \frac{d \Phi_{P P}^{\{H g \}{\rm{rest}}} (x_H, \vec{p}_H, n, \nu)}{d x_H d^2 \vec{p}_H} =  \frac{1}{f_g(x_H)} \frac{d \Phi_{P P}^{ \{H \}(0)} (x_H, \vec{p}_H, n, \nu)}{d x_H d^2 \vec{p}_H} \frac{ \sqrt{2} C_A}{(\vec{p}_H^{\; 2})^{i \nu - \frac{1}{2}} e^{i n \phi_H}}  \frac{\alpha_s}{2 \pi}  \int_{x_H}^1 \frac{d z_H}{z_H} f_g \left( \frac{x_H}{z_H} \right)
\end{equation*}
\begin{equation*}
   \times \left \{ \frac{2 m_H^2 \left[ z_H^2 - 2 (1-z_H) \epsilon \right]}{[(1-z_H) m_H^2 + \vec{p}_H^{\; 2}]} \left[ I_1(-1,1,n,\nu) - \frac{(\vec{p}_H^{\; 2})^{\frac{1}{2}}}{2} \left( e^{-i \phi_H} I_1 \left( -\frac{1}{2}, 1, n+1, \nu \right) \right. \right. \right.  
\end{equation*}
\begin{equation*}
 \left. \left. + e^{i \phi_H} I_1 \left( -\frac{1}{2}, 1, n-1, \nu \right) \right) \right] - 2 m_H^2 [z_H^2 - 2(1-z_H) \epsilon] \left. \Bigg[ I_2 \left( -1, n, \nu \right) \right. 
\end{equation*}
\begin{equation*}
- \frac{(\vec{p}_H^{\; 2})^{\frac{1}{2}}}{2} \left( e^{-i \phi_H} I_2 \left(-\frac{1}{2}, n+1, \nu \right) + e^{i \phi_H} I_2 \left( -\frac{1}{2},n-1,\nu \right) \right) \Bigg] - 4 z_H m_H^2 I_3 (0,1,n, \nu) 
\end{equation*}
\begin{equation*}
    + (1-\epsilon) z_H (1-z_H) m_H^4 \left[ I_3(0,2,n, \nu) + \left(2+\frac{4\epsilon}{1-\epsilon}\right) \frac{I_3 (0,1,n,\nu)}{[(1-z_H) m_H^2 + \vec{p}_H^{\; 2}]}\right] .
    \label{RestForFin1}
\end{equation*}
Using the explicit result for the integral $I_1$ and applying the same ideas as befor, after 
the $\epsilon$-expansion we end up with 
\begin{equation*}
    \frac{d \Phi_{P P}^{\{H g \}{\rm{rest}}} (x_H, \vec{p}_H, n, \nu)}{d x_H d^2 \vec{p}_H} =  \frac{1}{f_g(x_H)} \frac{d \Phi_{P P}^{ \{H \}(0)} (x_H, \vec{p}_H, n, \nu)}{d x_H d^2 \vec{p}_H} \frac{ \sqrt{2} C_A}{(\vec{p}_H^{\; 2})^{i \nu - \frac{1}{2}} e^{i n \phi_H}}  \frac{\alpha_s}{2 \pi}  \int_{x_H}^1 \frac{d z_H}{z_H} f_g \left( \frac{x_H}{z_H} \right)
\end{equation*}
\begin{equation*}
   \times \left \{ \frac{2 m_H^2 z_H^2 }{[(1-z_H) m_H^2 + \vec{p}_H^{\; 2}]} \left[ \frac{(\vec{p}_H^{\; 2})^{i \nu -\frac{1}{2} - \epsilon} e^{i n \phi_H}}{2 \sqrt{2}} \left( \frac{1}{\left( \frac{1}{2} + \frac{n}{2} - i \nu \right)} - \frac{1}{\left(-\frac{1}{2} + \frac{n}{2} + i \nu \right)} \right) \right] \right.  
\end{equation*}
\begin{equation*}
  - 2 m_H^2 z_H^2 \Bigg[ I_{2, \rm{reg}} \left( -1, n, \nu \right) \Biggr.
  \end{equation*}
  \begin{equation*}
      \left. - \frac{(\vec{p}_H^{\; 2})^{\frac{1}{2}}}{2} \left( e^{-i \phi_H} I_{2, \rm{reg}} \left( -\frac{1}{2}, n+1, \nu \right) + e^{i \phi_H} I_{2, \rm{reg}} \left( -\frac{1}{2}, n-1,\nu \right) \right) \Bigg] \right.
\end{equation*}
\begin{equation}
- 4 z_H m_H^2 I_3 (0,1,n, \nu) + z_H (1-z_H) m_H^4 \left[ I_3(0,2,n, \nu) + \frac{2 I_3 (0,1,n,\nu)}{[(1-z_H) m_H^2 + \vec{p}_H^{\; 2}]} \right] .
    \label{RestForFin}
\end{equation}

\subsection{Final result}
\label{ssec:projection_final}

In this subsection we present the final result. First of all, we observe that if we sum all the $\epsilon$-divergent contributions in Eqs.~(\ref{CoupliforDiv})-(\ref{PggforDiv}),~(\ref{BFKLforDiv}),~(\ref{VirforDiv}),~(\ref{QuarkforDiv})-(\ref{(1-xH)forDiv}),~(\ref{CollforDiv}), we get that they cancel completely. The finite parts in the same equations, together with the contributions in Eqs.~(\ref{(1-z_H)ForFin}), (\ref{RestForFin}) represent the final result. Setting $\epsilon=0$, we can cast the final result in the sum of three terms,
\begin{equation*}
    \frac{d \Phi_{P P}^{\{H \}, \rm{NLO}} (x_H, \vec{p}_H, n, \nu)}{d x_H d^2 \vec{p}_H} = \frac{d \Phi_{P P, 1}^{\{H \}, \rm{NLO}} (x_H, \vec{p}_H, n, \nu; s_0)}{d x_H d^2 \vec{p}_H} 
\end{equation*}
\begin{equation}
    + \frac{d \Phi_{P P, 2}^{\{H \}, \rm{NLO}} (x_H, \vec{p}_H, n, \nu)}{d x_H d^2 \vec{p}_H} + \frac{d \Phi_{P P, 3}^{\{H \}, \rm{NLO}} (x_H, \vec{p}_H, n, \nu)}{d x_H d^2 \vec{p}_H} \; ,
\end{equation}
where the first term is given by the sum of all contributions purely proportional to $f_g(x_H)$, {\it i.e.}
\begin{equation*}
    \frac{d \Phi_{P P, 1}^{\{H \}, \rm{NLO}} (x_H, \vec{p}_H, n, \nu; s_0)}{d x_H d^2 \vec{p}_H} \equiv \frac{d \Phi^{{\rm{BFKL}}}_{PP, {\rm{fin}}}(x_H, \vec{p}_H,n,\nu;s_0)}{d x_H d^2 p_H} 
\end{equation*}
\begin{equation}
    + \frac{d \Phi_{PP, {\rm{fin}}}^{ \{H \}} (x_H, \vec{p}_H, n, \nu)}{d x_H d^2 \vec{p}_H} \bigg|_{{\rm{coupling \; c.t.}}} + \frac{d \Phi_{PP,{\rm{fin}}}^{\{ H \}(1)} (x_H, \vec{p}_H, n, \nu)}{d x_H d^2 \vec{p}_H} \; ;
\end{equation}
the second is the sum of all contributions that are taken in convolution with $\sum_a f_a(x_H/z_H)$, {\it i.e.}
\begin{equation*}
    \frac{d \Phi_{P P, 2}^{\{H \}, \rm{NLO}} (x_H, \vec{p}_H, n, \nu; s_0)}{d x_H d^2 \vec{p}_H} \equiv \frac{d \Phi_{PP,{\rm{fin}}}^{ \{H \}}(x_H, \vec{p}_H, n, \nu)}{d x_H d^2 \vec{p}_H}\bigg|_{{\rm{P_{qg} \; c.t.}}} + \frac{d \Phi_{P P, {\rm{fin}}}^{\{H q \}} (x_H, \vec{p}_H, n, \nu)}{d x_H d^2 \vec{p}_H} \; ;
\end{equation*}
the third is the sum of all contributions that are taken in convolution with $f_g(x_H/z_H)$, {\it i.e.}
\begin{equation*}
    \frac{d \Phi_{P P, 3}^{\{H \}, \rm{NLO}} (x_H, \vec{p}_H, n, \nu; s_0)}{d x_H d^2 \vec{p}_H} \equiv \frac{d \Phi_{PP,{\rm{fin}}}^{ \{H \}}(x_H, \vec{p}_H, n, \nu)}{d x_H d^2 \vec{p}_H} \bigg|_{{\rm{P_{gg} \; c.t.}}} + \frac{d \Phi_{P P, {\rm{fin}}}^{\{H g\} {\rm{plus}}} (x_H, \vec{p}_H, n, \nu)}{d x_H d^2 \vec{p}_H} 
\end{equation*}
\begin{equation*}
    + \frac{d \Phi_{P P, {\rm{fin}}}^{\{H g\}(1-x_H)} (x_H, \vec{p}_H, n, \nu)}{d x_H d^2 \vec{p}_H} + \frac{d \Phi_{P P, {\rm{fin}}}^{\{H g\}{\rm{coll}}} (x_H, \vec{p}_H, n, \nu)}{d x_H d^2 \vec{p}_H} 
\end{equation*}
\begin{equation}
    + \frac{d \Phi_{P P}^{\{H g\}(1-z_H)} (x_H, \vec{p}_H, n, \nu)}{d x_H d^2 \vec{p}_H}
    + \frac{d \Phi_{P P}^{\{H g\}{\rm{rest}}} (x_H, \vec{p}_H, n, \nu)}{d x_H d^2 \vec{p}_H} \;.
\end{equation}

\section{Summary and outlook}
\label{sec:conclusions}

We calculated the full NLO correction to the impact factor for the production of a Higgs boson emitted in the forward rapidity region. 
Its analytic expression was obtained both in the momentum and in the Mellin representation. The latter is particularly relevant to clearly observe a complete cancellation of NLO singularities, and it is useful for future numeric studies.
We relied on the large top-mass limit approximation, thus we employed the gluon-Higgs effective field theory.
We have found that the Gribov trick (see Eq.~(\ref{Gribov})) cannot be applied to both the $t$-channel gluon legs connected to the effective gluon-Higgs vertex.
This prevents the use of the technique outlined in Refs.~\cite{Fadin:1999qc,Fadin:2001dc} to simplify calculations.
Formal studies on the generalization of that procedure to non-QCD vertices are underway~\cite{GribovNext}.
As a first step forward towards phenomenology, we plan to extend the analysis on high-energy resummed distributions for the inclusive Higgs-plus-jet hadroproduction done in Ref.~\cite{Celiberto:2020tmb}, having in mind a twofold goal.
On one hand, our forward-Higgs NLO impact factor is a key ingredient to conduct a precise BFKL versus fixed-order study, previously made in a partial NLL approximation.
On the other hand, it represents a core element to hunt for further and stronger signals that Higgs emissions in forward directions act as \emph{natural stabilizers} of the high-energy resummation.
Then, we will investigate inclusive single-forward emissions of Higgs bosons as direct probe channels of the proton structure in the small-$x$ regime.
These studies are relevant to constrain the proton UGD, thus complementing the information already gathered from forward light vector-meson~\cite{Bolognino:2018rhb,Celiberto:2019slj,Bolognino:2021niq} and Drell--Yan dilepton~\cite{Brzeminski:2016lwh,Celiberto:2018muu} analyses, as well as to explore the common ground between the high-energy and the \emph{transverse-momentum-dependent} (TMD) factorization (see Ref.~\cite{Collins:2011zzd} for a review).
The TMD formalism provides us with a tomographic three-dimensional description of the nucleon that accounts for effects stemming from the transverse motion and polarization of partons and from their interplay with the spin of the parent hadron.
Notably, the density of linearly-polarized gluons can give rise to spin effects even in collisions of unpolarized protons. They are known in the context of hadron-structure studies as Boer--Mulders effects. They were first observed for quarks~\cite{Boer:1997nt,Bacchetta:2008xw,Barone:2008tn}.
A striking outcome of model studies of twist-2 gluon TMDs~\cite{Bacchetta:2020vty,Celiberto:2021zww} it that, in the small-$x$ limit, the weight of these gluon Boer--Mulders effects grows with the same power as the unpolarized ones. Since the linearly-polarized gluon density can be easily accessed {\it via} inclusive detection of Higgs bosons in proton collisions (see, \emph{e.g.}, Refs.~\cite{Boer:2011kf,Boer:2014lka,Echevarria:2015uaa,Gutierrez-Reyes:2019rug}), analyses of such final states in forward-rapidity regions represent unique venues where to unveil the interplay between small-$x$ and TMD dynamics.
From a more formal perspective, a natural development of this work is the calculation of the Higgs impact factor {\it via} gluon fusion in the central-rapidity region.
At the LO level it takes the form of a doubly off-shell coefficient function~\cite{Pasechnik:2006du}, where the top-quark loop connects two incoming Reggeized gluons with the outgoing scalar-boson line.
Moving to NLO, the computation of contribution due to real emissions is not complicated, while technical issues are expected to emerge from the extraction of the vertex at 1-loop accuracy, due
to the presence of an additional scale in the off-shellness of the incoming Reggeon (which replaces the incoming real gluon).
Once calculated, this doubly off-shell impact factor will be employed in the description of the central inclusive Higgs production in a pure high-energy factorization scheme, given as a $\kappa_T$-convolution between two UGDs describing the incoming protons and the aforementioned $g^*g^*H$ vertex.
We plan to address this point in the near future, having in mind the immediate phenomenological application of comparing high-energy resummed predictions for transverse momentum and rapidity distributions with corresponding ones calculated {\it via} a small-$x$ improved collinear framework~\cite{Ball:2013bra,Bonvini:2018ixe,Bonvini:2018iwt,Ball:2017otu}.

\acknowledgments

We thank Saad Nabeebaccus, Samuel Wallon, Lech Szymanowski and Victor Fadin for useful discussions. We are also indebted to Maxim Nefedov for many discussions and for helpful hint in the use of \textsc{FeynCalc}.

F.G.C. acknowledges support from the INFN/NINPHA project and thanks the Universit\`a degli Studi di Pavia for the warm hospitality. M.F., M.M.A.M. and A.P. acknowledge support from the INFN/QFT@COLLIDERS project. The work of D.I. was carried out within the framework of the
state contract of the Sobolev Institute of Mathematics (Project No. 0314-2019-0021).

\appendix

\section{Feynman rules of the gluon-Higgs effective field theory and some common definitions}
\label{AppendixA}

  \begin{figure}
  \begin{center}
  \includegraphics[scale=0.65]{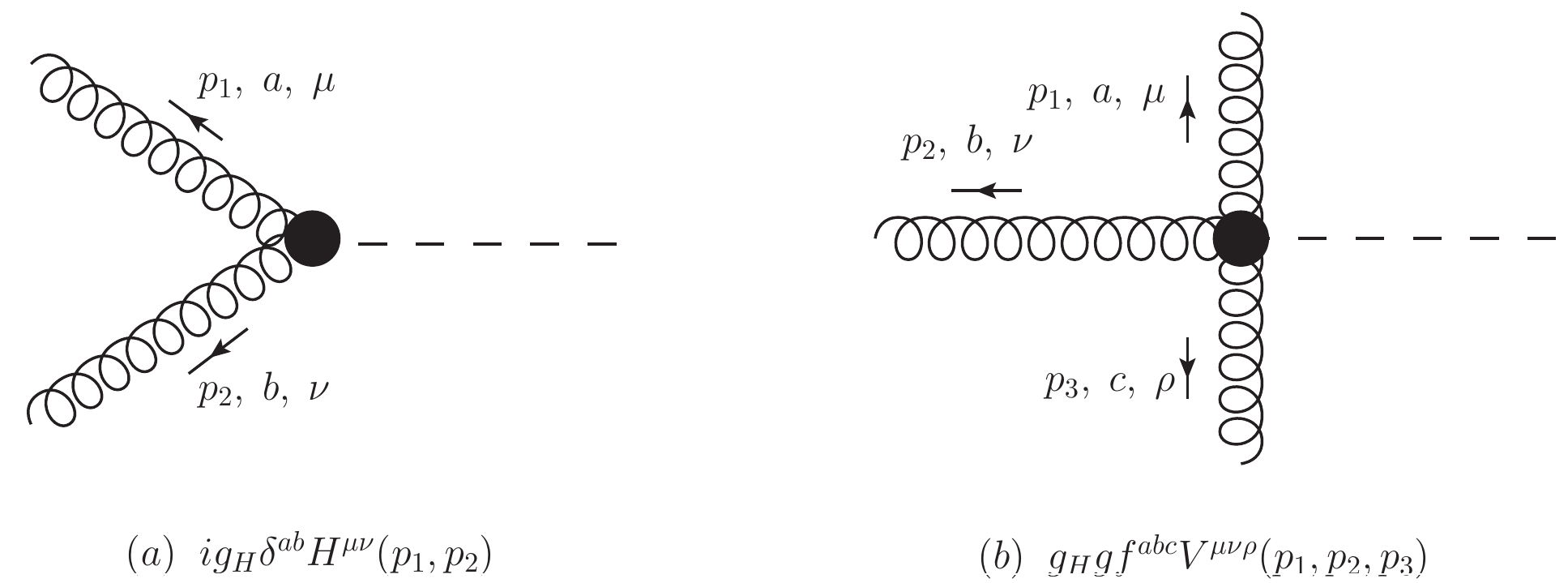}
  \end{center}
  \caption{Feynman rule for the (a) $ggH$ vertex and (b) $gggH$ vertex.}
  \label{HiggsDiagrams}
  \end{figure}

In this Appendix we give more details about the gluon-Higgs effective theory. The Feynman rules associated to the Lagrangian~(\ref{EffLagrangia}) and used in this work are shown in Fig.~\ref{HiggsDiagrams}. The tensor structures appearing in Fig.~\ref{HiggsDiagrams} are
\begin{equation}
    H^{\mu \nu} (p_1,p_2) = g^{\mu \nu} (p_1 \cdot p_2) - p_1^{\nu} p_2^{\mu} \;,
\label{ggH}
\end{equation}
\begin{equation}
    V^{\mu \nu \rho} (p_1, p_2, p_3) = (p_1 - p_2)^{\rho} g^{\mu \nu} + (p_2 - p_3)^{\mu} g^{\nu \rho} + (p_3 - p_1)^{\nu} g^{\rho \mu} \;.
\label{gggH}
\end{equation}
It is important to note that, as in QCD, in using Feynman rules for this theory, symmetry factors must be taken into account correctly. In particular, the first and second diagram in the Fig.~\ref{VirtualDiagrams} require a symmetry factor $S=1/2$. \\
We also define here some useful functions used in the main text:
\begin{equation}
    \psi (z) = \frac{\Gamma'(z)}{\Gamma(z)} \; , \hspace{0.5cm} \psi' (z) = \frac{d}{dz} \psi (z) \; ,\hspace{0.5 cm} H_{n} = \gamma_E + \psi (n+1) \; ,
\end{equation}
\begin{equation}
    {\rm{Li}}_2 (z) = - \int_0^z \frac{\ln (1-t)}{t} dt = - \int_0^1 \frac{\ln (1-zt)}{t} dt \; ,
\end{equation}
\begin{equation}
\, {_2}F_1(a,b,c,z) = \frac{1}{\mathcal{B}(b,c-b)} \int_0^1 dx \, x^{b-1} (1-x)^{c-b-1} (1-zx)^{-a} \;  \hspace{0.2 cm} \rm{for} \; \; \; \Re e{ \{c \} } > \Re e{ \{b \} } > 0 \; ,
\label{hypergeometric}
\end{equation}
where $\mathcal{B}$ is the Euler beta function and $z$ is not a real number such that it is greater than or equal to 1. 
It is useful to show the behaviour for $z \rightarrow 1^{-}$ of the hypergeometric function:
\begin{itemize}
    \item If $\Re e{ \{c \} } > \Re e{ \{ a+b \}}$, then
    \begin{equation}
    {_2}F_1(a,b,c,1) = \frac{\Gamma (c) \Gamma (c-a-b)}{\Gamma (c-a) \Gamma (c-b)} \; .
    \end{equation}
    \item If $\Re e{ \{c \} } = \Re e{ \{ a+b \}}$, then
    \begin{equation}
        \lim_{z \rightarrow 1^{-}} \frac{{_2}F_1(a,b,a+b,z)}{- \ln (1-z)} = \frac{\Gamma (a+b)}{\Gamma(a) \Gamma(b)} \hspace{0.6 cm} {\rm{for}} \; \;  c=a+b \; ,
    \end{equation}
    and
    \begin{equation}
        \lim_{z \rightarrow 1^{-}} (1-z)^{a+b-c} \left( {_2}F_1(a,b,c,z) - \frac{\Gamma (c) \Gamma (c-a-b)}{\Gamma (c-a) \Gamma (c-b)} \right) =  \frac{\Gamma (c) \Gamma (a+b-c)}{\Gamma (a) \Gamma (b)} \hspace{0.6 cm}
    \end{equation}
    for $c \neq a+b$.
    \item If $\Re e{ \{c \} } < \Re e{ \{ a+b \}}$, then
    \begin{equation}
        \lim_{z \rightarrow 1^{-}} \frac{{_2}F_1(a,b,c,z)}{(1-z)^{c-a-b}} = \frac{\Gamma (c) \Gamma (a+b-c)}{\Gamma (a) \Gamma (b)}
        \label{LimitHyper}
    \end{equation}
\end{itemize}

\section{Integrals for the virtual corrections}
\label{AppendixB}

We give here definition and result of Feynman integrals that appear in the calculation of the virtual corrections. The integrals $B_0$ and $C_0$ can be found also in the Appendix~A of Ref.~\cite{Fadin:2000yp}, the integral $D_0$ in Ref.~\cite{Bern:1993kr} (see also
Ref.~\cite{Zanderighi:2008on}).\footnote{In our notation the subscript $0$ means that all propagators appearing in the integral are massless. The arguments in the definitions of $B_0$ and $C_0$ integrals below represent the squared of the external momenta on which they depend. $D_0$, in addition to the four squared of the external momenta, depends also on the two independent Mandelstam variables typical of a $2 \rightarrow 2$ process.}
Concerning integrals with two denominators, we have
\begin{equation}
    B_0(-\vec{q}^{\; 2}) = \int \frac{d^D k}{i(2 \pi)^D} \frac{1}{k^2 (k+q)^2} = - \frac{1}{(4 \pi)^{2-\epsilon}} \frac{\Gamma
   (1+\epsilon) \Gamma^2 (-\epsilon)}{2 (1-2 \epsilon ) \Gamma (-2 \epsilon )} (\vec{q}^{\; 2})^{-\epsilon } \; ,
\end{equation}
\begin{equation}
    B_0(m_H^2) = \int \frac{d^D k}{i(2 \pi)^D} \frac{1}{k^2 (k+p_H)^2} = - \frac{1}{(4 \pi)^{2-\epsilon}} \frac{\Gamma
   (1+\epsilon) \Gamma^2 (-\epsilon)}{2 (1-2 \epsilon ) \Gamma (-2 \epsilon )} (-m_H^2)^{-\epsilon } \; .
\end{equation}
Independent integrals with three denominators appearing in the computation are
\begin{equation*}
   C_0(m_H^2,0, -\vec{q}^{\; 2}) = C_0(0, -\vec{q}^{\; 2}, m_H^2) = \int \frac{d^D k}{i(2 \pi)^D} \frac{1}{k^2 (k+q)^2 (k+p_H)^2} 
\end{equation*}
\begin{equation}
    = \frac{1}{(4 \pi)^{2-\epsilon}} \frac{ \Gamma (1+\epsilon) \Gamma^2(-\epsilon)}{2 \Gamma (-2 \epsilon )} \frac{1}{\epsilon} \frac{\left((\vec{q}^{\;2})^{-\epsilon }-\left(-m_H^2\right)^{-\epsilon }\right)}{m_H^2 + \vec{q}^{\; 2}} \; , 
\end{equation}
\begin{equation}
   C_0(0,0, -\vec{q}^{\; 2}) = \int \frac{d^D k}{i(2 \pi)^D} \frac{1}{k^2 (k+q)^2 (k+k_2)^2} = \frac{1}{(4 \pi)^{2-\epsilon}} \frac{ \Gamma (1+\epsilon) \Gamma^2(-\epsilon)}{2 \Gamma (-2 \epsilon )} \frac{1}{\epsilon} (\vec{q}^{\;2})^{-\epsilon-1} \; , 
\end{equation}
\begin{equation*}
   C_0(m_H^2,0,s) = \int \frac{d^D k}{i(2 \pi)^D} \frac{1}{k^2 (k+p_H)^2 (k+k_1+k_2)^2}
\end{equation*}
\begin{equation}
    \simeq -\frac{1}{(4 \pi)^{2-\epsilon}} \frac{ \Gamma (1+\epsilon) \Gamma^2(-\epsilon)}{2 \Gamma (-2 \epsilon )} \frac{1}{\epsilon} \frac{\left((-s)^{-\epsilon }-\left(-m_H^2\right)^{-\epsilon }\right)}{s} \; ,
\end{equation}
\begin{equation}
   C_0(0,0,s) = \int \frac{d^D k}{i(2 \pi)^D} \frac{1}{k^2 (k-k_1)^2 (k+k_2)^2}  = \frac{1}{(4 \pi)^{2-\epsilon}} \frac{ \Gamma (1+\epsilon) \Gamma^2(-\epsilon)}{2 \Gamma (-2 \epsilon )} \frac{1}{\epsilon} (-s)^{-\epsilon-1} \; , 
\end{equation}
There is only one relevant four-denominator integral,
\begin{equation*}
   D_0(m_H^2,0,0,0;-\vec{q}^{\; 2},s) = \int \frac{d^D k}{i(2 \pi)^D} \frac{1}{k^2 (k+p_H)^2 (k+q)^2 (k-k_2')^2} 
\end{equation*}
\begin{equation}   
= \frac{\Gamma (1+\epsilon) \Gamma^2(-\epsilon)}{(4 \pi)^{2-\epsilon} \vec{q}^{\; 2} s} 
\frac{1}{\epsilon \Gamma (-2 \epsilon )} \left[(\vec{q}^{\; 2})^{-\epsilon }+(-s)^{-\epsilon}-\left(-m_H^2\right)^{-\epsilon}
\right.\end{equation}
\begin{equation*}
\left. -\epsilon^2 \left(\text{Li}_2\left(1+\frac{m_H^2}{\vec{q}^{\; 2}}\right)+\frac{1}{2} \ln
   ^2\left(-\frac{\vec{q}^{\; 2}}{s}\right)+\frac{\pi ^2}{3}\right) \right]
   + {\cal O}(\epsilon) \; ,
\end{equation*}
where $k_2'=k_2-q$ and $(p_H+k_2')^2=(k_1+k_2)^2=s$.
The definition of the integrals $C_0(m_H^2,0,-s)$, $C_0(0,0,-s)$, $D_0(m_H^2,0,0,0;-\vec{q}^{\; 2},-s)$ can be obtained by exchanging $k_2$ and $-k_2'$; the result for these integrals is obviously obtained through the $s \rightarrow -s$ transformation. We observe that what said above is valid in the Regge limit; it allows us to neglect $m_H^2$ and $\vec{q}^{\; 2}$ when these appear summed to $s$.

\section{Master Integrals for the projection}
\label{AppendixC}

For the projection onto the eigenfunctions of the BFKL kernel we just need the following integrals:
\[
I_1(\gamma_1, \gamma_2, n, \nu) = \int \frac{d^{2-2 \epsilon} \vec{q}}{\pi \sqrt{2}} (\vec{q}^{\; 2})^{i \nu - \frac{3}{2}} e^{i n \phi} (\vec{q}^{\; 2})^{-\gamma_1} \left[ \left( \vec{q} - \vec{p}_H \right)^2 \right]^{-\gamma_2} \;,
\]    
\[
I_2(\gamma_1, n, \nu) = \int \frac{d^{2-2 \epsilon} \vec{q}}{\pi \sqrt{2}} (\vec{q}^{\; 2})^{i \nu - \frac{3}{2}} e^{i n \phi} (\vec{q}^{\; 2})^{-\gamma_1} \frac{1}{ \left[ (\vec{q}-\vec{p}_H)^2 \right] \left[ (1-z_H) m_H^2 + \left( \vec{p}_H - z_H \vec{q} \right)^2 \right]} \; , 
\]
\[
I_3(\gamma_1, \gamma_2, n, \nu) = \int \frac{d^{2-2 \epsilon} \vec{q}}{\pi \sqrt{2}} (\vec{q}^{\; 2})^{i \nu - \frac{3}{2}} e^{i n \phi} (\vec{q}^{\; 2})^{-\gamma_1} \left[(1-z_H) m_H^2 + \left( \vec{p}_H - z_H \vec{q} \right)^2 \right]^{-\gamma_2}\;.    
\]
We note that
\begin{equation}
    \lim_{z_H \rightarrow 1} I_2 (\gamma_1, n, \nu) = I_1 (\gamma_1, 2, n, \nu) \; ,
\label{First Limit}
\end{equation}
\begin{equation}
    \lim_{z_H \rightarrow 1} I_3 (\gamma_1, \gamma_2, n, \nu) = I_1 (\gamma_1, \gamma_2, n, \nu) \; .
\label{Second Limit}
\end{equation}

We show here the explicit calculation of the most complicated one, $I_2$ and just quote the result for the other two. First, we introduce a vector $l \equiv (1, i)$ and write
\begin{equation}
e^{in\phi}=\left(\cos{\phi}+i\sin{\phi}\right)^n=\left(\frac{q_x+i q_y}{|\vec{q}\;|}\right)^n=\frac{(\vec{q}\cdot\vec{l}\;)^n}{\left(\vec{q}^{\;2}\right)^{\frac{n}{2}}} \; .
\label{ExpTrick}
\end{equation}
Then, after Feynman parameterization and the shift
\begin{equation}
    \vec{q} \rightarrow \vec{q} + \left( x + \frac{y}{z_H} \right) \vec{p}_H
\end{equation}
one finds
\begin{equation*}
    I_2(\gamma_1, n, \nu) = \frac{1}{z_H^2} \frac{\Gamma \left(\frac{7}{2} + \gamma_1 + \frac{n}{2} -i \nu \right)}{\Gamma \left(\frac{3}{2} + \gamma_1 + \frac{n}{2} -i \nu \right)} \int \frac{d^{2-2 \epsilon} \vec{q}}{\pi \sqrt{2}} \int_0^1 dx \int_0^{1-x} dy (1-x-y)^{\frac{1}{2} + \gamma_1 + \frac{n}{2} -i \nu} 
\end{equation*}
\begin{equation}
    \frac{ \left[ \vec{q} \cdot \vec{l} + \left( x + \frac{y}{z_H} \right) \vec{p}_H \cdot \vec{l} \; \right]^n}{\left[ \vec{q}^{\; 2} - \left( x + \frac{y}{z_H} \right)^2 \vec{p}_H^{\; 2} + x \vec{p}_H^{\; 2} + y \left( \Tilde{m}_H^2 + \frac{\vec{p}_H^{\; 2}}{z_H^2} \right) \right]^{\frac{7}{2} + \gamma_1 + \frac{n}{2} -i \nu}} \; ,
\end{equation}
where $\Tilde{m}_H^2 = \frac{(1-z_H)m_H^2}{z_H^2}$. Now, we expand the binomial in the numerator as
\begin{equation}
\left[ \vec{q} \cdot \vec{l} + \left( x + \frac{y}{z_H} \right) \vec{p}_H \cdot \vec{l} \; \right]^n = \sum_{j=0}^{n} \binom{n}{j}  (\vec{q} \cdot \vec{l})^{j} \left( x + \frac{y}{z_H} \right)^{n-j}  (\vec{p}_H \cdot \vec{l})^{n-j} 
\end{equation}
and observe that only the term with $j=0$ survives. At this point, we are able to obtain the simple form,
\[
    I_2(\gamma_1, n, \nu) = \frac{(\vec{p}_H^{\; 2})^{\frac{n}{2}} e^{i n \phi_H}}{z_H^2} \frac{\Gamma \left(\frac{7}{2} + \gamma_1 + \frac{n}{2} -i \nu \right)}{\Gamma \left(\frac{3}{2} + \gamma_1 + \frac{n}{2} -i \nu \right)} 
\]
\begin{equation}
    \times \int_0^1 dx \int_0^{1-x} dy (1-x-y)^{\frac{1}{2} + \gamma_1 + \frac{n}{2} -i \nu} \left( x + \frac{y}{z_H} \right)^n \int \frac{d^{2-2 \epsilon} \vec{q}}{\pi \sqrt{2}} \frac{1}{ \left[ \vec{q}^{\; 2} + L \right]^{\frac{7}{2} + \gamma_1 + \frac{n}{2} -i \nu}} \; ,
\end{equation}
where
\begin{equation}
    L = x \vec{p}_H^{\; 2} + y \left( \Tilde{m}_H^2 + \frac{\vec{p}_H^{\; 2}}{z_H^2} \right) - \left( x + \frac{y}{z_H} \right)^2 \vec{p}_H^{\; 2} \; .
\end{equation}
Finally, we can integrate over $\vec{q}$ and then make the change of variables
\begin{equation}
    x = \lambda \Delta \; , \hspace{0.5 cm} y = \lambda (1-\Delta) \; ,
\end{equation}
to obtain
\begin{equation*}
    I_2(\gamma_1, n, \nu) = \frac{(\vec{p}_H^{\; 2})^{\frac{n}{2}} e^{i n \phi_H}}{z_H^2 \sqrt{2} \pi^{\epsilon}} \frac{\Gamma \left(\frac{5}{2} + \gamma_1 + \frac{n}{2} -i \nu + \epsilon \right)}{\Gamma \left(\frac{3}{2} + \gamma_1 + \frac{n}{2} -i \nu \right)} \int_0^1 d \Delta \left( \Delta + \frac{(1-\Delta)}{z_H} \right)^n \; ,
\end{equation*}
\begin{equation*}
    \left[ \left( \Delta + \frac{(1-\Delta)}{z_H^2} \right) \vec{p}_H^{\; 2} + (1-\Delta) \Tilde{m}_H^{\; 2} \right]^{-\frac{5}{2} - \gamma_1 - \frac{n}{2} +i \nu-\epsilon} \int_0^1 d \lambda \;  \lambda^{-\frac{3}{2} - \gamma_1 + \frac{n}{2} +i \nu-\epsilon}
\end{equation*}
\begin{equation}
   \times (1-\lambda)^{\frac{1}{2} + \gamma_1 + \frac{n}{2} -i \nu} (1-\lambda \zeta)^{-\frac{5}{2} - \gamma_1 -\frac{n}{2} +i \nu -\epsilon} \; ,
\end{equation}
where
\begin{equation}
    \zeta = \frac{\left(\Delta + \frac{(1-\Delta)}{z_H} \right)^2 \vec{p}_H^{\; 2}}{\left[ \left(\Delta + \frac{(1-\Delta)}{z_H^2} \right) \vec{p}_H^{\; 2} + \frac{(1-\Delta)(1-z_H) m_H^2}{z_H^2}  \right]} \; .
\end{equation}
It is easy to see that the integral over $\lambda$ gives a representation of the hypergeometric function. By using Eq.~(\ref{hypergeometric}) we arrive at the final form
\begin{equation*}
    I_2(\gamma_1, n, \nu) = \frac{(\vec{p}_H^{\; 2})^{\frac{n}{2}} e^{i n \phi_H}}{z_H^2 \sqrt{2} \pi^{\epsilon}} \left[ \frac{\Gamma \left( \frac{5}{2} + \gamma_1 + \frac{n}{2} - i \nu + \epsilon \right) \Gamma \left( -\frac{1}{2} - \gamma_1 + \frac{n}{2} + i \nu - \epsilon \right)}{\Gamma \left( 1+n-\epsilon \right)} \right] 
\end{equation*}
\begin{equation*}
    \times \int_0^1 d \Delta \left( \Delta + \frac{(1-\Delta)}{z_H} \right)^n  \left[ \left( \Delta + \frac{(1-\Delta)}{z_H^2} \right) \vec{p}_H^{\; 2} + \frac{(1-\Delta)(1-z_H) m_H^2}{z_H^2} \right]^{- \frac{5}{2} - \gamma_1 + i \nu - \frac{n}{2} - \epsilon } 
\end{equation*}
\begin{equation}
   \times \; _2 F_1 \left( - \frac{1}{2} - \gamma_1 + \frac{n}{2} + i \nu - \epsilon , \frac{5}{2} + \gamma_1 - i \nu + \frac{n}{2}+\epsilon, 1+n-\epsilon, \zeta \right) \; .
\end{equation}
It is very important to note that this integral is indeed divergent for every $\epsilon > -1$ (see the asymptotic behaviors of the hypergeometric function in Appendix~\ref{AppendixA}). It is important to extract the divergent contribution of the integral $I_2$ in the form of a pole in $\epsilon$. For this purpose, we take the limit $\Delta \to 1$ in the integrand of $I_2$. Using Eq.~(\ref{LimitHyper}), we have, up to terms ${\cal O}(\epsilon)$,
\begin{equation*}
  I_{2, \rm{as}}(\gamma_1, n, \nu) =  \frac{(\vec{p}_H^{\; 2})^{-\frac{3}{2}-\gamma_1+i \nu -\epsilon} e^{i n \phi_H} \Gamma(1+\epsilon)}{(1-z_H) \sqrt{2} \pi^{\epsilon}} \frac{1}{\left(m_H^2 + (1-z_H) \vec{p}_H^{\; 2} \right)} \int_{0}^{1} d \Delta (1-\Delta)^{-\epsilon-1}
\end{equation*}
\begin{equation}
     = - \frac{1}{\epsilon} \frac{(\vec{p}_H^{\; 2})^{-\frac{3}{2}-\gamma_1+i \nu -\epsilon} e^{i n \phi_H} \Gamma(1+\epsilon)}{(1-z_H) \sqrt{2} \pi^{\epsilon}} \frac{1}{\left(m_H^2 + (1-z_H) \vec{p}_H^{\; 2} \right)} 
\end{equation}
This simpler integral has the same singular behavior of $I_2$, therefore the combination
    \begin{equation*}
    I_{2, {\rm{reg}}} \equiv I_2 - I_{2,{\rm{as}}} = \frac{(\vec{p}_H^{\; 2})^{\frac{n}{2}} e^{i n \phi_H}}{z_H^2 \sqrt{2}} \left[ \frac{\Gamma \left( \frac{5}{2} + \gamma_1 + \frac{n}{2} - i \nu \right) \Gamma \left( -\frac{1}{2} - \gamma_1 + \frac{n}{2} + i \nu \right)}{\Gamma \left( 1+n \right)} \right] 
\end{equation*}
\begin{equation*}
    \times \int_0^1 d \Delta \left( \Delta + \frac{(1-\Delta)}{z_H} \right)^n  \left[ \left( \Delta + \frac{(1-\Delta)}{z_H^2} \right) \vec{p}_H^{\; 2} + \frac{(1-\Delta)(1-z_H) m_H^2}{z_H^2} \right]^{- \frac{5}{2} - \gamma_1 + i \nu - \frac{n}{2}} 
\end{equation*}
\begin{equation*}
   \times \left \{ _2 F_1 \left( - \frac{1}{2} - \gamma_1 + \frac{n}{2} + i \nu , \frac{5}{2} + \gamma_1 - i \nu + \frac{n}{2}, 1+n, \zeta \right) - \frac{z_H^2 (\vec{p}_H^{\; 2})^{-\frac{3}{2}-\gamma_1 - \frac{n}{2} +i \nu}}{\left(m_H^2 + (1-z_H) \vec{p}_H^{\; 2} \right)} \right. \; 
\end{equation*}
\begin{equation}
    \left. \times \frac{\Gamma(1+n)}{\Gamma(\frac{5}{2} + \gamma_1 + \frac{n}{2} - i \nu) \Gamma (-\frac{1}{2} - \gamma_1 + \frac{n}{2} + i \nu)} \frac{1}{(1-\Delta)(1-z_H)} \right \} \; ,
    \label{TrickAsy}
\end{equation}
is not divergent and has a finite $\epsilon\to 0$ limit.

The other two integrals can be computed by using the same technique:
\begin{equation*}
    I_1(\gamma_1, \gamma_2, n, \nu) = \int \frac{d^{2-2 \epsilon} \vec{q}}{\pi \sqrt{2}} (\vec{q}^{\; 2})^{i \nu - \frac{3}{2}} e^{i n \phi} (\vec{q}^{\; 2})^{-\gamma_1} \left[ \left( \vec{q} - \vec{p}_H \right)^2 \right]^{-\gamma_2} = \frac{(\vec{p}_H^{\; 2})^{-\frac{1}{2} + i \nu - \epsilon - \gamma_1 - \gamma_2} e^{i n \phi_H}}{\sqrt{2} \pi^{\epsilon}} 
\end{equation*}
\begin{equation}
    \times \left[ \frac{\Gamma \left( \frac{1}{2} + \gamma_1 + \gamma_2 + \frac{n}{2} - i \nu + \epsilon \right) \Gamma \left( -\frac{1}{2} - \gamma_1 + \frac{n}{2} + i \nu - \epsilon \right) \Gamma \left( 1- \gamma_2 -\epsilon \right)}{\Gamma \left( \frac{3}{2} + \gamma_1 + \frac{n}{2} - i \nu \right) \Gamma \left( \frac{1}{2} - \gamma_1 - \gamma_2 + \frac{n}{2} + i \nu - 2 \epsilon \right)\Gamma \left( \gamma_2 \right)} \right] \; ,
    \label{FirstMasterIntegral}
\end{equation}
note that for $\gamma_2 \geq 1$ and integer, we cannot put $\epsilon=0$ because the integral is divergent;
\vspace{0.2 cm}
\begin{equation*}
    I_3(\gamma_1, \gamma_2, n, \nu) = \int \frac{d^{2-2 \epsilon} \vec{q}}{\pi \sqrt{2}} (\vec{q}^{\; 2})^{i \nu - \frac{3}{2}} e^{i n \phi} (\vec{q}^{\; 2})^{-\gamma_1} \left[(1-z_H) m_H^2 + \left( \vec{p}_H - z_H \vec{q} \right)^2 \right]^{-\gamma_2}
\end{equation*}
\begin{equation}
    = \frac{(\vec{p}_H^{\; 2})^{\frac{n}{2}} e^{i n \phi_H}}{ (z_H^2)^{\gamma_2 + \frac{n}{2}} \sqrt{2} \pi^{\epsilon}} \left( \frac{\vec{p}_H^{\; 2}}{z_H^2} + \frac{(1-z_H) m_H^2}{z_H^2} \right)^{- \frac{1}{2} - \gamma_1 -\gamma_2 - \frac{n}{2} + i \nu - \epsilon } 
\end{equation}
\begin{equation*}
    \times \left[ \frac{\Gamma \left( \frac{1}{2} + \gamma_1 + \gamma_2 + \frac{n}{2} - i \nu + \epsilon \right) \Gamma (- \frac{1}{2} - \gamma_1 + \frac{n}{2} + i \nu - \epsilon) \Gamma ( \frac{3}{2} + \frac{n}{2} + \gamma_1 - i \nu)}{\Gamma \left( \frac{3}{2} + \gamma_1 + \frac{n}{2} - i \nu \right) \Gamma \left( \gamma_2 \right) \Gamma (1+n - \epsilon)} \right] 
\end{equation*}
\begin{equation*}
    \times \; _2 F_1 \left( - \frac{1}{2} - \gamma_1 + \frac{n}{2} + i \nu - \epsilon , \frac{1}{2} + \gamma_1 + \gamma_2 + \frac{n}{2} - i \nu + \epsilon, 1 + n -\epsilon, \xi \right) \; ,
\end{equation*}
where
\begin{equation}
    \xi = \frac{1}{1 + \frac{(1-z_H) m_H^2}{\vec{p}_H^{\; 2}}} \, .
\end{equation}
\bibliographystyle{apsrev}
\bibliography{references}

\end{document}